\documentclass{aa}

\usepackage{subcaption}

\usepackage{natbib,twoopt}
\usepackage{hyperref} %% to avoid \citeads line fills
\bibpunct{(}{)}{;}{a}{}{,} 

\usepackage{siunitx}

\usepackage{longtable}
\usepackage{multirow}
\usepackage{flushend}

\usepackage{booktabs}
\usepackage{graphicx}
\usepackage{txfonts}
\newcommand{\gaia}{\emph{Gaia}\xspace}

\begin{document} 

\title{Radio emission as a stellar activity indicator}

%\subtitle{I. Overviewing the $\kappa$-mechanism}

\author{T.~W.~H.~Yiu\inst{1,2}\fnmsep\thanks{Email: yiu@astron.nl}, H.~K.~Vedantham\inst{1,2}, J.~R.~Callingham\inst{1,3}, M.~N.~G\"{u}nther\inst{4}}

\institute{
ASTRON, The Netherlands Institute for Radio Astronomy, Oude Hoogeveensedijk 4, Dwingeloo, 7991 PD, The Netherlands
\and
Kapteyn Astronomical Institute, University of Groningen, PO Box 72, 97200 AB, Groningen, The Netherlands
\and
Leiden Observatory, Leiden University, PO Box 9513, 2300 RA Leiden, The Netherlands
\and
European Space Agency (ESA), European Space Research and Technology Centre (ESTEC), Keplerlaan 1, NL-2201 AZ Noordwijk, The Netherlands
}

\date{Received 4 August 2023 / Accepted 7 December 2023}

% \abstract{}{}{}{}{}
% 5 {} token are mandatory
 
\abstract{Radio observations of stars trace the plasma conditions and magnetic field properties of stellar magnetospheres and coronae. Depending on the plasma conditions at the emitter site, radio emission in the metre- and decimetre-wave bands is generated via different mechanisms such as gyrosynchrotron, electron cyclotron maser instability, and plasma radiation processes. The ongoing LOFAR Two-metre Sky Survey (LoTSS) and VLA Sky Survey (VLASS) are currently the most sensitive wide-field radio sky surveys ever conducted. Because these surveys are untargeted, they provide an opportunity to study the statistical properties of the radio-emitting stellar population in an unbiased manner. Here, we perform an untargeted search for stellar radio sources down to sub-mJy level using these radio surveys. We find that the population of radio-emitting stellar systems is mainly composed of two distinct categories: chromospherically active stellar (CAS) systems and M dwarfs. We also seek to identify signatures of a gradual transition within the M-dwarf population from chromospheric/coronal acceleration close to the stellar surface similar to that observed on the Sun, to magnetospheric acceleration occurring far from the stellar surface similar to that observed on Jupiter. We determine that radio detectability evolves with spectral type, and we identify a transition in radio detectability around spectral type M4, where stars become fully convective. Furthermore, we compare the radio detectability versus spectra type with X-ray and optical flare (observed by TESS) incidence statistics. We find that the radio efficiency of X-ray/optical flares, which is the fraction of flare energy channelled into radio-emitting charges, increases with spectral type. These results motivate us to conjecture that the emergence of large-scale magnetic fields in CAS systems and later M dwarfs leads to an increase in radio efficiency. 
}

   \keywords{
             Radio continuum: stars --
             Stars: flare --
             Stars: statistics --
             Radiation mechanisms: non-thermal --
             Catalogs
               }

\maketitle
\section{Introduction}
\label{sec:intro}
Radio observations are excellent tracers of supra-thermal plasma in the coronae of stars, magnetic activity of ultracool dwarfs (UCDs; including brown dwarfs), and the mangetospheres of planets. This is because plasma oscillation and electric charges moving in a magnetic field emit at radio frequencies, implying a radio detection of a stellar system is sensitive to the magnetic fields and plasma dynamics of the source. Therefore, radio observations probe phenomena on these objects that are otherwise not easily accessible in other wavelengths \citep{bastian_bible, gudel_bible}. For example, radio emission from exoplanets provides the only known technique to directly measure the planet's magnetic field strength and topology, which is an important aspect of exo-habitability (e.g. \citealp{driscoll2015, airapetian2020, lopez-morales2011, khodachenko2007, see2014, griessmeier2015}). Additionally, plasma emission from stars can be used to constrain fundamental coronal plasma parameters such as plasma density and the radial structure and dynamics of coronal plasma \citep{rscvn-toet2021}, giving insights on the impact of stellar plasma on exoplanet atmospheres.

Depending on the plasma parameters at the emitter, radio emission in the metre- and decametre-wave bands can be generated via a variety of mechanisms: from incoherent ones such as free-free, gyromagnetic, and gyrosynchrotron processes \citep{incoherent-nindos2020} to coherent ones such as plasma radiation \citep{dulk1985} and electron cyclotron maser instability (ECMI; \citealp{wu1979,melrose1982,melrose1984,treumann2006}).
For the Sun, there exist two principal radio components: the unpolarised or weakly polarised broadband solar flares at gigahertz (GHz) frequencies, and the narrowband coherent solar bursts at megahertz (MHz) frequencies. 
The unpolarised radio emission from solar flares have generally been attributed to gyrosynchrotron radiation \citep{gyro-alissandrakis1986}, produced by mildly relativistic electrons ($\gamma \lesssim 2-3$) either arising from the Maxwellian tail of a thermal distribution \citep{dulk1985}, or from a non-thermal energy distribution produced via flare acceleration and/or magnetic reconnection \citep{parker_nanoflare1988, sweet_reconnection1958a, sweet_reconnection1958b, from_khz_to_thz}. Many stellar radio flares on isolated main-sequence stars (e.g. \citealp{bastian1990, gb_gb1993}) as well as very close binaries (e.g. \citealp{drake1989,mutel1985}) are often observed to be broadband and unpolarised. On the contrary, coherent solar bursts with narrow instantaneous bandwidths are much more luminous and more circularly polarised compared to gyrosynchrotron flares. These bursts are typically produced by plasma radiation processes \citep{dulk1985, melrose_book, melrose2009, melrose2017, gudel_bible}, thus emitting at plasma frequencies or its harmonics that are typically below the GHz regime.
Regardless of the mechanism, all solar radio-emitting charges are expected to be created via chromospheric/coronal acceleration close to the stellar surface due to magnetic reconnection, driven by immense shear in the solar radiative-convective interface (known as tachocline; \citealp{tachocline_spiegel1992}).
We refer to this broad class of acceleration mechanisms collectively as the ``Sun-like radio engine''.

In comparison, radio observations of Jupiter show that Jupiter can generate luminous auroral radio bursts with strong circular polarisation (reaching 100\%; \citealp{aurora_zarka1998, zarka_planet2004, zarka_bible}) at the ambient cyclotron frequency and its harmonics through the ECMI mechanism. Therefore, Jupiter's plasma dynamics are fundamentally different to the Sun due to its fast rotation rate, strong magnetic field, and low-density magnetosphere. Jovian radio emission is thus magneto-rotation-driven instead, with the magnetospheric acceleration occurring far from the planetary surface as a result of a breakdown of co-rotation between the magnetic field and plasma. An alternate acceleration mechanism known as centrifugal breakout, which also operates far from the surface, has been suggested for radio emission from hot magnetic stars with dipole-dominated magnetopspheres \citep{shultz2020, owocki2020, owocki2022}. We shall use the umbrella term ``Jupiter-like engine'' for acceleration in a magnetosphere far from the surface of the central body.

In contrast to the Sun, the coherent radio emission of UCDs are generally interpreted as electron cyclotron maser (ECM) emission powered by Jupiter-like engines (e.g. \citealp{hallinan2007, hallinan2008, kao_LT2016, turnpenney2017}), since their highly circular-polarised radio periodic bursts share many properties with that of Jupiter and other magnetised solar-system planets \citep{zarka_bible}. 
The change in radio emission mechanism between solar-like stars and UCDs suggests that there exists a transition in radio-emitting engine of stellar systems -- from the Sun-like paradigm to the Jupiter-like paradigm -- somewhere in the M-dwarf sequence. Finding this transition is vital in understanding the radio detectability of stars and the predominant emission mechanism of radio-bright stellar systems. It may also shed light on the effect of the transition from partially convective to fully convective interiors (i.e. the tachocline no longer exists) on stellar activity (around spectral type M4; \citealp{chabrier2000, dorman1989, m4-stassun2011, williams2014}).

There have been previous studies that investigate and characterise the radio properties of stellar systems,  from the hottest stars to the coldest brown dwarfs, in order to determine trends (see e.g. \citealp{bieging1989, gb_gb1993, williams2014, kao_LT2016, villadsen2019}).
However, all of the above studies employed some target selection strategy, and thus could be implicitly biased towards special cases of stars and UCDs that may not reflect the general stellar population as a whole. 
One benefit of searching for radio-bright stellar systems in wide-field radio surveys is therefore to bypass these biases. However, this method is not without some major challenges.
\cite{helfand1999} performed the first large unbiased study of radio stars using the Faint Images of the Radio Sky at Twenty-Centimeters (FIRST; \citealp{FIRST-survey}), and concluded that FIRST astrometry alone was insufficient to avoid crippling chance-coincidence rates, especially when good proper-motion information was unavailable. \cite{kimball2009} conducted a search for radio stars by combining FIRST with optical data from the Sloan Digital Sky Survey (SDSS), but they estimated that $108 \pm 13$ out of 112 candidate radio stars were contamination, i.e. optical-faint radio quasars in chance alignment with a foreground star. Thus, they concluded a radio survey with a much higher sensitivity and resolution compared to FIRST was necessary to confidently identify radio stars. \cite{antonova2013} conducted a 4.9-GHz volume-limited radio survey of 32 nearby UCDs with spectral types M7--T8, but failed to detect any radio emission from them. They thus highlighted that low rotation rates and long-term variability of these targets might be the cause of the non-detections. \cite{mwa-lenc2018} conducted the first low-frequency circular-polarised all-sky survey using the Murchison Widefield Array (MWA; \citealp{tingay2013}). However, with its relatively low sensitivity of $\approx 3 \si{mJy~beam^{-1}}$, the survey only detected pulsars which were previously known to be radio-bright.

Studies with a newer crop of radio surveys have been more successful. \cite{pritchard2021} presented results from a circular polarisation survey for radio stars in the Rapid Australian Square Kilometre Array Pathfinder (ASKAP) Continuum Survey (RACS; \citealp{racs_mcconnell2020}), and they identified M dwarfs, close binaries, young stellar objects, and chemically peculiar A- and B-type stars in the radio sample.
\cite{joe_nature} detected coherent radio emission of M dwarfs from an untargeted flux-limited low-frequency survey using LOw-Frequency ARray (LOFAR; \citealp{lofar2013}). They highlighted the benefit of utilising circular polarisation (Stokes V) information to suppress false associations \citep{false-positive}. More recently, \cite{driessen2023} identified radio stellar sources from multiple radio surveys using their proper motions provided by \gaia Data Release 3 (DR3; \citealp{gaia-dr3}), and showed that the transient nature of radio stellar sources makes surveys with multiple epochs more preferable for searching them.

Having a wide-field radio survey with high sensitivity and angular resolution is therefore paramount in order to identify radio-bright stellar systems from the radio data, lest non-detections and extragalactic contamination dominate the sample. The ongoing LOFAR Two-metre Sky Survey (LoTSS; \citealp{lotss-white-paper2017}) and the Karl G. Jansky Very Large Array (VLA; \citealp{vla-perley2011}) Sky Survey (VLASS; \citealp{vlass-white-paper}) are some of the most sensitive wide-field radio sky surveys ever conducted. Because the surveys are untargeted, they provide an opportunity to study the statistical properties of the radio-emitting stellar population in an unbiased manner. Moreover, \cite{v-lotss} recently published V-LoTSS: the circular-polarised component of LoTSS. This provides yet another advantage as the Stokes V radio sky is orders of magnitudes sparser than its Stokes I counterpart \citep{v-lotss}.
In this paper, we present our latest efforts to identify and study the radio emission from stellar systems in these surveys as a way to understand if and how radio activity evolves with spectral type, and see whether a transition in acceleration mechanism (the so-called radio-emitting engine) exists, specifically from the aforementioned Sun-like to Jupiter-like engine as one goes from earlier to later spectral types. 
Additionally, to understand how stellar activity impacts radio luminosity and thus detectability, we also compare the radio detection rate in our sample to canonical activity indicators such as optical and X-ray flares.

The paper is structured as follows: in Sect.~\ref{sec:radio-optical-crossmatch-and-flare}, we present details of the radio survey catalogues, \gaia Catalogue of Nearby Stars (GCNS) and the TESS/X-ray flare statistics used in our analysis. Sect.~\ref{sec:methods} contains the description of our methodology, including the radio $\times$ GCNS crossmatching and TESS-flare-rate debiasing procedure. We present our results in Sect.~\ref{sec:results_discussion} and conclude in Sect.~\ref{sec:conclusion}. This paper contains many acronyms, and thus we include a table of acronyms (Table~\ref{tab:acronyms}) in the Appendix for clarity.

\begin{table*}
    \caption{Parameters of the radio surveys utilised for identifying radio-bright stellar systems.}
    \centering
    \begin{tabular}{c|c|c|c|c|c|c|c}
        \toprule
        Survey name & $\nu_{\rm survey}$ & Resolution & Median rms & AA & $t_{\rm exp}$ & Source density & $\Omega_{\rm survey}$ \\
        \midrule
        LoTSS-DR2 & 120--168\,MHz & $6\arcsec$ & $83~\si{\mu Jy/beam}$ & $0.2\arcsec$ & 3451 hrs & $780~\si{sources/deg}^{2}$ & $5634~\si{deg}^{2}$\\ 
        V-LoTSS & 120--168\,MHz & 20\arcsec & $140~\si{\mu Jy/beam}$ & 0.2\arcsec & 3451 hrs & $0.012~\si{sources/deg}^{2}$ & $5634~\si{deg}^{2}$\\
        VLASS (per epoch) & 2--4\,GHz & $2.5\arcsec$ & $130~\si{\mu Jy/beam}$ & $1\arcsec$ & 1833 hrs & $100~\si{sources/deg}^{2}$ & $33885~\si{deg}^{2}$\\
        % VLASS Epoch 2 & &\\
        % GCNS
        \bottomrule
    \end{tabular}
    \tablefoot{The symbols $\nu_{\rm survey}$, AA, $t_{\rm exp}$, and $\Omega_{\rm survey}$ represent radio survey frequency coverage, absolute astrometric accuracy, exposure time, and the sky coverage respectively. Note that the final uncertainty in the radio position must also factor the astrometric precision that depends on the signal-to-noise ratio (SNR) of the detection. For example, LoTSS source with $<$1\,mJy has an average astrometric precision of $0.5\arcsec$ \citep{false-positive}. The $\Omega_{\rm survey}$ of V-LoTSS is slightly larger since \cite{v-lotss} searched for sources further down the primary beam of single LoTSS pointing, where the noise would have been higher than if the data was mosaicked.}
    \label{tab:surveys}
\end{table*}

\section{Datasets}
\label{sec:radio-optical-crossmatch-and-flare}
\subsection{Radio sky surveys}
\label{sec:radio-surveys}
In order to draw conclusions regarding the radio-bright stellar population as a whole, we begin by crossmatching different radio catalogues to optical positions of known stars in the solar neighbourhood.
The following section details the radio sky surveys used in this work. A summary of the important parameters of these surveys can be found in Table~\ref{tab:surveys}.

\subsubsection{LOFAR Two-metre Sky Survey}
\label{sec:lotss}
The LOFAR Two-metre Sky Survey (LoTSS; \citealp{lotss-white-paper2017}) is an ongoing low-frequency (120--168\,MHz) radio survey of the Northern Sky. Each LoTSS pointing is observed for 8 hours, reaching a median $1\sigma$ rms noise of around $83\,\mu$Jy/beam.
Here we use the LoTSS radio catalogue from the second data release (DR2; \citealp{lotss-dr2}) covering 27\% of the Northern Sky (5634 square degrees) containing around 4 million Stokes I sources.
The unprecedented depth of LoTSS allows us to reveal a low-frequency radio stellar and substellar population never-before-seen as previous searches below $\sim$300\,MHz generally lacked the required sub-mJy sensitivity necessary to detect the general population \citep{joe_nature, firstBD-vedantham2020, j1019-vedantham2023}.

However, using the LoTSS Stokes I catalogue alone to identify radio stellar system candidates is not straightforward due to the potentially crippling rate of chance-coincidence associations (i.e. the so called false-positive matches; more details in Sect.~\ref{sec:false-positive}). To circumvent this, we also include the available circular polarisation (Stokes V) information of LOFAR-detected stellar sources, described in the following section.

\subsubsection{Circular-polarised sky of LoTSS}
\label{sec:v-lotss}
The source density of the Stokes V radio sky is $>5$ orders of magnitude lower than the Stokes I radio sky because most radio sources are extragalactic objects powered by the synchrotron mechanism \citep{begelman1984-synchrotron}, and therefore do not have a significant degree of circular polarisation \citep{cp-extragalactic_rayner2000, cp-extragalactic-beckert2002}. As anticipated, only 1 extragalactic source (an active galactic nucleus) has a detectable circular polarisation ($\approx 1\%$) in LoTSS-DR2 \citep{v-lotss}.
In rare cases where extragalactic radio sources are circularly polarised, the known radio-emission mechanisms of these objects would only allow them to have at most $\approx 1\%$ circular polarised fraction \citep{cp-extragalatic_saikia1988, cp-extragalatic_valtaoja1984, cp-extragalactic_wardle2003}, which matches with observations (e.g. \citealp{cp-agn_macquart2003, cp-agn_agudo2022}).

On the other hand, the circular polarisation of some stellar radio sources are known to be able to reach $\approx 100 \%$ (e.g. \citealp{hallinan2006, hallinan2007, hallinan2008, williams-berger2015}) owing to plasma radiation processes and ECMI \citep{dulk1985, radio-vedantham2021}.
\cite{v-lotss} presented V-LoTSS, which consists of Stokes V maps of LoTSS-DR2 with median $140~\si{\mu Jy} \si{beam}^{-1}$ and a resolution of $20\arcsec$.
As we expect the radio emission of stellar systems to be time variable, an overall mosaic of the fields may wash out genuine detections as these sources may not be emitting in Stokes V in an adjoining field that was observed on a different date. Therefore, using V-LoTSS is again advantageous since unlike LoTSS-DR2, V-LoTSS performs the Stokes V search on individual LoTSS pointings rather than on a mosaicked image.

\subsubsection{VLA Sky Survey}
\label{sec:vlass}
As the aim of this paper is to determine how much radio emission is indicative of stellar activity, we are also interested in the radio population that has a higher frequency than the LOFAR band, as different radio frequencies probe different radio emission mechanisms \citep{gudel_bible}. 
For example, a flaring star can emit intense gyrosynchrotron radiation which typically falls under the decimetre-wave regime \citep{incoherent-nindos2020}, a frequency range observable by the VLA. However, such a star might not have detectable emission in the LOFAR band due to self-absorption and brightness temperature limitations: the brightness temperature of gyrosynchrotron emissions cannot exceed inverse Compton limit of $T_B \sim 10^{12}$\,K \citep{inverse_compton}. Since $T_B \propto \nu^{-2}$, it can be trivially shown that gyrosynchrotron cannot produce emission in the LOFAR band detectable with the LoTSS sensitivity. 
Conversely, stellar systems detected in the metre-wave regime are much more likely to be due to plasma radiation processes and/or ECMI \citep{radio-vedantham2021}. These stars may have a large-scale magnetic field of few hundred gauss which can produce strong coherent ECM radiation that peaks at LOFAR band but decays sharply at VLA band.

Therefore, we also utilise the VLA Sky Survey (VLASS; \citealp{vlass-white-paper}) to select radio-bright stellar systems, and see how different the VLA-detected stellar population is when compared to the LOFAR-detected population.
VLASS is an ongoing multi-epoch S-band (2--4\,GHz) continuum radio survey covering the whole sky visible to the VLA (i.e. $\delta > -40^{\circ}$; 33885 square degrees) and aims to produce images with high angular resolution ($\approx 2.5\arcsec$) and $1\sigma$ rms noise of $\approx 130 \si{\mu Jy}$.
Three epochs of observation (with two sub-epochs each) are planned and currently the first two epochs (VLASS1 \& VLASS2) have been fully observed and processed.
We use the latest release of VLASS which contains Epochs 1 and 2 `Quick Look' (QL) Catalogues\footnote{\url{https://cirada.ca/catalogues}}, each of them containing around 3 million sources.
We note that currently the VLASS data releases do not include Stokes V information.

\subsection{\gaia Catalogue of Nearby Stars}
\label{sec:gcns}
To identify radio-bright stellar systems, we crossmatch the aforementioned radio catalogues with \gaia, the largest available star catalogue with detailed astrometric information.
Previously, \cite{joe_nature} examined the LoTSS Stokes V maps for $\geq 4 \sigma$ sources and crossmatched those detections to sources in the \gaia Data Release 2 (DR2) to search for radio counterparts to stellar systems. However, the \gaia DR2 catalogue becomes significantly incomplete for spectral types later than $\sim$M7 \citep{gaia-incomplete_kiman2019}.
Therefore, in this work, we instead use the \gaia Catalogue of Nearby Stars (GCNS; \citealp{gcns}). The GCNS is a catalogue of well-characterised objects within 100\,pc of the Sun from the \gaia Early Data Release 3 (EDR3; \citealp{gaia-edr3}). 
The GCNS contains 331312 objects, 40234 of which are within 50\,pc. The fact that the catalogue is volume-complete for all objects earlier than M8 at the nominal $G = 20.7$ magnitude limit\footnote{Note that \gaia (and consequently GCNS) avoids the very brightest stars such as Alpha Centauri.} of \gaia \citep{gcns} is essential as our aim is to see whether radio detection rates of stellar systems evolves with spectral type across the stellar main sequence.

\subsection{Flare statistics}
\label{sec:flare-stat}
As we aim to understand the correlation between the radio emission of stars and stellar activity, statistics of stellar flares occurrence are also needed in order to analyse if a higher flaring activity implies a higher chance of radio detection. The motivation for searching such a correlation is that flares may be the essential in accelerating the radio-emitting electrons.
To do this, we use the study by \cite{tess_flare} for optical stellar flares from the stars observed by NASA's Transiting Exoplanet Survey Satellite (TESS; \citealp{tess}), and by \cite{x-ray_flare_johnstone2021} for X-ray stellar flares statistics using the NEXXUS database \citep{nexxus-schmitt2004}.

\subsubsection{TESS flares}
\label{sec:tess_flare}
TESS is an all-sky survey mission designed to discover exoplanets orbiting bright nearby main-sequence stars using transit photometry. As TESS has observed numerous stars (continuously for days, some even months) and measured their light curve in order to find exoplanet transits, TESS data also contains valuable information on the incidence of stellar flares. \cite{tess_flare} performed a study of optical stellar flares for the 24809 stars observed with 2-minute cadence during the first two months of the TESS mission. Most importantly, they studied the flare rate and energy as a function of stellar type and rotation period.

As seen in Figs. 4 and 5 reported by \cite{tess_flare}, flares are most commonly detected on M dwarfs, especially on mid to late spectral type (beyond M4), where more than 40\% of the stars showing observable flares lie. Stars of spectral type earlier than M-type have significantly lower observable flare occurrences of less than 10\%. This general picture is also consistent with past findings from e.g. Kepler \citep{davenport2016, vandoorsselaere2017} and MEarth \citep{mondrik2019} catalogues of stellar flares: stars of later spectral types (especially fast-rotating, young M dwarfs) are the most likely to flare, and that their flare amplitude is independent of the rotation period (cf. \citealp{maehara2012}).

Although \cite{tess_flare} only present findings derived from the first two months (i.e. Sectors 1 and 2) of TESS data, here we use the data from the first 2 years (Sectors 1--26) of TESS mission instead (G\"{u}nther et al., in prep.). Such a database increases the number of flaring stars detected by a factor of $\approx 13$, greatly improving the flare statistics. The data for the TESS short-cadence targets from Sectors 1--26 is publicly available\footnote{\url{https://github.com/MNGuenther/tess_infos}}. 

In addition, for our analysis in Sect.~\ref{sec:tess_flare_cdf}, we now also account for the average TESS flare rate $\nu_{\rm TESS}$ (i.e. number of stellar flares per unit time) in each spectral-type bin, instead of merely considering the fraction of flaring stars provided by \cite{tess_flare}. The motivation is to compare optical flare rate to the radio detectability as a function of spectral type. If stellar activity has direct correlation to radio detectability, then it would be expected that the more frequently a star flares, the more likely it is detectable in radio within any given observation window. We describe the definition of a flaring star and the method of how TESS flares are counted in Sect.~\ref{sec:counting-tess-flare}.

The additional data obtained from year 1 and 2 further reinforce the original conclusion by \cite{tess_flare}, where they found M dwarfs of type M4--M6 dominate the TESS sample of flaring stars than any other star (G\"{u}nther et al., in prep.). Note that, however, \cite{tess_flare} pointed out the several biases in their TESS flare study as a consequence of TESS only being able to observe in relative flux units. And so, a common flare energy threshold for all stars is thus necessary for an unbiased comparison of flare rates between different spectral types of stars. The several TESS flare biases and subsequently our method for debiasing the TESS flare statistics are described in Sect.~\ref{sec:debiasing-tess-flare}.

\subsubsection{X-ray flares}
\label{sec:x-ray_flares}
Besides optical stellar flares like the ones observed by TESS, X-ray flares are also indicative of stellar activity. The X-ray luminosity of a star is empirically determined by its mass, age, and rotation rate \citep{pizzolato2003}. This empirical correlation implies stellar activity evolution is linked to its rotational evolution, since many activity parameters, such as magnetic dipole field strength and mass loss rate, are tightly correlated with the Rossby number $R_{\rm o} = P_{\rm rot}/\tau_{\rm c}$, where $P_{\rm rot}$ is the rotation period and $\tau_{\rm c}$ is the convective turnover time \citep{wright2011}.

Therefore, we also utilise the model by \cite{x-ray_flare_johnstone2021} for X-ray (0.1--2.4 keV) flare rates of F, G, K, and M dwarfs, with masses between 0.1 and 1.2 $M_{\odot}$. As seen in Fig. 19 in their work, the rates of X-ray flares above a fixed energy threshold monotonically decline with declining stellar effective surface temperature. Such a conclusion is the exact opposite of that from TESS flare study by \cite{tess_flare}, in which lower mass stars seemingly flare more than higher mass ones.
One reason behind such discrepancy may stem from the different definition of flares in these two studies. \cite{x-ray_flare_johnstone2021} only includes flares of energy above an energy threshold of $10^{32}$ erg in their analysis, unlike the TESS flare statistics where all impulsive changes in relative flux are classified as flares based on an inference framework called \texttt{allesfitter} \citep{allesfitter_gunther2021} and complementary criteria such as high signal-to-noise ratio (SNR). Therefore, although debiasing is required to compare TESS stellar flare rates of different spectral types (more details in Sect.~\ref{sec:debiasing-tess-flare}), there is no such need for any debiasing in the X-ray flare statistics owing to the fact that \cite{x-ray_flare_johnstone2021} only counts flares above an energy threshold.

\section{Methodology}
\label{sec:methods}
\subsection{Crossmatching method}
\label{sec:crossmatching-method}
Now that we have introduced the different all-sky surveys used in this study, we outline the method we applied for crossmatching these catalogues.

\subsubsection{False alarm rate}
\label{sec:false-positive}
As mentioned in Sects.~\ref{sec:intro} and \ref{sec:lotss}, one key challenge of using the Stokes I catalogue with \gaia alone to identify radio stellar systems is that the radio catalogue is composed mostly of galaxies. With a high density of radio sources ($\approx 780$ sources per square degree on average), the LoTSS sky is far denser than any previous wide-area radio survey such as the NRAO VLA Sky Survey (NVSS; \citealp{nvss_1998condon}), FIRST \citep{FIRST-survey}, TIFR GMRT Sky Survey first alternative data release (TGSS ADR1; \citealp{tgss}), and GaLactic and Extragalactic All-sky MWA (GLEAM; \citealp{gleam}) survey. 
Therefore, despite the LoTSS's high astrometric precision of $0.2\arcsec$ to $0.5\arcsec$ \citep{lotss-dr2}, true associations between dense optical surveys and radio sources cannot be confidently done by simple blind crossmatching; 
\cite{false-positive} showed that a blind search for radio-bright stellar systems in \gaia\ and LoTSS is dominated by false positives, and thus either additional observational or physically motivated information is needed in order to form a reliable sample of Galactic \gaia-LoTSS counterparts. For example, assuming the radio sources in the sky surveys are homogeneous, the following equation gives us an approximate estimate of false matches with the \gaia catalogue:
\begin{equation}
    N_{\rm false} = n_{\rm radio} \times n_{\gaia} \times \pi \theta^2 \times \Omega_{\rm radio},
    \label{eq:false-positive}
\end{equation}
where $N_{\rm false}$ is the number of chance-coincidence associations, $n_{\rm radio}$ and $\Omega_{\rm radio}$ are the source density and sky coverage of the radio survey respectively, and $\theta$ is the crossmatching radius.
For LoTSS-\gaia DR3 crossmatching, this gives us $N_{\rm false} \sim 10^5$, hence highlighting the difficulty of confidently associating stellar systems with radio emission from LoTSS alone. Similarly, the VLASS catalogue also faces the same issue of false positive as it has a source density only a factor of $\approx 8$ lower than that of LoTSS.

\subsubsection{LoTSS $\times$ GCNS match}
\label{sec:lotss-gcns}
Nevertheless, identifying radio-bright stellar systems in the LoTSS Stokes I catalogue while keeping chance-coincidence associations small is possible. We achieve this by crossmatching with GCNS within 50\,pc and by setting a small crossmatching radius of 1.4\arcsec with proper motion correction based on \gaia information. This crossmatching radius corresponds to $\approx 1$ chance-coincidence association on average.

\subsubsection{V-LoTSS $\times$ GCNS match}
\label{sec:vlotss-gcns}
On the other hand, as mentioned in Sect.~\ref{sec:v-lotss}, the circular-polarised sky of LoTSS allows us to be significantly more confident in the radio sources' association with stellar systems. Since the number of detected sources in V-LoTSS is less than a hundred, we need not restrict the crossmatching with GCNS to be within 50\,pc. Moreover, the low source density also allows us to set the crossmatching radius generously to be 6\arcsec (which corresponds to a false association rate of $10^{-2}-10^{-3}$ sources in the LoTSS sky) with \gaia proper motion correction. However, since the resolution of LoTSS's Stokes V images is lower than that of Stokes I images (20\arcsec vs 6\arcsec), this generally corresponds to nearly a fourfold worsening of astrometric precision of a Stokes V source compared to its Stokes I counterpart if the signal-to-noise of the source in Stokes V and Stokes I is identical \citep{v-lotss}. To avoid this, we therefore still use the Stokes I position of the source instead of Stokes V when performing the crossmatching.

\subsubsection{VLASS $\times$ GCNS match}
\label{sec:vlass-gcns}
Owing to the lack of Stokes V information and the relatively high source density ($\approx$ 100 sources/$\si{deg}^2$ per epoch) in VLASS catalogue, we follow a similar procedure as Sect.~\ref{sec:lotss-gcns} for the crossmatching between VLASS and GCNS: we set the crossmatching radius to be 1.4\arcsec, and only consider GCNS sources within 50\,pc, in order to attain minimal chance-coincidence associations. 
Also, we use the recommended criteria by \cite{vlass-ql-gordon2021} and consider only sources that satisfy \textit{Duplicate\_flag < 2} and \textit{Quality\_flag == 0}. This further reduces the amount of VLASS sources in each epoch to around 1.7 million and ensures the radio data is of high reliability.
The expected number of false associations is $\approx 0.5$ for each VLASS epoch.

\begin{figure*}
    \centering
    \includegraphics[width=\linewidth]{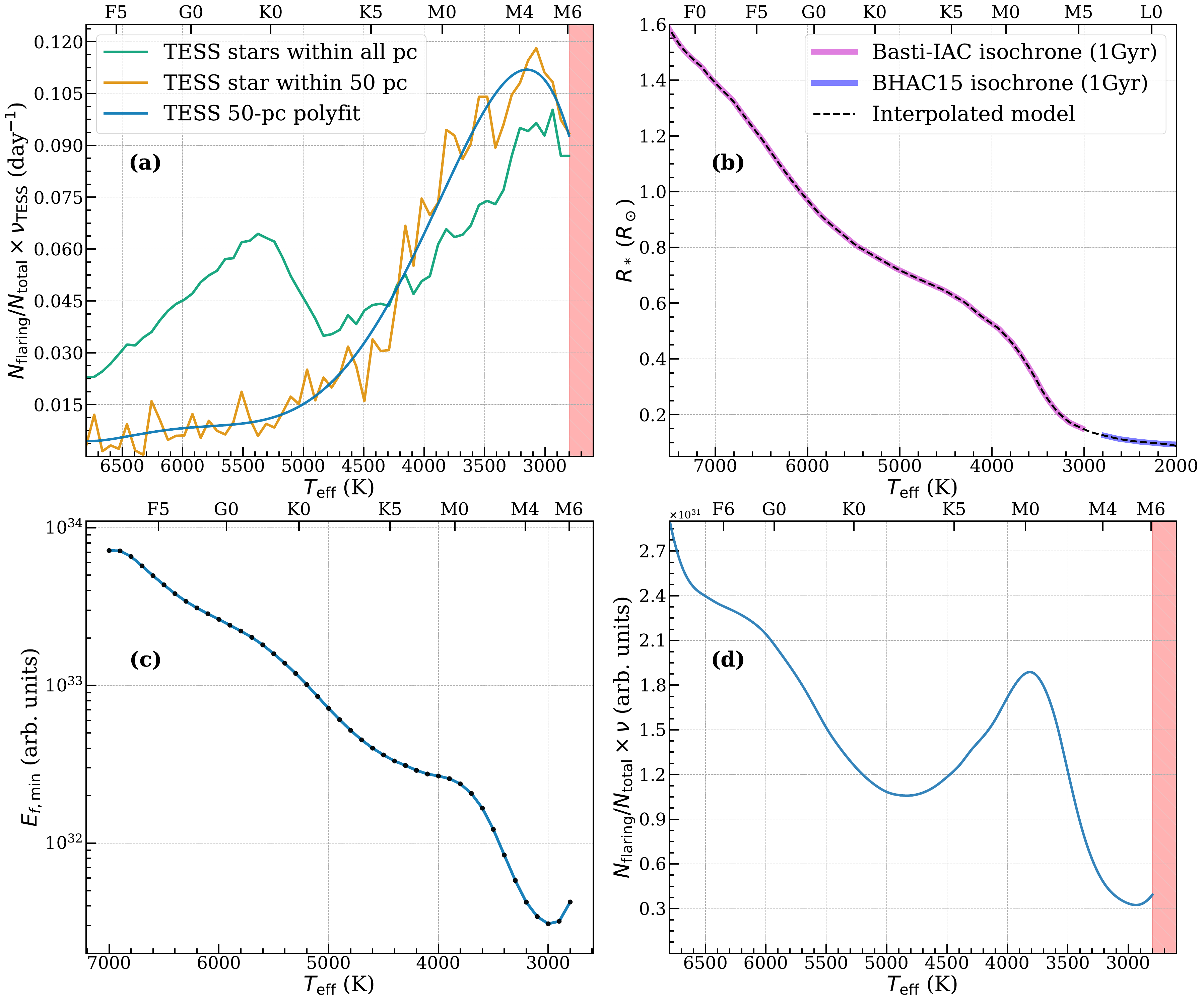}
    \caption{TESS flare statistics and debiasing. The top horizontal axis of each subfigure indicates the nominal stellar spectral types \citep{star_colour}. 
    \textbf{(a):} The average TESS ensemble flare rate -- defined as the fraction of flaring stars $N_{\rm flaring}/N_{\rm total}$ in the TESS short-cadence year 1 \& 2 observations, weighted by the average flare rate $\nu_{\rm TESS}$ -- as a function of stellar effective temperature $T_{\rm eff}$. $\nu_{\rm TESS}$ follows the definition from Equation~\ref{eq:flare_rate_tess}. The green curve shows the average TESS ensemble flare rate using the entire TESS flare statistics which suffers greatly from TESS sampling bias (see Sect.~\ref{sec:tess-sampling-bias}), whereas the orange curve only considers TESS stars within 50\,pc, with the blue curve as its polynomial fit (from \texttt{numpy.polyfit; deg = 5}) used for subsequent analysis.
    M dwarfs later than $\sim$M6 were not observed in a high enough sample size due to TESS target selection (which avoids very faint stars) and are thus excluded (represented by the red region). 
    \textbf{(b):} Stellar radius $R_{*}$ as a function of $T_{\rm eff}$. The Basti-IAC isochrone model \citep{solar-isochrone_hidalgo2018} and the BHAC15 isochrone model \citep{low-mass-isochrone_baraffe2015} are combined to create our interpolated model.
    \textbf{(c):} Minimum flare energy $E_{f, \min}$ detectable by TESS on a particular star with effective temperature $T_{\rm eff}$. The unit for the flare energy is arbitrary since we are only interested in the shape of the $E_{f, \min}$ versus spectral type curve. 
    \textbf{(d):} Same as Fig.~\ref{fig:all_TESS_debiasing_analysis}a, but with the debiased flare rate $\nu$ instead of TESS-observed flare rate $\nu_{\rm TESS}$. This debiased average-TESS-ensemble-flare-rate curve is obtained by multiplying the two blue curves in Figs.~\ref{fig:all_TESS_debiasing_analysis}a and \ref{fig:all_TESS_debiasing_analysis}c.}
    \label{fig:all_TESS_debiasing_analysis}
\end{figure*}

\subsection{Average TESS ensemble flare rates}
\label{sec:counting-tess-flare}
We use the catalogue for all individual flares found in TESS year 1 and 2 (\citealp{tess_flare} and in prep.). Each entry contains information on a candidate flare from a TESS short-cadence-targeted star, such as the flare's peak time, full width at half maximum (FWHM), and flare amplitude (see \citealp{tess_flare}). The new catalogue also contains extra information on quality-control filters and a probability of the candidate being a true flare, which is computed by the convolutional neural network \texttt{stella} \citep{stella_feinstein2020, cnn_feinstein2020}. This extra information is also detailed in \cite{tess_feinstein2022}.
In this work, we follow the definition of TESS flare rate per star $\nu_{\rm TESS}$ by \cite{tess_feinstein2022}:
\begin{equation}
    \nu_{\rm TESS} = \frac{1}{t_{\rm obs}} \sum^N_i p_i,
    \label{eq:flare_rate_tess}
\end{equation}
where $N$ is the number of flares for a given star, $p_i$ is the
\texttt{stella} probability for each flare candidate from that star, and $t_{\rm obs}$ is the total observed time of the star by TESS. By binning the flare rates into effective temperature $T_{\rm eff}$ bins, we obtain a relationship between stellar spectral type and the average $\nu_{\rm TESS}$. This curve, when multiplied by the fraction of flaring stars in each $T_{\rm eff}$ bin, gives us the average likelihood of TESS detecting a flare on a star of a particular spectral type (see Fig.~\ref{fig:all_TESS_debiasing_analysis}a). We shall refer to this quantity, i.e. $N_{\rm flaring}/N_{\rm total} \times \nu_{\rm TESS}$, as the ``average TESS ensemble flare rate''. Intuitively, this quantity tells us that if one were to pick a star at random and observe it with TESS, how many flares on average would be seen on that star given its particular spectral type.

\subsection{Debiasing TESS flare statistics}
\label{sec:debiasing-tess-flare}
As mentioned in Sect.~\ref{sec:tess_flare}, the TESS flare statistics used here are inherently biased since they defined ``flaring'' as a sufficient increase in flux amplitude compared to its quiescent level (i.e. sufficient increase in relative flux units), rather than a lower energy threshold above which all flares are counted. These TESS flare statistics biases are described in depth in the following sections.

\subsubsection{TESS sampling bias}
\label{sec:tess-sampling-bias}
Firstly, TESS' short-cadence observations are biased towards brighter stars to ensure follow-up spectroscopic observations of transiting exoplanets \citep{tess_collins2018}.
Indeed, TESS is essentially a targeted survey with a complicated target selection (for light curves) determined by the myriad of short-cadence proposals submitted by the scientific community.
We can minimise this TESS sampling bias by excluding stars beyond a certain distance such that TESS has observed every star within such volume; around 15\,pc is where the TESS sample is volume-complete (\citealp{tess_flare} and in prep.). However, the number of flaring stars with early spectral type within 15\,pc is too small to perform proper statistics. In particular, there is almost no star earlier than K0 within 15\,pc in the TESS data. Therefore, we compromise on considering TESS stars within 50\,pc instead, as this achieves a good balance between sample completeness and a sufficient sample size for proper statistical analysis. As shown in Fig.~\ref{fig:all_TESS_debiasing_analysis}a, the orange curve -- representing the average-TESS-ensemble-flare-rate that excludes TESS stars beyond 50\,pc -- has a very different shape when compared to the green curve which does not exclude any star. In particular, hotter stars with spectral type earlier than $\sim$K5 have significant amount of flaring events in the green curve compared to the orange curve. In addition, along with the usual mid-M peak, there is an additional peak around K0 which is absent in the orange curve. However, M dwarfs (especially around M4) still show the highest average TESS ensemble flare rate ($N_{\rm flaring}/N_{\rm total} \times \nu_{\rm TESS}$) in both curves.

The deviations between the curves are the manifestation of the TESS sampling bias from which the full TESS flare dataset (i.e. green curve) suffers. For example, Fig.~\ref{fig:all_TESS_debiasing_analysis}a suggests that flaring solar-type stars are oversampled by TESS. One of the possible reason for this oversampling is that the scientific community is interested in studying the superflares of solar-type stars in order to obtain insights on space weather around solar analogs (e.g. \citealp{namekata2017, maehara2017, tu2020}). 
In any case, we shall only consider TESS stars within 50\,pc in subsequent analysis, as such a 50-pc sample achieves a good balance between minimising the TESS sampling bias and having enough flaring stars to avoid small-number statistics.

\subsubsection{TESS flare detection biases}
\label{sec:tess-flare-detection-bias}
Secondly, there exists an interplay of two opposing flare-detection biases: (i) since TESS detects flares in relative flux units, a flare of a given energy is more readily detected on cooler stars because of a larger contrast between the flare and the stellar quiescent flux density; and (ii) a higher photometric noise in cooler stars decreases the SNR of their relative flare flux densities. As shown in Fig. 2f of \cite{tess_flare}, the recovery rates of flares in injection tests is significantly lower for cooler stars in the M-dwarf range compared to F/G/K dwarfs. And so, in order to have a fair flare-rate comparison between stars of different spectral types, these biases must be taken into account. In the following paragraphs, we describe the procedure for the debiasing of the TESS flare statistics. Here, we start by converting the known flare completeness in relative flux units (\citealp{tess_flare}; in Fig. 2f), to a flare completeness in absolute flux units:

Let $A_S$ be the relative flux of a stellar flare observed by TESS with respect to the corresponding quiescent stellar flux. We define $A_{\min} = L'_{f, \min}/L'_*$ as the minimum relative flux of which a stellar flare of luminosity $L_{f, \min}$ can still be detectable from a star with an effective temperature $T_{\rm eff}$ and a stellar quiescent luminosity $L_*$, i.e. $A_{\min} = A_{\min}(T_{\rm eff})$, with the prime symbol $'$ denoting the luminosity in the TESS observing bandpass. This function encapsulates the interplay of the two aforementioned detection biases of TESS flares. The value of $A_{\min}(T_{\rm eff})$ is already known from Fig. 2f by \cite{tess_flare}. The stellar quiescent luminosity $L'_*$ can be expressed as
\begin{equation}
    L'_* = \pi R_*^2 \int_{0}^{\infty} B_*(\lambda, T_{\rm eff})~S_{\rm TESS}(\lambda) d\lambda,
    \label{eq:L_*}
\end{equation}
where $R_{*}$ is the stellar radius, $B_*$ is the Planck function at the $T_{\rm eff}$ of the star, and $S_{\rm TESS}$ is the TESS spectral response function which is defined as the product of the long-pass filter transmission curve and the detector quantum efficiency curve \citep{tess}. To relate the stellar radius $R_{*}$ to its $T_{\rm eff}$, we use the Basti-IAC isochrone model \citep{solar-isochrone_hidalgo2018} with a solar-type metallicity for stars with $T_{\rm eff} \gtrsim 3000$~K and the BHAC15 isochrone model \citep{low-mass-isochrone_baraffe2015} for main sequence low-mass stars down to $T_{\rm eff} = 2600$~K. We assume a stellar age of 1 Gyr (around the same order of magnitude of most stars in our galaxy, e.g. \citealp{haywood2013, snaith2015}) for these models. Fig.~\ref{fig:all_TESS_debiasing_analysis}b shows $R_{*}$ as a function of $T_{\rm eff}$ and the corresponding smooth spline interpolation.

We can determine the minimum flare luminosity observable by TESS:
\begin{equation}
    L'_{f, \min} = A_{\min}(T_{\rm eff})~\pi R_*^2 \int_{0}^{\infty} B_*(\lambda, T_{\rm eff})~S_{\rm TESS}(\lambda) d\lambda.
    \label{eq:L_f}
\end{equation}
Now, to relate this quantity to its unprimed counterpart, we assume that flares have a blackbody spectrum with the same effective temperature no matter the spectral type of the star (e.g. \citealp{shibayama2013,davenport2016,yang2018,tess_flare,flare-rate-gao2022}). The precise value of effective temperature does not matter in subsequent analysis\footnote{However, it is typical to choose the value of $\approx$9000 K (e.g. \citealp{shibayama2013,chang2018,tess_flare,goodarzi2019}) since the spectral energy distribution of flares can be fitted by the blackbody spectrum with this temperature \citep{hawley-fisher1992,kretzschmar2011}.} as we only are only interested in the ratio between the flare luminosities from stars of different spectral types, not their absolute values. This assumption implies that for two stars with stellar effective temperatures $T_1$ and $T_2$ respectively, we have $L_f(T_1)/L_f(T_2) = L'_f(T_1)/L'_f(T_2)$.
Therefore, we now have $L_{f, \min}$ as a function of minimum relative flux $A_{\min}$ and $T_{\rm eff}$. Moreover, we assume the average duration of a stellar flare is the same for stars of all spectral type, thus $E_f(T_1)/E_f(T_2) =  L_f(T_1)/L_f(T_2)$, where $E_f(T)$ is the flare energy from a star with $T_{\rm eff} = T$. Since the curve of $A_{\min}$ vs $T_{\rm eff}$ is already given by \cite{tess_flare}, we can determine the minimum flare energy $E_{f, \min}$ detectable by TESS on a particular star with $T_{\rm eff} = T$. As seen in Fig.~\ref{fig:all_TESS_debiasing_analysis}c, the weakest TESS-detected flare on a early-type star is much stronger than its late-type star counterpart. For example, the $E_{f, \min}$ of a solar-like G star is around $\sim$2 orders of magnitude stronger than the $E_{f, \min}$ of an M dwarf.

Lastly, we need to convert this flare energy threshold into a flare rate $\nu$. To do this, we utilise the study by \cite{flare-rate-gao2022} regarding cumulative flare frequency distributions (FFDs; e.g. \citealp{lacy1976,hawley2014,tess_flare,jackman2021}) detected by TESS and Kepler. The cumulative FFD represents the number of flares in unit time with an energy greater than a particular value $E$, i.e. how often a flare of energy $E_f \geq E$ occurs on a star. Mathematically,
\begin{equation}
    \log_{10}(\nu) = \alpha_{\rm cum} \log_{10}(E) + \beta,
    \label{eq:alpha_cum}
\end{equation}
where $\alpha_{\rm cum}< 0$ and $\beta$ are empirically determined parameters. The debiased flare rate for a star with spectral type $S$ is then
\begin{equation}
    \log_{10}(\nu_S/\nu_{\rm TESS, S}) = \alpha_{\rm cum}\log_{10}(E_0/E_{f, \min, S}),
    \label{eq:alpha_cum_tess}
\end{equation}
where $E_0$ is the common energy threshold above which all flares are counted for all stars, $\nu_{\rm TESS, S}$ and $E_{f, \min, S}$ are the TESS-observed flare rate and the minimum TESS-detected flare energy on star of type $S$ respectively. \citet[in Fig. 12]{flare-rate-gao2022} empirically determined that $\alpha_{\rm cum}$ from solar-type stars to mid M dwarfs is approximately consistent with a value of $-1$, and so we shall assume $\alpha_{\rm cum} = -1$ in subsequent calculations. 
We can now convert the ratio of observed flare rates for stars of types $S_1$ and $S_2$ into a ratio of their true flare rates above some energy threshold as
\begin{multline}
    \log_{10}(\nu_{S_1}/\nu_{\rm TESS, S_1}) - \log_{10}(\nu_{S_2}/\nu_{\rm TESS, S_2}) \\ 
    = - \log_{10}(E_0/E_{f, \min, S_1}) + \log_{10}(E_0/E_{f, \min, S_2}).
\end{multline}
Hence,
\begin{equation}
    \frac{\nu_{S_1}}{\nu_{S_2}} = \frac{\nu_{\rm TESS, S_1}}{\nu_{\rm TESS, S_2}} \times \frac{E_{f, \min, S_1}}{E_{f, \min, S_2}}.
    \label{eq:true_flare_rate}
\end{equation}

Finally, multiplying the two blue curves in Figs.~\ref{fig:all_TESS_debiasing_analysis}a and \ref{fig:all_TESS_debiasing_analysis}c gives us Fig.~\ref{fig:all_TESS_debiasing_analysis}d, which shows the the debiased average TESS ensemble flare rate, i.e. the fraction of flaring stars $N_{\rm flaring}/N_{\rm total}$ weighted by the debiased flare rate $\nu$. Here, given a particular energy threshold above which flares are counted, we can see a general trend of the ensemble flare rate decreasing as we move from early-type stars towards K dwarfs. Then, at $T_{\rm eff} \approx 4800$~K (around spectral type K3), the ensemble flare rate increases, peaking at spectral type M0 before decreasing again in the range of mid-M dwarfs.
Specifically, early-type ($\lesssim$G0) stars are actually around $2-4$ times more likely to flare than early-K and mid-M dwarfs, while the ensemble flare rate around spectral type M0 is quite comparable.
This debiased TESS flare trend is not exactly in line with the X-ray flare rate statistics by \cite{x-ray_flare_johnstone2021}.
Regardless, we have now obtained a debiased average TESS ensemble flare rate curve, which shall be used in the TESS flare statistics analysis in Sect.~\ref{sec:tess_flare_cdf}.

\begin{figure*}
    \centering
    \begin{subfigure}{.5\textwidth}
          \centering
          \includegraphics[width=\linewidth]{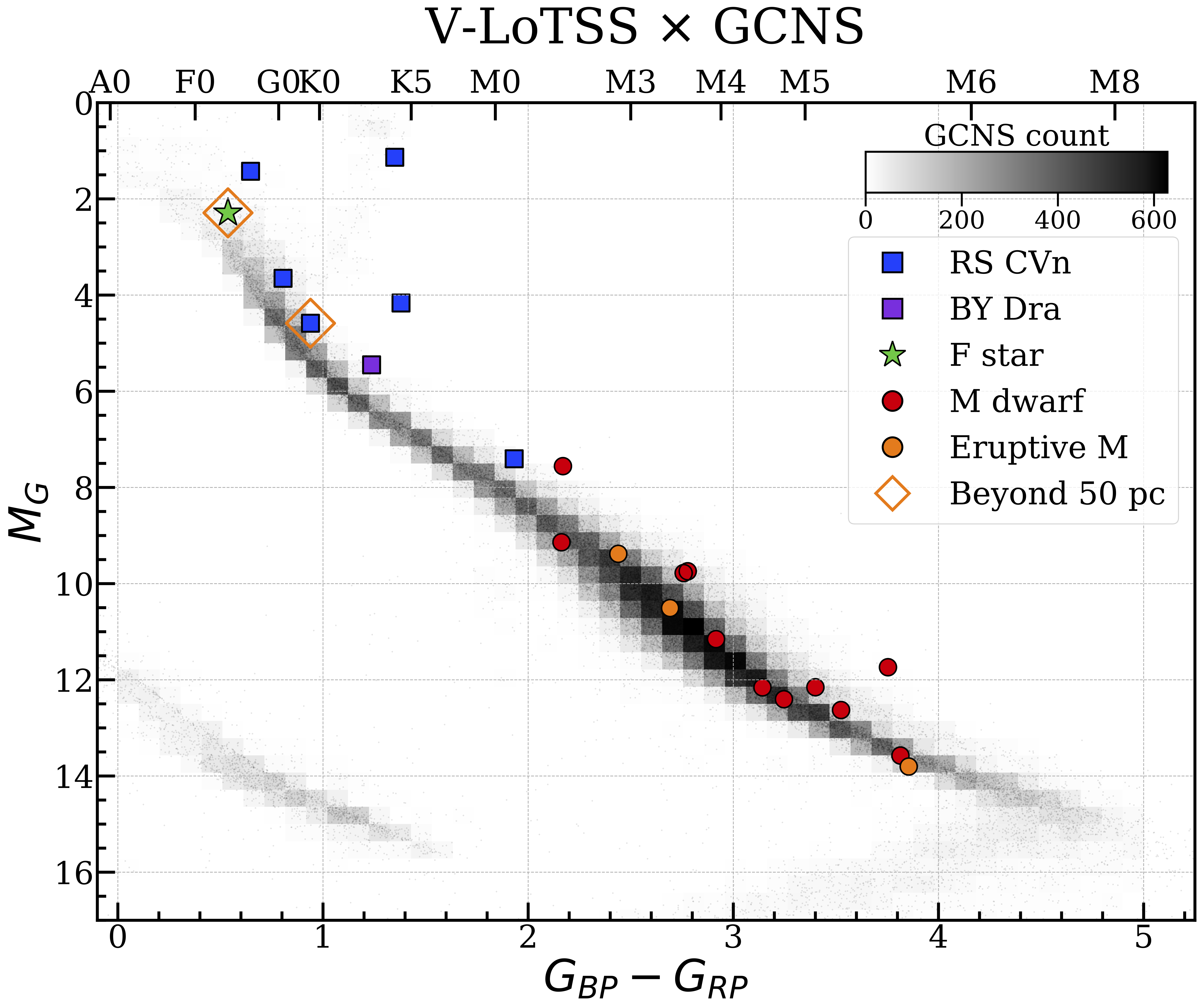}
    \end{subfigure}%
    \begin{subfigure}{.5\textwidth}
          \centering
          \includegraphics[width=\linewidth]{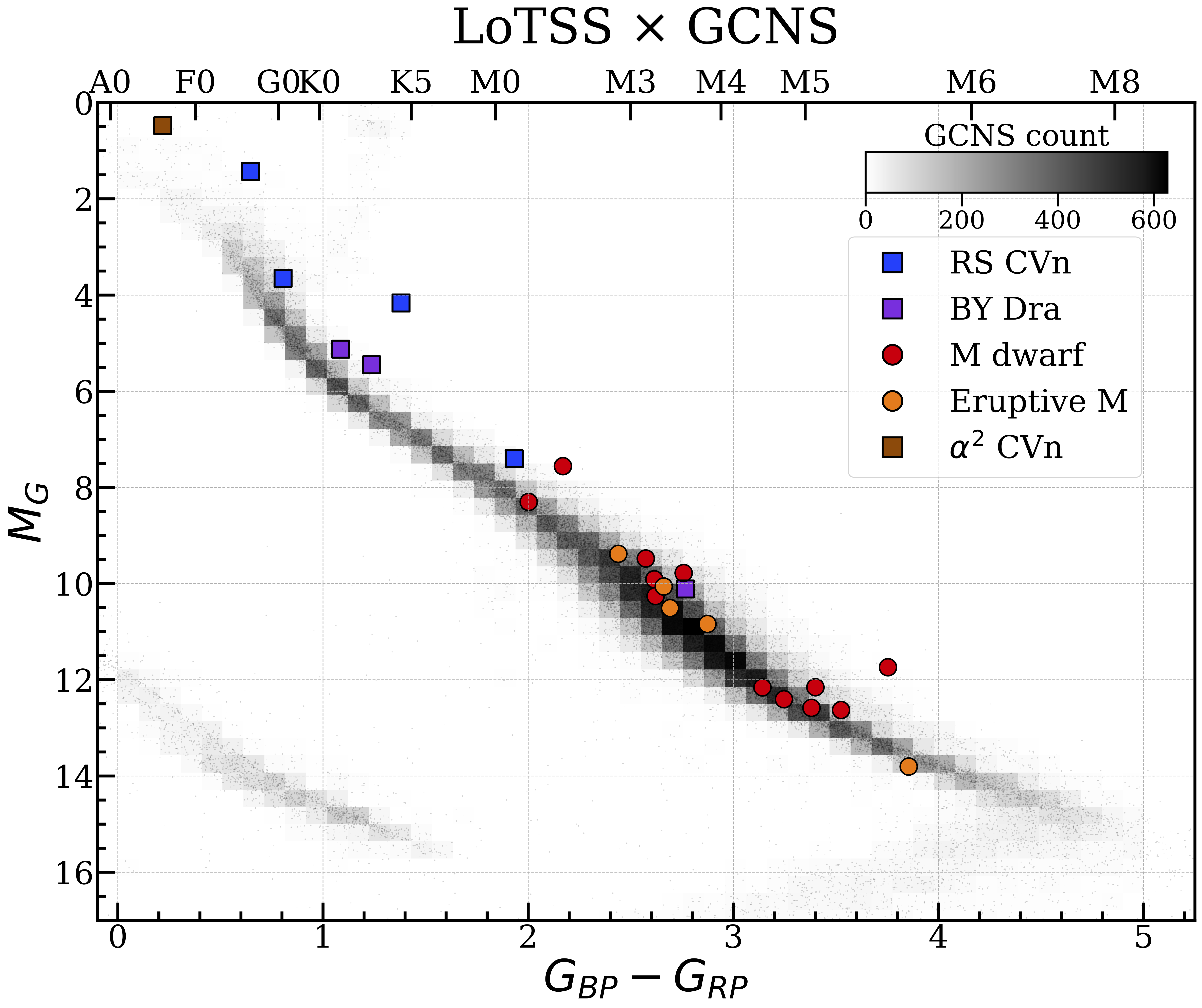}
    \end{subfigure}
    \caption{Hertzsprung-Russell (HR) diagrams for the V-LoTSS $\times$ GCNS sample (left panel) and LoTSS $\times$ GCNS sample (right panel), according to \gaia $G_{BP}-G_{RP}$ colour and \gaia $G$ absolute magnitude. The radio-bright stellar systems are represented by different colours and symbols according to object classification, as shown in the legends. There exist two sources in the V-LoTSS $\times$ GCNS population that are beyond 50\,pc, whereas only radio detections within 50\,pc are included in the LoTSS $\times$ GCNS population to suppress chance-coincidence associations. The top axis of the HR diagrams indicates the nominal stellar spectral types \citep{star_colour}.
    }
    \label{fig:hr-lotss}
\end{figure*}

\begin{figure*}
    \centering
    \includegraphics[width=\linewidth]{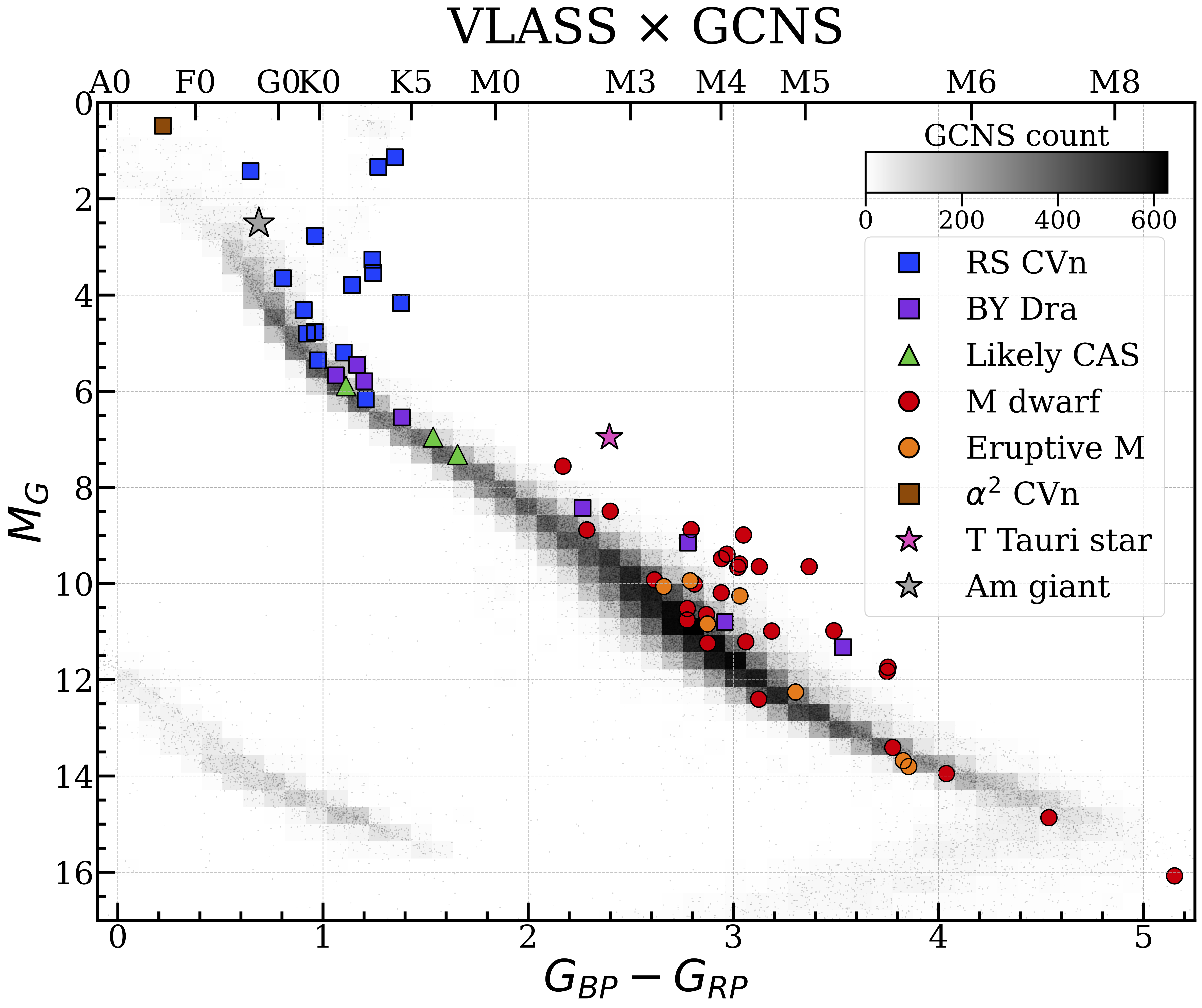}
    \caption{Hertzsprung-Russell (HR) diagram for the VLASS $\times$ GCNS sample according to \gaia $G_{BP}-G_{RP}$ colour and \gaia $G$ absolute magnitude. The radio-bright stellar systems are represented by different colours and symbols according to object classification, as shown in the legends. Green triangle represents stellar systems that are likely to be chromospherically active from literature. The radio sources here are all within 50\,pc, as crossmatching the entire GCNS would lead to significant coincidence associations with VLASS sources. The top axis of the HR diagrams indicates the nominal stellar spectral types \citep{star_colour}.}
    \label{fig:hr-vlass}
\end{figure*}

\section{Results and Discussion}
\label{sec:results_discussion}

\subsection{Crossmatching Results}
\label{sec:crossmatching-results}
In Tables~\ref{tab:lotss-table}, \ref{tab:v-lotss-table}, and \ref{tab:vlass-table} (all shown in Appendix~\ref{app:tables}), we present the radio $\times$ GCNS crossmatching samples from the three aforementioned radio surveys. There are 22 V-LoTSS-detected sources that have optical counterparts in the \gaia Catalogue for Nearby Star (GCNS), while there are 25 LoTSS-detected sources and 65 VLASS-detected sources that have an optical counterpart in GCNS within 50\,pc. Figures~\ref{fig:hr-lotss} and \ref{fig:hr-vlass} show each of three radio-detected stellar populations in the Hertzsprung-Russell (HR) diagram.

For the V-LoTSS $\times$ GCNS sample, note that out of the 22 matches, all except two have an angular separation between the V-LoTSS source (based on Stokes I astrometry) and \gaia counterpart of $\lesssim 1\arcsec$. The remaining two sources, HAT 182-00605 and II Pegasi, have an angular separation of 1.22\arcsec and 1.83\arcsec respectively. However, since both of them are already known to be radio-bright \citep{joe_nature, rscvn-toet2021}, we are confident that none of the matches in our V-LoTSS $\times$ GCNS sample are false positives. 
Moreover, all of these matches are consistent with the stellar sample in V-LoTSS, as \cite{v-lotss} also crossmatched V-LoTSS sources to the \gaia Data Release 2 \& 3 (DR2 \& DR3; \citealp{gaia-dr2, gaia-dr3}) catalogues. However, there are two stellar systems in V-LoTSS that are not in our radio sample despite being within 100 pc: DG CVn and i Bo\"{o}tis. They are not present in the GCNS likely due to unreliable astrometry and thus their \gaia DR2 source identifier are no longer available in \gaia DR3. Therefore, we choose to ignore these stellar systems in our analysis.

Not every source in the V-LoTSS $\times$ GCNS sample makes an appearance in the LoTSS $\times$ GCNS sample, and vice versa. The latter is expected since some radio-bright stellar systems may have fractional polarisation values that makes them drop below the detection threshold of V-LoTSS. In addition, due to time variability of stellar radio sources, a genuine V-LoTSS source that appears in a particular LoTSS field may get ``washed out'' during mosaicing of the fields in LoTSS-DR2 \citep{v-lotss}. The mosaicing also explains why even when a stellar system shows up in both the Stokes I and Stokes V catalogues, the quoted Stokes I flux is slightly different in each catalogue.

As for the VLASS $\times$ GCNS sample, despite the fact that VLASS and LoTSS have very similar sensitivity and that the VLASS sky fully overlaps with that of LoTSS, very few LoTSS/V-LoTSS detections can be actually found in the VLASS $\times$ GCNS sample, and vice versa. The reasons for this discrepancy include the transient nature of stellar radio emission and the vastly different frequencies of the two radio surveys (see Sect.~\ref{sec:vlass} on the importance of survey frequency coverage). 
Therefore, there is no guarantee that a stellar system that emit in LOFAR band must also emit in VLA band, and vice versa.

There is one source that we remove from the VLASS sample, despite a crossmatching association with GCNS. The source is a M2 dwarf named HD 9770C (\gaia DR3 5022972468944971648), which is part of the visual triple system HD 9770 (also known as BB Scl). Only the C component of this system is in the GCNS, as both the primary A star and secondary B star are not present in the GCNS due to incomplete astrometry in \gaia EDR3 and DR3 (e.g. missing proper motion value). Moreover, the \gaia colour and \gaia absolute magnitude of HD 9770C makes it deviate from the HR main sequence by $>4$ mag. All of these peculiar \gaia properties are probably caused by the angular proximity of the three stars: the two stars A and B are in a well-defined 4.559-year orbit with a semi-major axis of 0.171\arcsec, and the AB $\times$ C system is in a 111.8-year orbit with a semi-major axis of 1.419\arcsec \citep{hirshfield-sinnott1985}. Moreover, both A and B are themselves binaries, with B being an eclipsing binary of the BY Dra type \citep{hd9770_watson2001}. Using \gaia DR2 astrometry, we found that the radio emission is actually closer to the AB system than the C component (around 0.3\arcsec vs 1.4\arcsec). Thus, we conclude that the radio emission most likely stems from the BY Dra variable from the secondary B component of HD 9770, and we remove HD 9770C from our analysis.

As shown in Figs.~\ref{fig:hr-lotss} and \ref{fig:hr-vlass}, both the LoTSS (V-LoTSS included) and VLASS populations can be generally classified into two categories: chromospherically active stellar (CAS) systems and M dwarfs.
The CAS systems can be further subdivided into RS Canum Venaticorum (RS CVn) variables and BY Draconis (BY Dra) variables. These are close stellar binaries\footnote{BY Dra does not necessarily have to be in a close binary system, but in most cases (such as the only BY Dra in our V-LoTSS detections: BF Lyn) they are.} that consist of late spectral types (F to M). Indicated by the presence of strong Ca\,II H and K emission lines, the stars in these systems have active chromospheres due to strong magnetic field generated by rapid stellar rotation ($P_{\rm rot}\sim$~days) twisting magnetic flux loops, as the binaries are generally in very close proximity ($\lesssim 0.01$\,AU) and thus tidally locked. This causes extreme degrees of solar-type activity in these systems, manifested in the form of large stellar spots and magnetic interactions. In the case of the latter, they coherently accelerate electrons \citep{rscvn-toet2021}, hence the circularly polarised radio emission from RS CVn and BY Dra.

On the other hand, M dwarfs compose the majority of the radio detections, with their spectral type ranging from M1.5\footnote{Technically, from our radio detections, there are three RS CVn systems -- YY Geminorum, II Pegasi, and V1044 Scorpii -- that consist of a early-M-dwarf ($\lesssim$M1.5) component. However, since the radio and X-ray luminosity of these RS CVn systems are typically much higher than that of isolated M dwarfs (see Sect.~\ref{sec:gb}), the origin of these two stellar systems' radio emission likely originates from their active chromospheric interactions rather than Jupiter-like behaviour exhibited by some M dwarfs.} to M6 for the LOFAR sample, and M1.5 to M8.5 for the VLA sample. All of these M dwarfs in our V-LoTSS sample were already presented in previous publication \citep{joe_nature}, where it was shown that coherent radio emission is ubiquitous across the M-dwarf main sequence, as each spectral type has an equal probability of detection. This implies V-LoTSS has detected both partially convective M dwarfs and fully convective ones. The same applies to both LoTSS and VLASS as well. In particular, VLASS dives into the realm of ultracool dwarfs (UCDs) as the survey detects a few late-M dwarfs as well. This includes the M8.5V ultracool dwarf LSR J1835+3259, which is recently discovered to have a Jupiter-like radiation belt \citep{climent2023, kao2023}.

The underlying reason why M dwarfs are more radio-bright than other isolated stars could be that M dwarfs generally exhibit not just Sun-like activity, but also Jupiter-like properties such as fast rotation, strong large-scale magnetic fields, and auroral radio emission (e.g. \citealp{aurora_hallinan2015, cr-dra_callingham2021}). Therefore, unlike earlier spectral types, their radio emission could be predicated on the existence of their strong large-scale stellar magnetic fields. The M dwarfs should then have a higher efficiency of converting the available energy to radio emission compared to early-type stars, thus making them easier to detect in an untargeted survey.

The only LoTSS/V-LoTSS $\times$ GCNS objects that do not fall into these two categories are the prototypical $\alpha^2$ Canum Venaticorum ($\alpha^2$ CVn) and HD 220242, the latter of which is an isolated F5 star situated 69\,pc away. This makes HD 220242 by far the furthest stellar radio detection from our sample. It is the only main-sequence star other than an M dwarf without known companion stars and is detected with high circular polarisation ($78 \pm 16$). Generally, one does not expect a main-sequence F-type star to harbour a large scale field strong enough to generate cyclotron maser emission at 144\,MHz. As for solar-type bursts powered by plasma emission, the brightest solar burst ever recorded would only be detectable up to $\approx$10\,pc \citep{vedantham2020}. Therefore, the strong circularly polarised radio emission from such a faraway F-type star is unexpected and more details about this peculiar star are discussed in Appendix~\ref{app:hd220242}.

However, unlike the LoTSS/V-LoTSS $\times$ GCNS population, where we mostly detect either CAS systems or M dwarfs, we see a larger variety of stellar systems in the VLASS detections (see Fig.~\ref{fig:hr-vlass}). Granted, M dwarfs are still the majority of the detections, along with the chromospherically active stars such as RS CVn and BY Dra systems. However, the VLASS population also consists some unique stellar systems, including a few UCDs, one T Tauri star (TTS), the prototypical $\alpha^2$ CVn, one chemically peculiar Am star, and a few K-type main-sequence stars which are probably chromospherically active. 
We describe more details regarding some of the more interesting stellar systems in Appendix~\ref{app:peculiar-matches}.

One peculiarity regarding the V-LoTSS $\times$ GCNS sample is the small amount of V-LoTSS stellar detections outside of the 50-parsec sphere; only 2 out of the 22 sources are beyond 50\,pc: EV Draconis which is a RS CVn system and the isolated F star HD 220242. See Appendix~\ref{app:incompleteness} for more details on the lack of detections and on the analysis of the V-LoTSS incompleteness. In short, there may be be some incompleteness in the V-LoTSS M-dwarfs population, but the statistical evidence for this is marginal.

\subsection{Pre-main-sequence stars in the VLASS $\times$ GCNS sample}
\label{sec:pre-main-sequence}
\begin{figure}
    \centering
    \includegraphics[width=\linewidth]{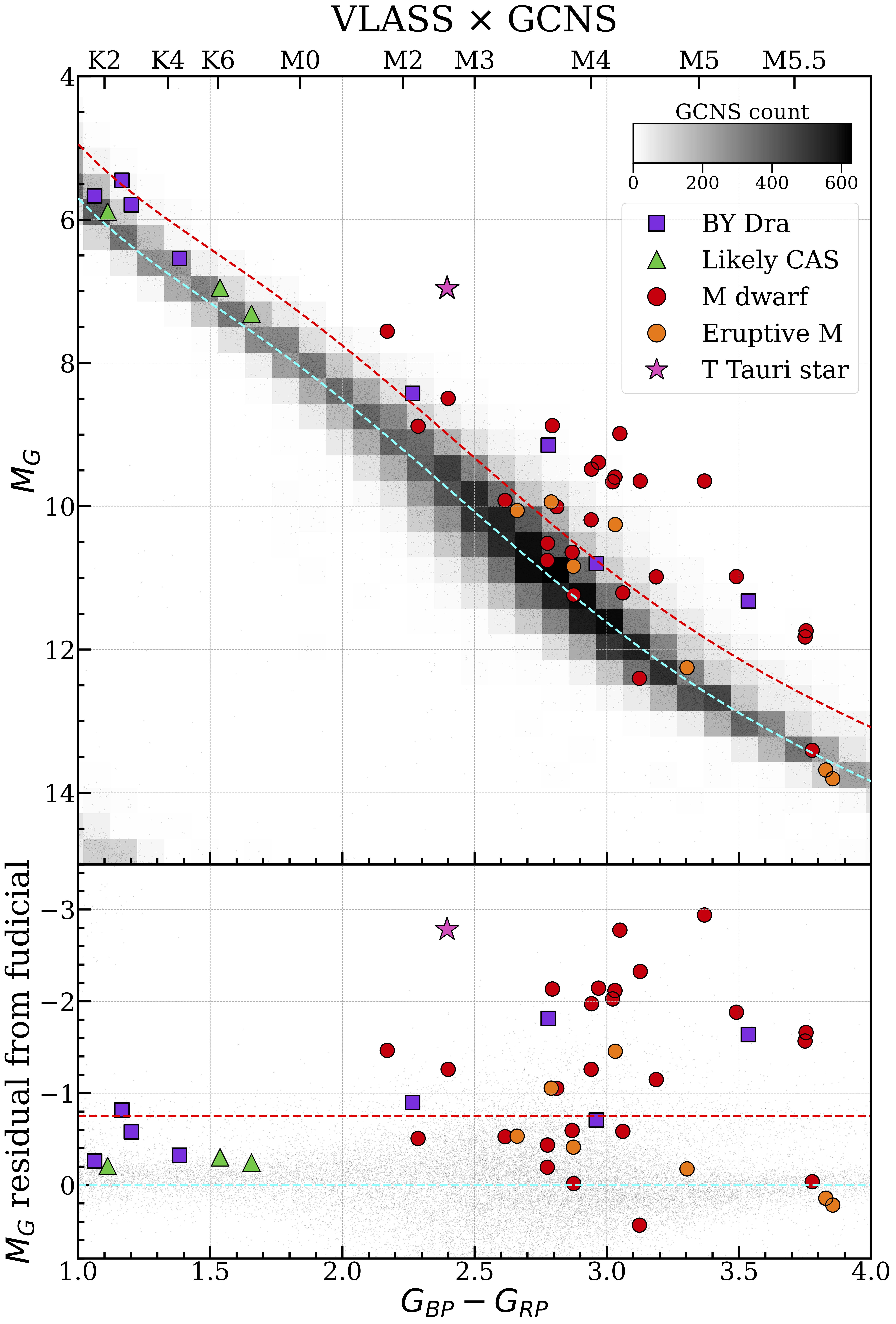}
    \caption{Top panel: Zoomed-in version of Fig.~\ref{fig:hr-vlass} with RS CVn variables removed. The cyan line represents the median fiducial and the red line shows the same fiducial shifted by -0.753 mag, which corresponds to an unresolved binary system of two identical stars. Therefore, stellar systems that significantly above the green line cannot be explained simply by binarity. The top axis of the HR diagrams indicates the nominal stellar spectral types \citep{star_colour}.
    Bottom panel: Residuals of $M_G$ relative to the median fiducial (cyan).}
    \label{fig:hr-vlass-fiducial}
\end{figure}

Another peculiarity regarding the VLASS $\times$ GCNS sample is the large number of stellar systems that deviate from the main sequence. Shown in Fig.~\ref{fig:hr-vlass}, some stellar systems such as RS CVn variables, BY Dra variables, and M dwarfs (including a T Tauri star) are away from the general GCNS population by a few magnitudes of brightness. For the RS CVn systems, this is not surprising since these are all binary systems and some RS CVn systems even host subgiants or giants as one of the components. These make them much more luminous than their main-sequence non-binary counterparts, thus branching off the main sequence. Yet, the fact that so many M dwarfs also exhibit the same phenomenon is unexpected, considering that nothing of such can be seen in the LOFAR sample. A naive explanation could be that these M dwarf systems are unresolved binaries with another star of very similar spectral type, causing the observed luminosity to be up to twice of a typical isolated M dwarf. This alone, however, cannot explain more than half of the M dwarfs with a deviation from the main sequence by more than 0.753 mag, which corresponds to an unresolved binary of two identical stars (see Fig.~\ref{fig:hr-vlass-fiducial}). Since these are M dwarfs, they also cannot be in an evolved state as a typical M-dwarf lifespan far exceeds the current age of the Universe \citep{laughlin1997}.

An alternative possibility is that these are all young active stars similar to a T Tauri star, and belong to known kinematic groups such as stellar associations, stellar nurseries, and young moving groups (YMGs). Ultimately, we found that around 80\% of the stars above the red line shown in Fig.~\ref{fig:hr-vlass-fiducial} are indeed known members of YMGs and stellar associations, including Beta Pictoris ($\beta$ Pic) Moving Group (e.g. \citealp{beta-pic_shkolnik2017}), Octans-Near Association (e.g. \citealp{octans-near_zuckerman2013}), AB Doradus (AB Dor) moving group (e.g. \citealp{ab-dor_malo2013}), and TW Hydrae association (TWA; e.g. \citealp{twa-neuh2010}). These stars are all very young ($<$100 Myr) and are considered to be in the stage when they have not yet reached the main sequence, i.e. pre-main-sequence stars. Therefore, these young stellar objects (YSOs) most likely have extremely fast rotation as they have yet to experience significant rotational spin-down through stellar winds, making them even more magnetically and chromospherically active than the typical M dwarfs due to their convective envelope enabling dynamo processes \citep{bouvier2014}. This leads to highly variable yet consistent radio emission from YSOs via mechanisms such as gyrosychrotron and ECMI.

The remaining question is why such a YSO population seen in the VLA sample is nowhere to be found in the LOFAR samples. The most likely explanation is that most of the nearby young moving groups (AB Dor, TWA, $\beta$ Pic, etc.) are located in the Southern Sky and/or the galactic plane, both of which lie outside of the current LoTSS footprint. Another possible explanation is that the predominant radio emission mechanism for these YSOs have most of its power emitted in the decimetre-wave (VLA band) regime rather than metre-wave (LOFAR band). This could happen in the case where most of their radio emissions are gyrosynchrotron emission stemming from flares \citep{incoherent-nindos2020}.

\subsection{Radio evolution with spectral type}
\label{sec:cdf_bprp}
As mentioned in Sects.~\ref{sec:vlotss-gcns} and \ref{sec:vlass-gcns}, the vast majority of isolated stars detected in the radio surveys are M dwarfs, so one might be tempted to claim that there is a radio evolution with spectral type and the radio-engine (Sun-like vs Jupiter-like) transition occurs in the realm of M dwarfs.
However, M dwarfs are also the most numerous stellar type, and so a proper statistical test is necessary to ascertain this apparent evolution of radio detectability with spectral type. 
As a statistical test, we compute the cumulative distribution functions (CDFs) with respect to \gaia colours for the 4 populations: the GCNS ``background'' population, the LoTSS $\times$ GCNS population, the V-LoTSS $\times$ GCNS population, and the VLASS $\times$ GCNS population. We only consider sources within 50\,pc of our Solar System in the following analyses since spectral types later than M9 start to become incomplete in GCNS beyond 50\,pc \citep{gcns}. The number of sources in GCNS thus decreases to $\approx 40$k, and the two sources beyond 50\,pc from V-LoTSS $\times$ GCNS are removed. The LoTSS $\times$ GCNS and VLASS $\times$ GCNS samples remain unchanged as we only crossmatch VLASS catalogue with GCNS within 50\,pc.

There are many statistical tests widely used for quantifying the discrepancy between two CDFs.
Here, we choose the Cramér-von Mises test (CvM test; \citealp{cvm-test_anderson1962}) to test the null hypothesis: the two samples being compared are drawn from the same distribution. 
The CvM test computes a $p$-value in a two-sample statistical test, which represents how likely the null hypothesis is true. For our analysis, a small $p$-value signifies a deviation between the radio $\times$ GCNS population and the background population, suggesting a transition of radio detection rate with spectral type.
We choose the classic threshold of $p$-value $<0.05$ as the condition for a rejection of the null hypothesis. 

In the following section, we consider the cases for both the inclusion and exclusion of CAS systems in our CDF analyses.

\subsubsection{GCNS background vs radio-detected samples}
\label{sec:background-vs-radio-cdf}
\begin{figure*}
    \centering
    \includegraphics[width=\linewidth]{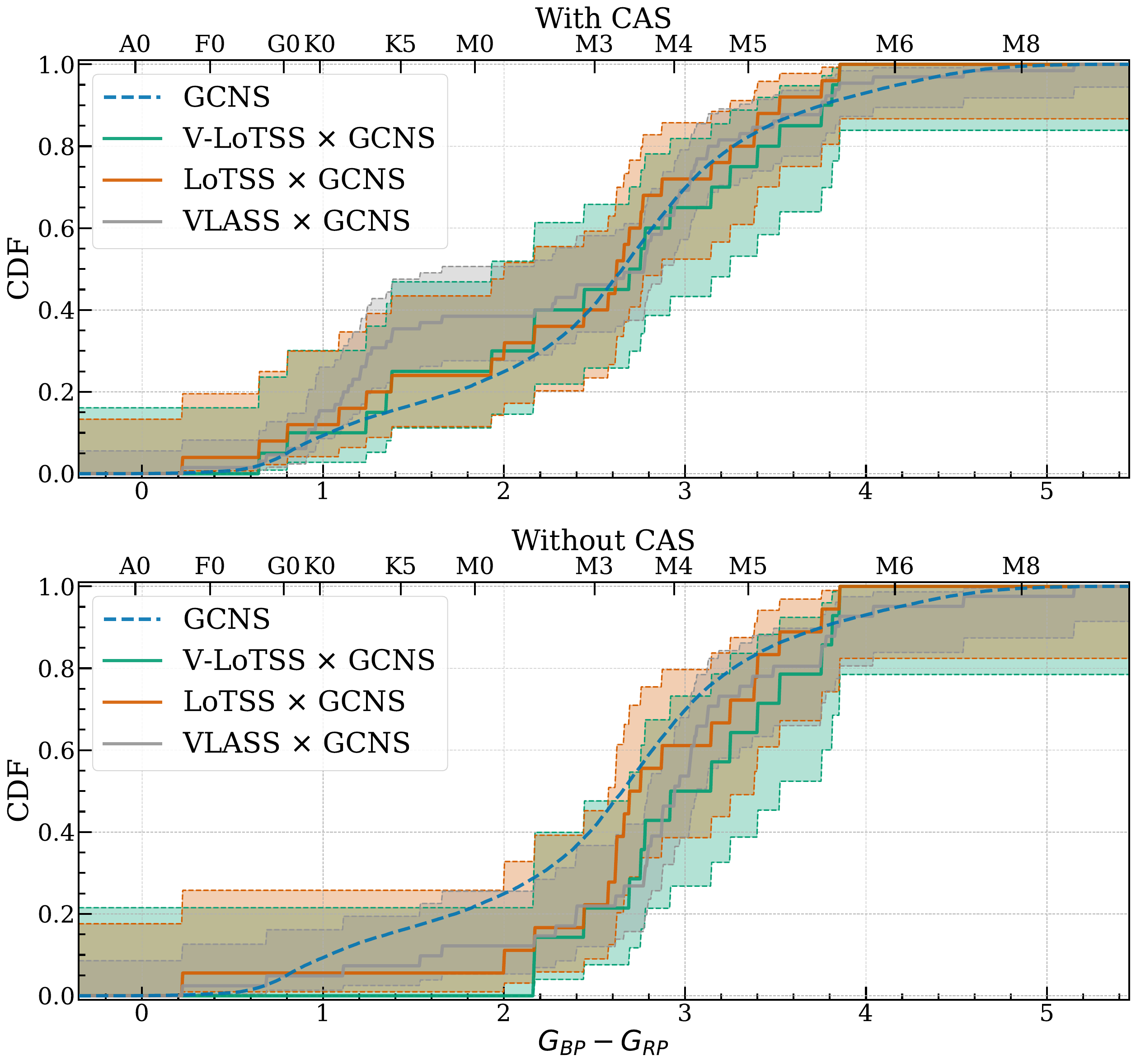}
    \caption{Cumulative distribution functions (CDFs) for the GCNS background distribution (dashed blue), the V-LoTSS $\times$ GCNS sample (green), the LoTSS $\times$ GCNS sample (orange), and VLASS $\times$ GCNS sample (grey), shown as a function of \gaia colour $G_{BP}-G_{RP}$. The top panel shows the two radio-detected populations with the inclusion of chromospherically active stellar (CAS) systems. The bottom panel shows the population without them. The shaded regions correspond to the 95\% confidence level based on a binomial distribution. As the number of sources in GCNS is much larger than the radio-detected populations, the confidence interval of GCNS is negligible. The top axis of the CDF plots indicates the nominal stellar spectral types \citep{star_colour}.}
    \label{fig:cdf_bprp}
\end{figure*}

In Fig.~\ref{fig:cdf_bprp}, we show the CDFs of the three radio-bright stellar populations (V-LoTSS, LoTSS, and VLASS) compared to the GCNS background CDF, both with and without inclusion of the chromospherically active stars (i.e. RS CVn and BY Dra). The justification for removing the CAS systems is that they likely have a different mechanism powering the radio emission as opposed to isolated stars. We note that with this removal, the GNCS background is practically unchanged due to the rarity of such systems.

We see from the top panel of Fig.~\ref{fig:cdf_bprp} that the VLASS $\times$ GCNS sample is inconsistent ($>3\sigma$) with the GCNS background distribution, specifically between the spectral types G0 and M3. This shows that RS CVn and BY Dra systems are much more luminous in radio wavelengths compared to any other stellar systems of similar spectral type, likely due to magnetic interactions, tidal locking, and high chromospheric activity \citep{dulk1985, slee2008, rscvn-toet2021}. Indeed, despite constituting less than 1\% of the stellar population \citep{cab_eker2008}, these CAS systems are so radio-luminous that they dominate the radio source counts. On the other hand, there is agreement between the GCNS background distribution and the LOFAR-detected population within uncertainties. However, since LoTSS and V-LoTSS both detect a significant amount of CAS systems despite their small number ($\lesssim$200 in our solar neighbourhood), this agreement should only be treated as a mere coincidence due to large uncertainties from the small LOFAR-detected sample size. We conclude that in terms of stellar systems, radio surveys are much more likely to detect these CAS systems than any other stellar systems.

The bottom panel of Fig.~\ref{fig:cdf_bprp} shows the radio populations after removal of the known CAS systems. Now, the conspicuous ``G0$-$M3 bump'' in the VLASS $\times$ GCNS sample is absent, leaving instead a sudden rise in radio detections around the spectral type M3 to M5. The same rise can be seen in the LOFAR sample, albeit with larger uncertainty. Visually, this region is where the VLASS population deviates from the background distribution the most. The CvM test agrees with the discrepancy as well; the $p$-value for VLASS vs GCNS is $\num{0.0039}$, whereas the $p$-values for V-LoTSS vs GCNS and for LoTSS vs GCNS are $0.039$ and $0.198$ respectively.

\subsubsection{M4 transition in stellar radio engine}
\label{sec:m4-transition}
\begin{figure}
    \centering
    \includegraphics[width=\linewidth]{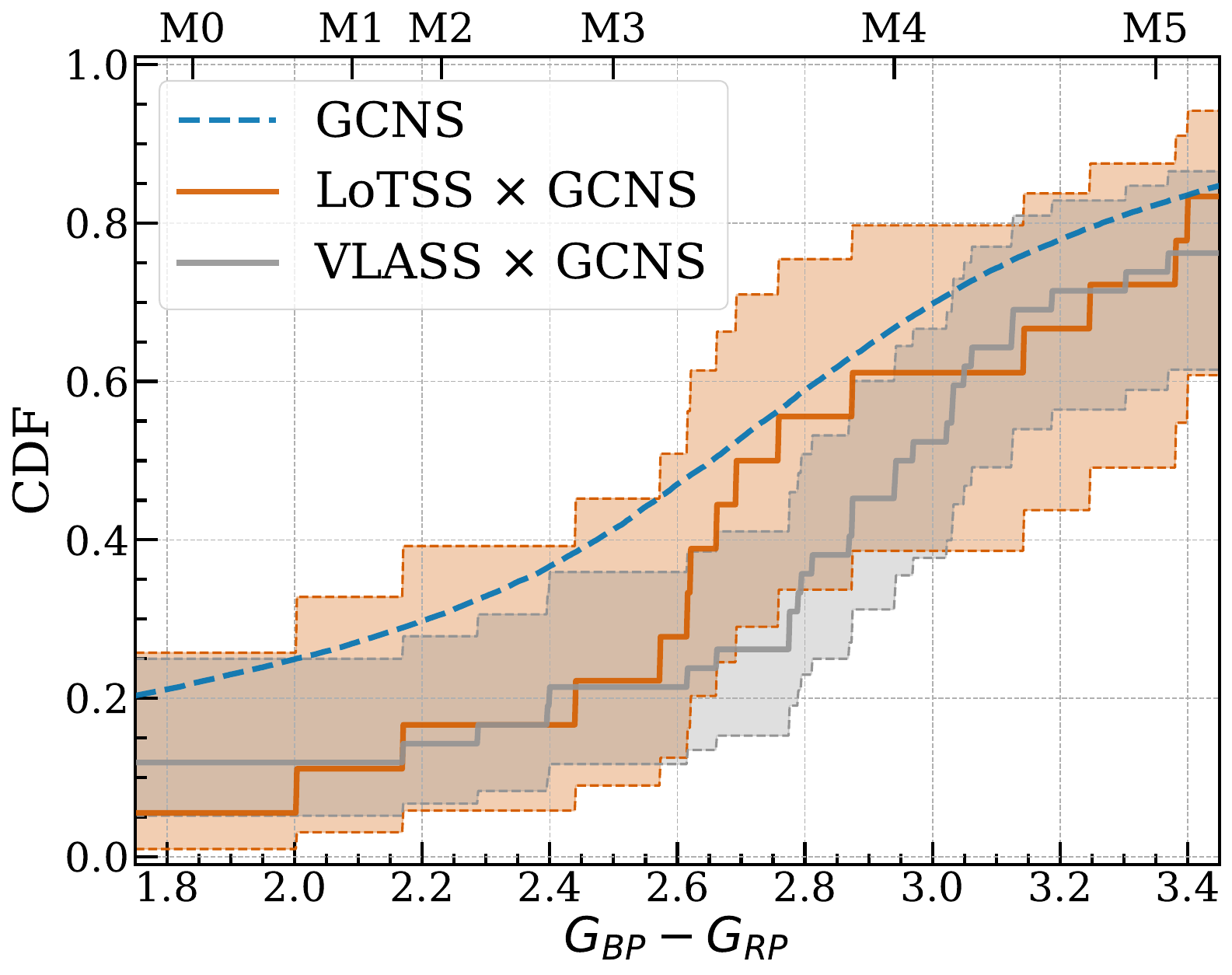}
    \caption{Zoomed-in version of the bottom panel of Fig.~\ref{fig:cdf_bprp}, with V-LoTSS $\times$ GCNS sample omitted for the sake of clarity.}
    \label{fig:cdf_bprp_M4}
\end{figure}

The low $p$-values demonstrate an evolution in radio properties with spectral type. Specifically, Fig.~\ref{fig:cdf_bprp_M4} shows where the largest discrepancy between the VLASS (grey line) and background population (blue line) is located. As one can see, while the error bar for the LoTSS population (orange region) is too large to make any definite statements, the grey line deviates from the blue line the most around spectral type M3--M4. This is approximately the spectral type at which most theoretical stellar evolution models predict the transition from partially convective regime (i.e. with a radiative-convective interface akin to the solar tachocline) into the fully convective regime (e.g. \citealp{dorman1989, clemens1998, ribas2006, morales2009}). This so-called ``M4 transition'' (see \citealp{m4-stassun2011}) is also supported by observational evidence from the properties of M dwarfs -- stellar parameters, activity lifetime, rotation, magnetic field strength, and topology (e.g. \citealp{west2008, donati2008, morin2008}).
As such, the radio evolution with spectral type may also be a direct cause of such a regime transition. This motivates us to conjecture that the M4 transition also leads to a gradual shift from a Sun-like radio engine to a Jupiter-like radio engine.

Originally, we also expect the LOFAR-detected and VLA-detected populations would be inconsistent with each other, with reasons previously mentioned in Sect.~\ref{sec:vlass}. However, in both cases where CAS systems are included/excluded, we cannot claim any significant deviations between the radio samples as they all overlap each other within 95\% confidence interval.

\subsection{Comparison to multiwavelength flare rates}
\label{sec:comparison-to-multiwavelength-flare-rates}
One natural question that arises from the detected radio evolution with spectral type is whether the evolution follows other known stellar activity indicators. This leads us to the investigation on the impact of stellar flare occurrence on radio detectability.

\subsubsection{GCNS $\times$ TESS flare CDF}
\label{sec:tess_flare_cdf}
If flares are ultimately the prime source of radio energy, and if the ratio between the flare energy that goes into the optical band and that into the radio band does not change with spectral type, then we expect the radio detections and the TESS flare detection to have the same CDF shapes. By comparing the two CDF curves, we can therefore determine if stars of some spectral type are more efficient than others at generating radio emission from flaring events.

To compare the CDFs of the optical flare rate and radio detectability, we start with the debiased average TESS flare rates versus $T_{\rm eff}$ from Sect.~\ref{sec:tess-flare-detection-bias}. The CDF of TESS optical flares as a function of spectral type is proportional to the multiplication of two CDF curves: the debiased average-TESS-ensemble-flare-rate CDF curve (recall Sects.~\ref{sec:counting-tess-flare} and \ref{sec:debiasing-tess-flare}), and the CDF of the GCNS background distribution.

To cast every CDF curve as a function of $T_{\rm eff}$, we determine the values of $T_{\rm eff}$ for each star in our sample from the TESS Input Catalog version 8 (TICv8; \citealp{ticv8_stassun2018}) catalogue if available, or from \gaia DR3 catalogue otherwise.
If neither catalogue provides a valid $T_{\rm eff}$, then we estimate $T_{\rm eff}$ from the \gaia colour $G_{BP}-G_{RP}$ using a polynomial fit (from \texttt{numpy.polyfit; deg = 10}) between colour and temperature to the data from \cite{star_colour}\footnote{\url{http://www.pas.rochester.edu/∼emamajek/EEM dwarf UBVIJHK colors Teff.txt}, Version 2022.04.16}.

\begin{figure}
    \centering
    \includegraphics[width=\linewidth]{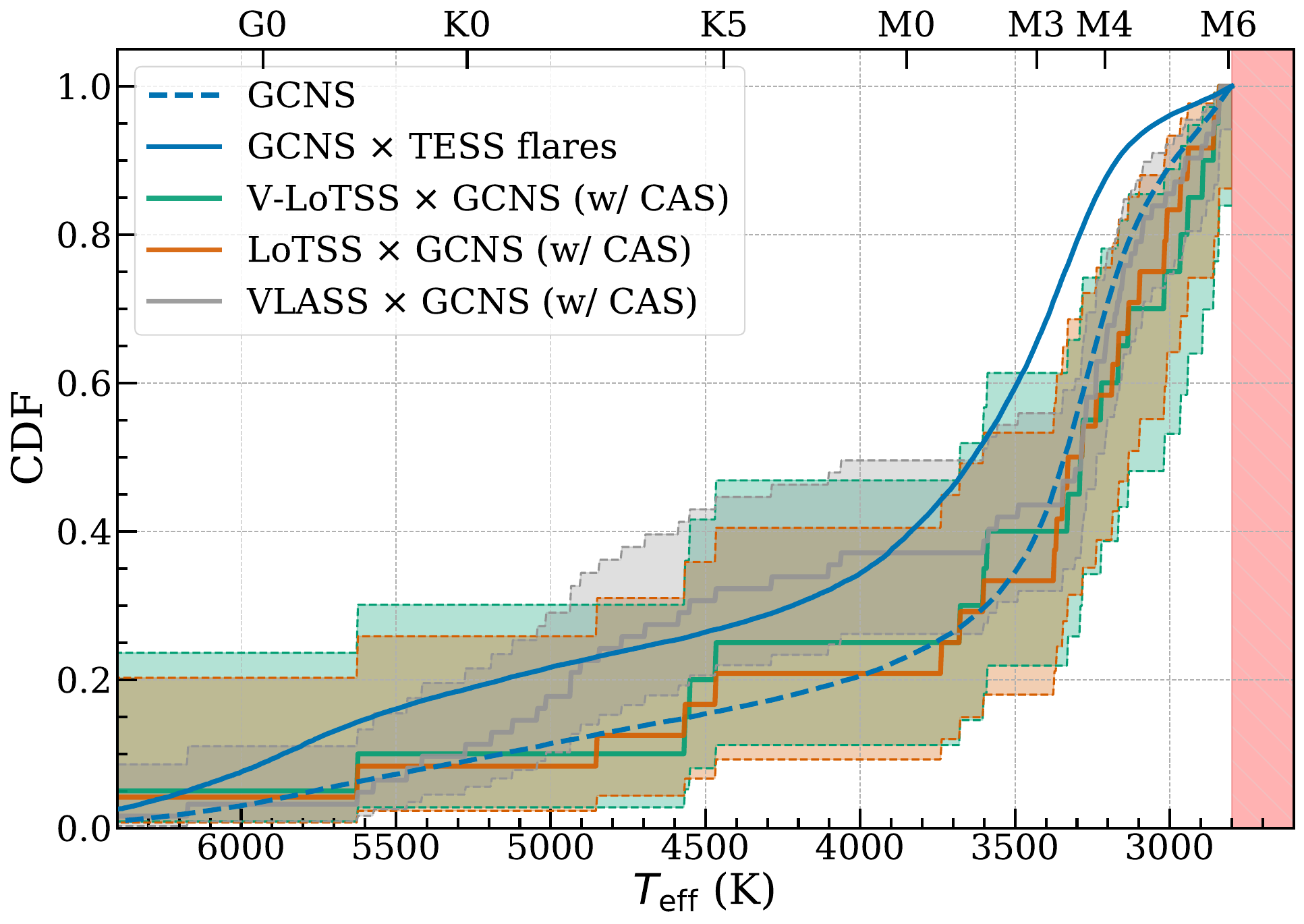}
    \caption{Cumulative distribution functions (CDFs) for the GCNS background distribution (dashed blue), the GCNS $\times$ TESS flare distribution (solid blue), the V-LoTSS $\times$ GCNS sample (green), the LoTSS $\times$ GCNS sample (orange), and VLASS $\times$ GCNS sample (grey), shown as a function of effective temperature $T_{\rm eff}$. The solid blue line is obtained from the product of the dashed blue line and TESS flare rate curve according to Fig.~\ref{fig:all_TESS_debiasing_analysis}d.
    The shaded regions correspond to the 95\% confidence level based on a binomial distribution. Note that the stars with $T_{\rm eff} < 2800$~K (around spectral type M6; shaded red box) are excluded in our analysis due to their incompleteness in TESS. The top axis of the CDF plots indicates the nominal stellar spectral types \citep{star_colour}.}
    \label{fig:cdf_tess_flare}
\end{figure}

The result is shown in Fig.~\ref{fig:cdf_tess_flare}. Note that here we did not plot the CDF without CAS systems (RS CVn and BY Dra variables) since the TESS targeting strategy is not expected to discriminate against close binaries or active flaring stars, thus there is no compelling reason to exclude them in our analysis.
As mentioned during the discussion of Fig.~\ref{fig:all_TESS_debiasing_analysis}d in Sect.~\ref{sec:tess-flare-detection-bias}, early-type stars (earlier than $\sim$G0) and late-K/early-M dwarfs are much more likely to flare compared to early- to mid-K dwarfs and mid-M dwarfs.
Conversely, late-type stars (later than $\sim$M3) are much more numerous in the \gaia stellar population. Hence, the interplay of these two factors yields a CDF (solid blue curve) not too dissimilar to the CDF of the GCNS population (dashed blue curve), as illustrated in Fig.~\ref{fig:cdf_tess_flare}.

Here, the GCNS $\times$ TESS flare CDF curve (solid blue line) is visually consistent with all three radio-detected populations at the $> 95\%$ confidence level for $T_{\rm eff} \gtrsim 3500$~K, i.e. most of the solid blue line is within the shaded regions of the radio $\times$ GCNS samples for $T_{\rm eff} \gtrsim 3500$~K.
More interestingly, it is only within the realm of M dwarfs (later than spectral type $\sim$M3 in particular) where the GCNS $\times$ TESS flare curve CDF starts to significant deviate from all three radio-detected populations. This inconsistency is confirmed by the CvM test as well; the $p$-values for GCNS $\times$ TESS flare vs V-LoTSS, LoTSS, and VLASS are $0.00770$, $0.00398$, and $0.00013$ respectively.

\begin{figure}
    \centering
    \includegraphics[width=\linewidth]{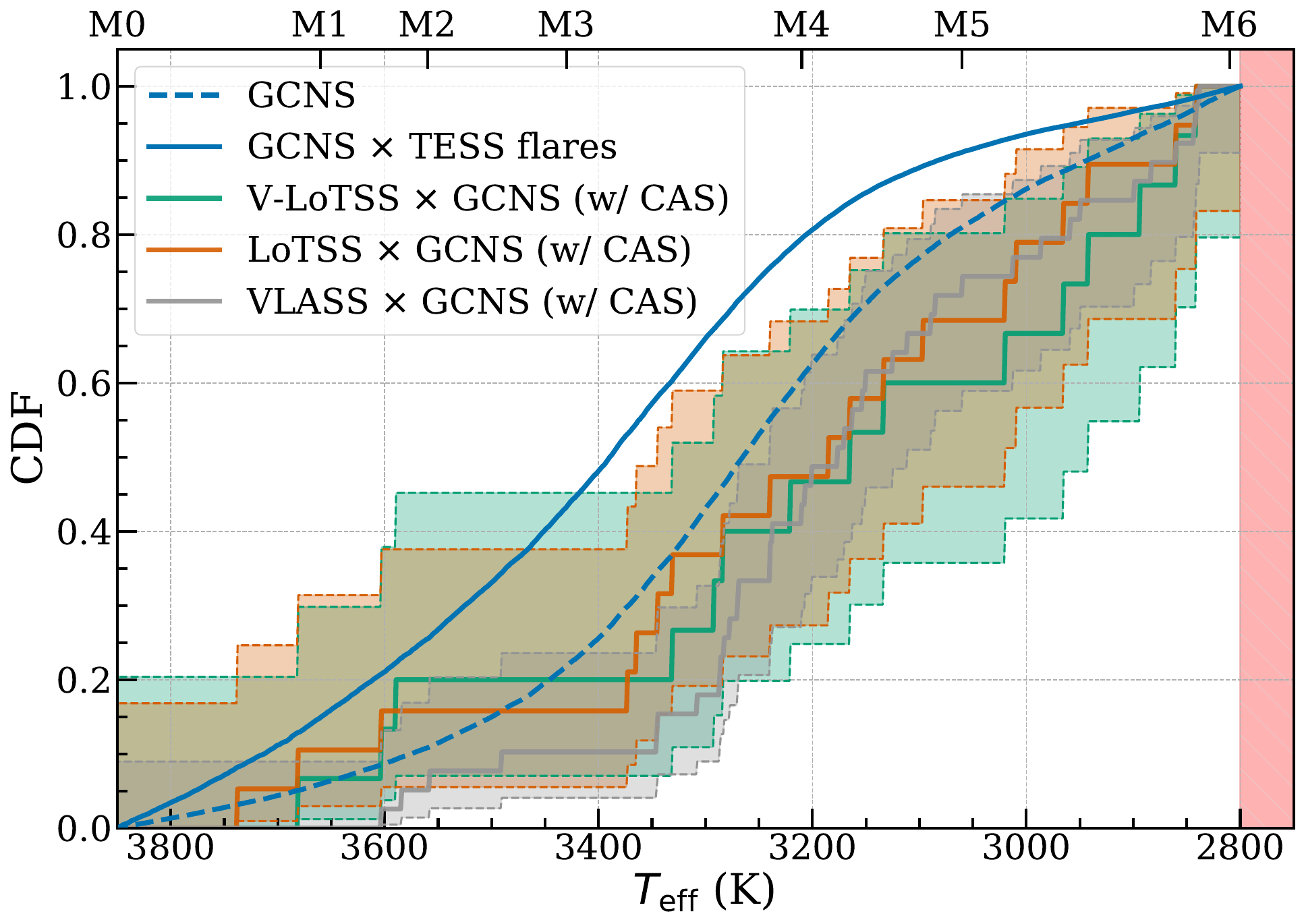}
    \caption{Same as Fig.~\ref{fig:cdf_tess_flare}, but with only M dwarfs considered. Stars before spectral type M0 are excluded from this specific analysis.}
    \label{fig:cdf_tess_flare_Monly}
\end{figure}

To confirm that the deviation is indeed the most significant in the M-dwarf population seen in Fig.~\ref{fig:cdf_bprp_M4}, we also compare the CDFs in the $T_{\rm eff}$ range of 3850K to 2800K only, as illustrated in Fig.~\ref{fig:cdf_tess_flare_Monly}. Such a $T_{\rm eff}$ range corresponds to M0--M6 dwarfs. Now, we can clearly see that the discrepancy between GCNS $\times$ TESS flare curve and the radio-detection population is the largest around spectral type M4--M5. Here, the CvM-test $p$-values are 0.00300, 0.00252, and $\num{3.66e-9}$, for V-LoTSS, LoTSS, and VLASS respectively. These $p$-values are all much smaller than their counterparts that consider the entire $T_{\rm eff}$ range.

Hence, we can confidently conclude that all three radio-detected population do not fully follow the TESS optical flare statistics. Specifically, the radio samples are consistent with TESS flare CDF for stars with $T_{\rm} \gtrsim 3500$~K, but significantly inconsistent for $T_{\rm} \lesssim 3500$~K.
This is somewhat unexpected, because one would expect that flares occur due to a universal underlying process which leads to a consistent fraction of energy channelled into radio-emitting accelerated charges. Our results instead suggest that the radio efficiency of optical flares -- the fraction of flare energy channelled into radio-emitting charges  -- evolves with spectral type. In particular, Figs.~\ref{fig:cdf_tess_flare} and \ref{fig:cdf_tess_flare_Monly} suggest that the radio efficiency of optical flares remains roughly constant throughout the stellar population until the M dwarfs, where the M4 transition (mentioned in Sect.~\ref{sec:m4-transition}) seems to manifest itself once again, this time in the form of an increase in radio efficiency of optical flares.
We hypothesise that this evolution is related to the evolution of magnetic field properties with spectral type owing to the transition from a Sun-like engine to a Jupiter-like engine.

We are aware that, despite our best debiasing efforts described in Sect.~\ref{sec:debiasing-tess-flare}, the TESS target selection is still inherently biased since the TESS 50-pc sample used in our analysis is not volume-complete, and thus the degree of biases depends on the topics of interest in the scientific community back during TESS year 1 and 2. To the best of our knowledge, however, a full analysis on the TESS target selection biases does not exist. Therefore, we assume that TESS target selection bias no longer plays a significant role when we only consider the TESS 50-pc sample.

\subsubsection{GCNS $\times$ X-ray flare CDF}
\label{sec:x-ray-flare-cdf}
Similarly, we follow the X-ray flare rate trend presented by \cite{x-ray_flare_johnstone2021} to construct a correlated CDF from the GCNS background distribution and X-ray flare statistics. Specifically, \citet[in bottom-most panel of Fig.~19]{x-ray_flare_johnstone2021} showed the evolution of flare rate with total emitted X-ray energies above $10^{32}$ erg vs stellar mass for stars older than 5 billion years old. This gives us the linear relation: $N(>10^{32} \si{erg}) = 1.11 M - 0.12$, where $N$ is the number of stellar flares per day and $M$ is the stellar mass in units of solar mass. 
From this relation, we simply construct the CDF of GCNS background distribution $\times$ X-ray flare statistics with very similar procedure from Sect.~\ref{sec:tess_flare} for the GCNS $\times$ TESS flare, but in stellar mass rather than effective temperature. In fact, the procedure is even simpler as there is no need to debias the X-ray flare statistics as previously stated in Sect.~\ref{sec:x-ray_flares}.

\begin{figure}
    \centering
    \includegraphics[width=\linewidth]{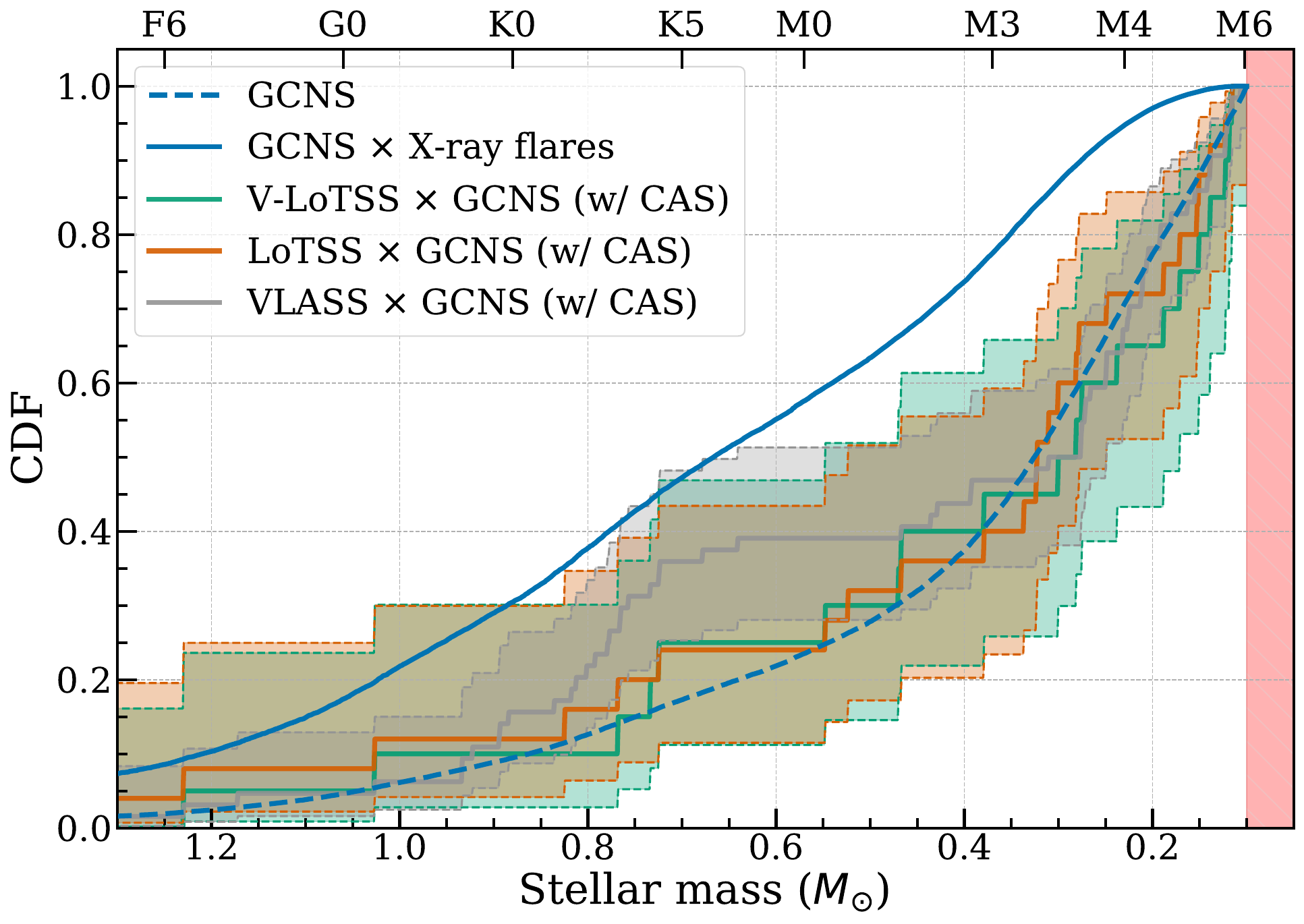}
    \caption{Cumulative distribution functions (CDFs) for the GCNS background distribution (dashed blue), the GCNS $\times$ X-ray flare distribution (solid blue) according to the study by \cite{x-ray_flare_johnstone2021} , the V-LoTSS $\times$ GCNS sample (green), the LoTSS $\times$ GCNS sample (orange), and the VLASS $\times$ GCNS sample (grey), shown as a function of stellar mass in units of solar mass $M_\odot$. The shaded regions correspond to the 95\% confidence level based on a binomial distribution. Note that \cite{x-ray_flare_johnstone2021} only consider stars of stellar mass between 0.1 and 1.2~$M_\odot$, and so we exclude stars with less than 0.1~$M_\odot$ (around spectral type M6) in this specific analysis. The top axis of the CDF plots indicates the nominal stellar spectral types \citep{star_colour}.}
    \label{fig:cdf_x-ray_flare}
\end{figure}

\begin{figure*}
    \centering
    \includegraphics[width=\linewidth]{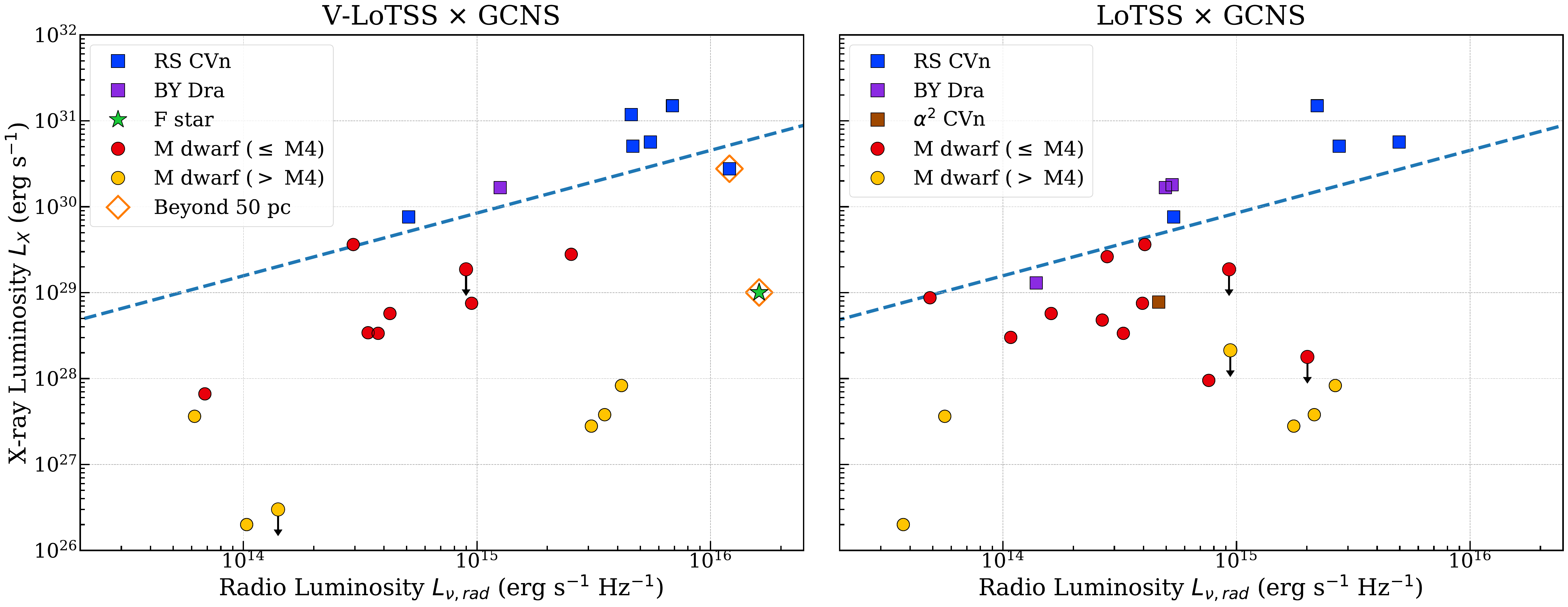}
    \caption{Radio-detected populations (V-LoTSS on the left panel, LoTSS on the right panel) plotted against the canonical G\"{u}del-Benz relationship (GBR) represented by the dashed blue line: $L_X = \num{9.48e18} L_{\nu, rad}^{0.73}$ \citep{williams2014}, where $L_X$ is the soft (0.1-2.4 keV) X-ray luminosity and $L_{\nu, \rm{rad}}$ is the 144-MHz total-flux radio luminosity for the LOFAR population. The radio sources without a detectable X-ray luminosity are indicated with a downward-pointing arrow and shown as $3\sigma$ upper limits. Colours and symbols are as in Fig.~\ref{fig:hr-lotss}, with the additional classification of M dwarfs beyond M4 (represented by yellow circles) motivated by the M4 transition mentioned in Sect.~\ref{sec:m4-transition}.}
    \label{fig:gb-lotss}
\end{figure*}

We present the result in Fig.~\ref{fig:cdf_x-ray_flare}. As expected, the GCNS $\times$ X-ray flare CDF has quite a different shape compared to the GCNS $\times$ TESS flare CDF, i.e. the X-ray flare CDF follows a more gently rising trend not seen in the TESS flare CDF. One reason for such a discrepancy could be of astrophysical origin. \cite{audard2000} found that the rate of X-ray flares is almost linearly proportional to X-ray luminosity among stars of all spectral types. Therefore, solar-mass stars appear to flare in X-ray more frequently than low-mass stars as the former are significantly more X-ray luminous on average. On the other hand, stellar activity is also strongly correlated to rotation period. M dwarfs generally rotate much faster than early-type stars since the latter experience more angular momentum loss owing to stronger stellar winds, causing a spin down of the star. This may imply that there is no universal flare energy partition between optical and X-ray bands, thus leading to disparate observed flare rates between the two wavelengths. 

Back to the comparison of X-ray flare CDF and radio-population CDF: Fig.~\ref{fig:cdf_x-ray_flare} shows that early-type stars get significantly boosted in the CDF since early-type stars have higher rate of X-ray flares than the low-mass stars. We can see that visually the GCNS $\times$ X-ray flare sample is completely inconsistent with all three radio-detected populations. Using the CvM test, we determine the $p$-values for GCNS $\times$ X-ray flare vs V-LoTSS, LoTSS, and VLASS are $\num{2.72e-4}, \num{1.07e-4}$ and $\num{8.03e-8}$ respectively. Therefore, we can conclude that the radio efficiency of X-ray flares -- the fraction of X-ray flare energy channelled into radio-emitting charges -- evolves with spectral type as well, similar to the conclusion drawn from the TESS flare CDF analysis in Sect.~\ref{sec:tess_flare_cdf}.

\begin{figure*}
    \centering
    \includegraphics[width=\linewidth]{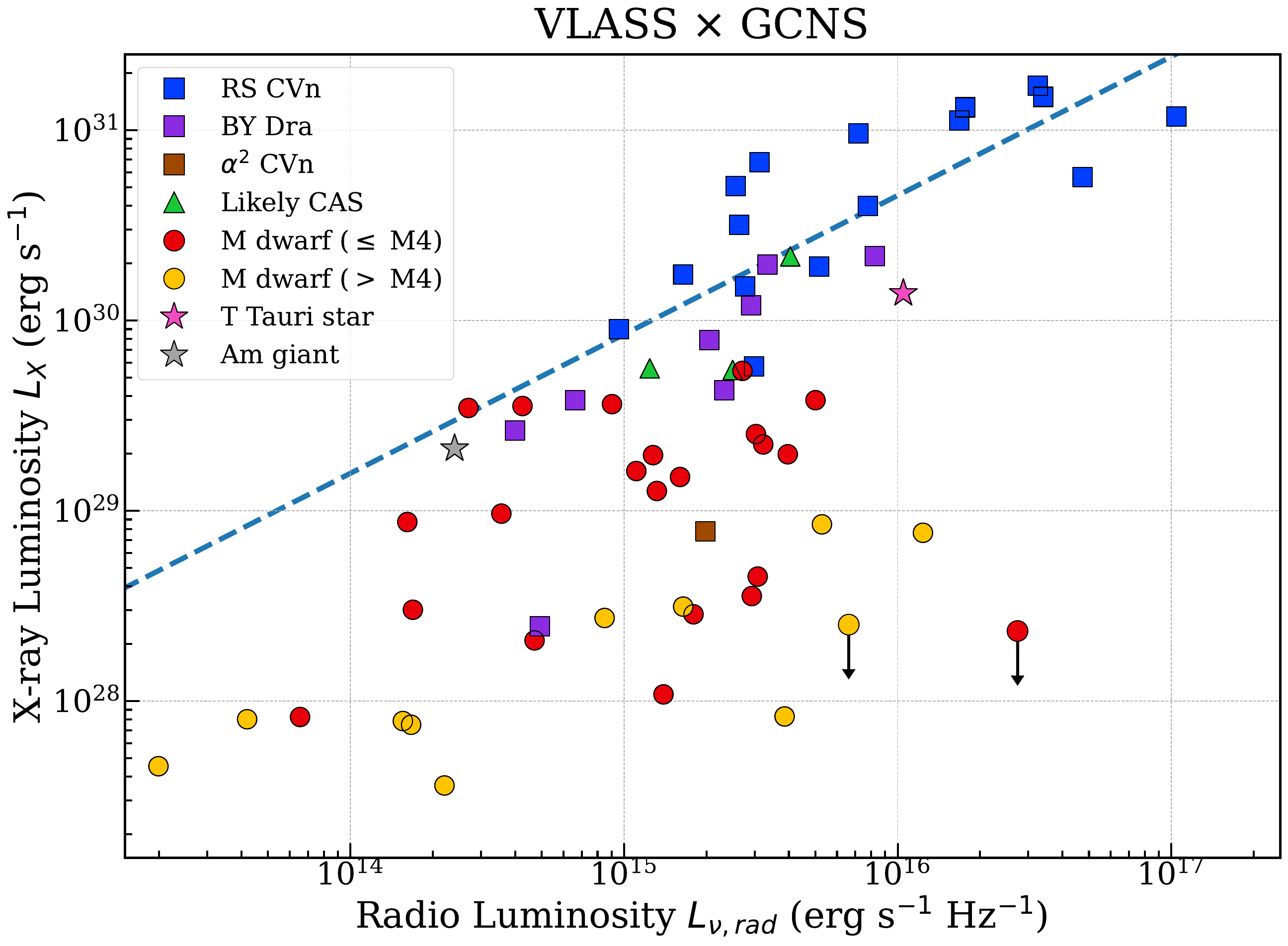}
    \caption{Same as Fig.~\ref{fig:gb-lotss}, but with the VLASS $\times$ GCNS population instead. Colours and symbols are as in Fig.~\ref{fig:hr-vlass}, with the additional classification of M dwarfs beyond M4 (represented by yellow circles) motivated by the M4 transition mentioned in Sect.~\ref{sec:m4-transition}. The one CAS system that significantly deviates from GBR is a BY Dra system called V1274 Her, which is a mid-M-dwarf binary.
    Note that two M dwarfs from the VLASS sample are not shown in this figure for the sake of clarity, since they have an X-ray luminosity of $\ll 10^{26} \si{erg~s^{-1}}$. These are LSR J1835+3259 and HD 43162C, with spectral type M8.5 and M5 respectively.}
    \label{fig:gb-vlass}
\end{figure*}

\subsection{G\"{u}del-Benz relationship}
\label{sec:gb}
We also investigate if our radio-detected samples adhere to the empirical quasi-linear relationship which correlates quiescent gyrosynchrotron 5-GHz radio spectral luminosity $L_{\nu, \rm{rad}}$ and soft X-ray luminosity $L_X$. This is known as the canonical G\"{u}del-Benz relationship (GBR; \citealp{gb_gb1993, gb_bg1994}), and the $L_X \propto L_{\nu, rad}^{0.73}$ law holds for over 10 orders of magnitude in $L_{\nu, \rm{rad}}$, from the energetic RS CVn systems all the way down to nanoflares from the Sun. Despite their vastly different emission mechanisms, where thermal Bremsstrahlung in the coronal plasma is responsible for the X-ray emission, and incoherent gyrosynchrotron emission from relativistic electrons gyrating in the coronal magnetic field for the radio emission, the presence of this relationship implies there must exist a mechanism that deposits a consistent fraction of flare energy into accelerating charges that emit in the radio and X-ray bands.

However, there are also stellar systems that violate GBR. For example, \cite{hallinan2008} found that the 5-GHz mildly circularly polarised quasi-quiescent radio emission of UCDs, including brown dwarfs, do not obey GBR. They argued that the radio emission for these systems stems from ECMI instead of gyrosynchrotron, making them radio-overluminous (with respect to GBR) as ECMI is more efficient and capable of creating luminous radio bursts (e.g. \citealp{yu2011}). Alternatively, it could be that the energetics of these systems or their magnetospheric structure do not support a stable thermal corona to form, which leads to their X-ray underluminosity \citep{berger2010, williams2014}.
\cite{joe_nature} found that the highly polarised radio emission at 144\,MHz from M dwarfs also violates GBR; these objects are very unlikely to be operating in an incoherent mechanism that is gyrosynchrotron due to their high degree of circular polarisation, and thus a coherent emission mechanism must be responsible for 144-MHz radio emission of these objects. 

And so, if one were to accept the explanation that the 5-GHz radio emission from the stellar systems that established the original GBR stems from gyrosynchrotron in the coronal magnetic field, one must expect our V-LoTSS sample to depart from the GBR. Conversely, the VLASS sample should obey GBR, except for stars of very late spectral type -- into the realm of UCDs -- where their corona starts disappearing. 

\subsubsection{GBR vs radio-detected samples}
\label{sec:gb-vs-radio}
Here, we plot the soft X-ray luminosity $L_X$ against radio spectral luminosity $L_{\nu, \rm{rad}}$ of the LOFAR-detected and VLA-detected populations, as shown in Figs.~\ref{fig:gb-lotss} and \ref{fig:gb-vlass} respectively, to see which stellar system obeys GBR and which deviates from it. As previously stated, none of the V-LoTSS sample should obey GBR as these stellar systems operate in coherent mechanisms rather than gyrosynchrotron processes. As expected, such is the case for most M dwarfs and the F-type star; most of these stellar systems deviate from the relationship by a few orders of magnitude.
However, the chromospherically active stellar systems in our V-LoTSS sample surprisingly adhere to GBR. This peculiar phenomenon was already discussed by \cite{harish-gb}, where they report that highly polarised 144-MHz radio emission from RS CVn/BY Dra systems and other high-activity stars unexpectedly follows the GBR. In the other two populations (LoTSS and VLASS), CAS systems also consistently obey GBR, though this is less surprising as previously mentioned.

\begin{figure}
    \centering
    \includegraphics[width=\linewidth]{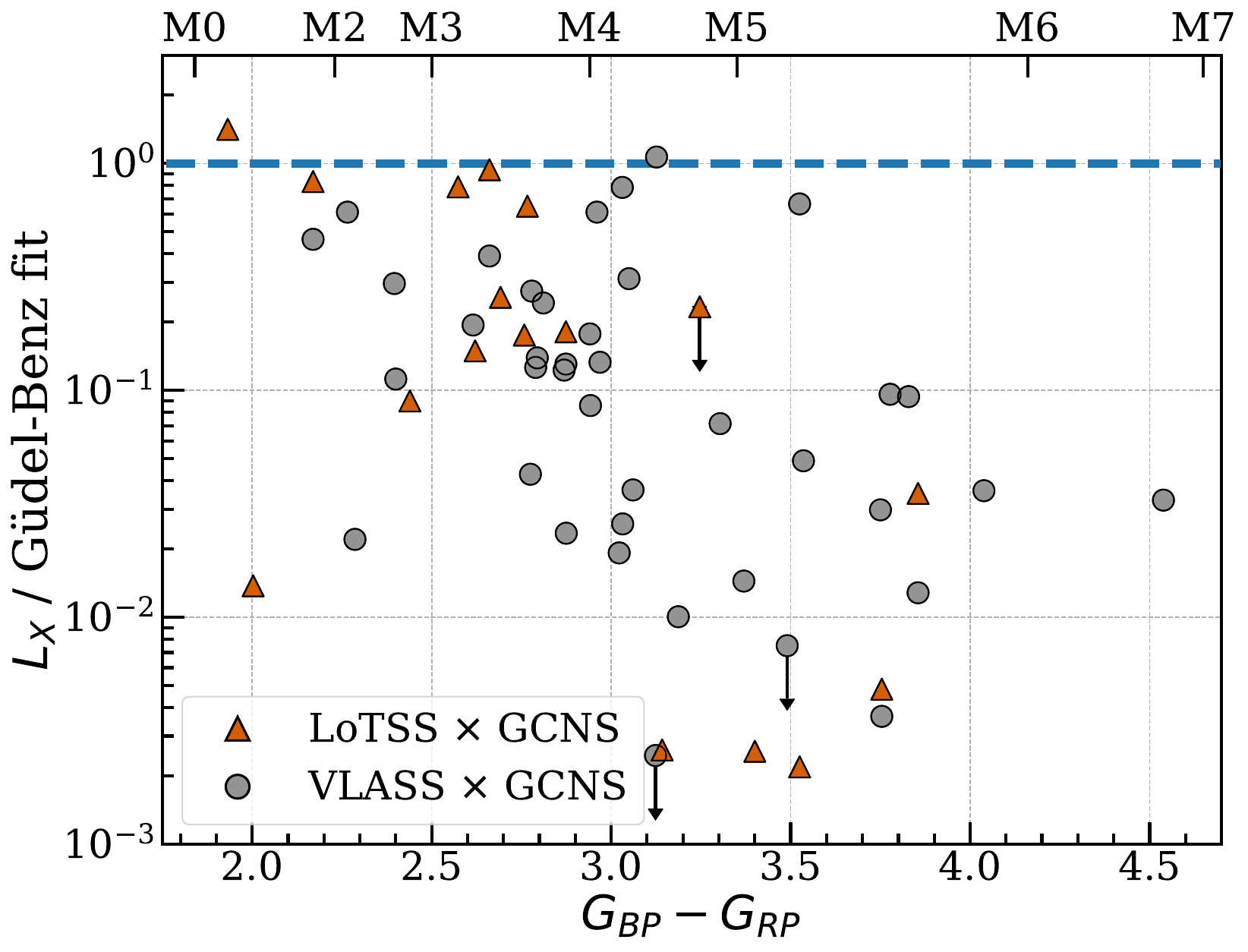}
    \caption{Residuals from the G\"{u}del–Benz relationship versus \gaia $G_{BP}-G_{RP}$ colour. For the sake of clarity, the V-LoTSS $\times$ GCNS sample is omitted, as well as the two M dwarfs from the VLASS sample (grey circle) with residuals $\ll 10^{-3}$. They are LSR J1835+3259 and HD 43162C, with spectral type M8.5 and M5 respectively. The radio sources without a detectable X-ray luminosity are indicated with a downward-pointing arrow and shown as $3\sigma$ upper limits.
    The top axis of the CDF plots indicates the nominal stellar spectral types \citep{star_colour}.}
    \label{fig:gb-residual}
\end{figure}

As for the G\"{u}del-Benz (GB) deviators, we can see from Figs.~\ref{fig:gb-lotss} and \ref{fig:gb-vlass} that virtually every single M dwarf in all the radio-survey population are consistently radio-bright/X-ray-dim, and most of them by a few orders of magnitude as well. 
In particular, the M4 transition also seems to manifests itself in the GBR plot, as M dwarfs beyond $\sim$M4 significantly deviate from the rest of the population on average. 
Intriguingly, there exist two stellar sources (LSR J1835+3259 and HD 43162C) that deviate from GBR by more than 5 orders of magnitude, and both of them are later than M4 as well.
In Fig.~\ref{fig:gb-lotss}, there is a clear gap between the early-M-dwarf and the late-M-dwarf populations, especially in V-LoTSS. The deviation from the GBR with respect to spectral type is better illustrated in Fig~\ref{fig:gb-residual}. Here, we can see that the radio detection becomes significantly more radio-overluminous/X-ray-underluminous than the values predicted by GBR in the region of M4--M6. This gap might also signify a bimodality in the dynamo of the M-dwarf population (e.g. \citealp{mclean2012, cook2014, williams2014}), supported by geodynamo simulations and Zeeman-Doppler imaging (ZDI) analysis \citep{morin2010,morin2011}.

To confirm the monotonic inverse relationship between the $L_X$ residual from GBR and spectral type for each of our radio M-dwarf sample in a statistical manner, we calculate the Kendall's $\tau$ \citep{kendall1938}, which is a measure of the relationship between two variables -- $L_X /~$GB fit and $G_{BP}-G_{RP}$ in our case. This allows us to perform a $\tau$ hypothesis test on each of our radio-detected samples, where the alternative hypothesis is that the rank correlation is less than zero, i.e. a negative correlation between $L_X /~$GB fit and $G_{BP}-G_{RP}$. To be succinct, we obtain a $p$-value much smaller than 0.05 and a negative $\tau$ coefficient (around $-0.3$ to $-0.4$) for all three radio M-dwarf samples.
We thus conclude there exists a negative correlation between goodness of GB fit to the radio data and spectral type in M dwarfs from all three radio surveys.

\subsubsection{Rotation-related activities}
\label{sec:rotation-related-activities}

Even though the M dwarfs are consistently radio-bright or X-ray-dim with respect to G\"{u}del-Benz relationship, there seems to be no obvious new scaling law that can be established as these GBR violators generally scatter around the $L_X$ against $L_{\nu, \rm{rad}}$ plot in Figs.~\ref{fig:gb-lotss} and \ref{fig:gb-vlass}. Such was the original conclusion from \cite{joe_nature}; the highly polarised radio-emitting M dwarfs do not establish a clear new relationship even when accounting for chromospheric activity indicators such as H$\alpha$ luminosity, rotation period, and Rossby number. Therefore, they suggest that the chromospheric activity has limited influence on the detected radio luminosity. In contrast, previous GHz-frequency studies discovered that the fastest rotating M dwarfs have a higher radio surface flux and radio detection fraction \citep{mclean2012}.
Here, we attempt to apply similar investigation in our VLA-detected sample.

As mentioned in Sect.~\ref{sec:pre-main-sequence}, the majority of the non-CAS stellar objects are members of young moving groups (YMGs). This already indicates that they are most likely very fast-rotating stars since they have yet to lose significant angular momentum due to stellar wind. The implication, therefore, is that perhaps the fastest rotators do have a higher radio detection fraction in the 3-GHz band, similar to previous findings by \cite{mclean2012}. Unfortunately, finding literature values of rotation period $P_{\rm rot}$ for the VLASS-detected M dwarfs proves to be difficult, as the studies of most of these M dwarfs are lacking in general since they are otherwise run-of-the-mill objects of little interest to the scientific community. Additionally, most of them are too distant and/or too active to be traced by detailed M-dwarf surveys such as CARMENES\footnote{The Calar Alto high-Resolution search for M dwarfs with Exoearths with Near-infrared and optical Échelle Spectrographs; \url{https://carmenes.caha.es}} and MEarth \citep{mearth_irwin2009}. However, for the stellar systems which have measured rotation period (e.g. GJ 3237, GJ 4185, V374 Peg, and V1274 Her), they indeed rotate very fast (typically $P_{\rm rot} < 1$ day). Therefore, although it is currently not possible to make any definite statements, we believe that rotation period plays an important role on the radio detectability of these stellar objects.

\section{Conclusion}
\label{sec:conclusion}
We conducted a search for radio-bright stellar systems using existing radio sky survey catalogues -- LoTSS, V-LoTSS, and VLASS -- and crossmatched them with the \gaia data along with proper-motion correction. The high sensitivities and resolutions of these radio surveys provided datasets to isolate an untargeted sample of radio stellar systems such that the statistical properties of the population could be determined accurately. We ascertained our stellar radio samples are not dominated by chance-coincidence associations by using the \gaia Catalogue of Nearby Stars.
To the best of our knowledge, this is the first untargeted search for ALL stellar radio sources down to sub-mJy level, and thus the first statistical analysis of radio activity trend with spectral type using an unbiased radio-emitting stellar population.

We found that both the LoTSS (V-LoTSS included) and VLASS populations can be generally classified into two categories: CAS systems and M dwarfs. In particular, the VLA-detected sample also contains many more peculiar stellar systems such as YSOs, seemingly isolated K dwarfs, and chemically peculiar stars. The reason why these systems are not present in the LOFAR-detected samples is most likely due to the 3 different survey parameters between LoTSS and VLASS. Firstly, the survey sky coverage: VLASS covers much more of the sky than LoTSS-DR2 (33885 deg$^2$ vs 5634 deg$^2$), thus it is currently easier to find rare stellar objects in VLASS compared to LoTSS. In addition, VLASS also covers the entire galactic plane visible to the VLA ($\delta > -40^\circ$), while LoTSS-DR2 focused on imaging regions at high Galactic latitude and thus deliberately avoids the galactic plane, in which most of the nearby young moving groups (AB Dor, TWA, $\beta$ Pic, etc.) are located. 
Secondly, the survey frequency coverage: these two surveys are probing very different radio regimes, as LoTSS is more likely to capture coherent emission compared to VLASS, as mentioned in Sects.~\ref{sec:vlass} and \ref{sec:pre-main-sequence}. The predominant radio emission mechanism for these peculiar systems may have most of its power emitted in the decimetre-wave (VLA band) regime rather than metre-wave (LOFAR band). This could happen in the case where most of their radio emissions are gyrosynchrotron emission stemming from flares \citep{incoherent-nindos2020}.
Thirdly, the survey exposure time: each LOFAR pointing in LoTSS is observed for a total of 8 consecutive hours, whereas VLASS creates 3-minute images. Hence, VLASS is much more susceptible to stellar radio sources with high variability.

Using CDF visualisation and the CvM statistical test, we demonstrated an evolution in radio properties with spectral type. The discrepancy between the radio-detected and the GCNS CDF curves is the largest for the VLASS $\times$ GCNS population, simply due to the fact that the VLASS sample size is larger -- more than twice the sample size of LoTSS/V-LoTSS. In particular, we showed that radio-detected populations deviate the most from the GCNS background population around spectral type M3--M4. This is remarkable due to its location at the so-called ``M4 transition'', where most theoretical stellar evolution models predict the transition from partially convective into the fully convective regime. This transition drastically changes the dynamo of stars that is responsible for the generation of magnetic field. As such, we suggest that the radio evolution with spectral type may be related to the changing nature of the stellar dynamo.

We also studied the evolution of radio efficiency with spectral type via comparison to other known stellar activity indicators. In particular, we tested the impact of stellar flare occurrence on radio detectability using the TESS and X-ray flare statistics. 
The flare statistics in these two wavelengths do not show the same trend: given a particular energy threshold above which flares were counted, we saw the general trend of flare rate decreasing as we moved towards stars of later spectral type in the X-ray flare statistics, and yet the trend seen in TESS flare statistics is much more complicated. Despite that, the radio detectability does not seem to follow neither of these trends.
To test if the ratio between the flare energy that goes into the optical/X-ray band and that into the radio band also changes with spectral type, we compared the CDF curves of the radio detection and the TESS/X-ray flare detection. We discovered that the GCNS $\times$ X-ray flare curve is clearly inconsistent with all three radio-detected populations at the $\gg 95\%$ confidence level, both visually and statistically. The inconsistency in the GCNS $\times$ X-ray flare curve is much more subtle: the GCNS $\times$ TESS flare curve agrees with the radio populations throughout most of the stellar population until we reached the M-dwarf population. There, the TESS flare CDF significantly deviates from the radio-population CDFs, especially around spectral type M4--M5. 
Therefore, the radio efficiency of flares in both optical and X-ray -- the fraction of flare energy channelled into radio-emitting charges -- evolves with spectral type, with later M-dwarfs being more radio-efficient than the rest of the stellar population. We hypothesise that such an evolution may be owing to two factors (or a combination thereof): (i) the efficiency with which charges are accelerated and/or the lifetime of accelerated charges may itself increase for stars of later spectral types; and (ii) the emergence of large-scale axisymmetric magnetic field in fully convective stars may assist more efficient radio emission mechanisms such as ECMI and/or provide new magnetospheric acceleration mechanisms in addition to flare-related processes.

We also found that nearly every single M dwarf in all the radio-survey population are consistently and significantly more radio-overluminous or X-ray-underluminous with respect to GBR. In particular, the M4 transition is alluded once again, as M dwarfs beyond $\sim$M4 significantly deviate from the rest of the population on average. This gap might be signifying a bimodality in the dynamo of the M-dwarf population (e.g. \citealp{mclean2012, cook2014, williams2014}), supported by geodynamo simulations and ZDI analysis \citep{morin2010, morin2011}. We also confirmed a negative correlation between goodness of GB fit to the radio data and spectral type in M dwarfs from all three radio surveys, i.e. late-M dwarfs deviate from GBR more than early M dwarfs. This was done by performing the Kendall's $\tau$ test. We propose follow-up ZDI observations on our radio stellar population to see if their radio emission is dominated by the large-scale axisymmetric magnetic field. Additional light curve analysis of the M dwarfs mentioned in Sect.~\ref{sec:pre-main-sequence} to get their rotation periods is also vital in helping us prove our conjecture of rotation period playing an important role on the radio detectability of the stellar objects in VLA band.

The fact that radio-detected stellar population is dominated by the CAS systems and M dwarfs regardless of the radio surveys implies there must exist some similarities between these two classes of stars such that they are consistently radio bright. Therefore, we postulate that all of these radio-bright stellar systems exhibit strong large-scale stellar magnetic field. CAS systems and M dwarfs (especially young ones) can generate magnetic field strength up to kilogauss (e.g. \citealp{giampapa1983, saur1986, petit2004, kochukhov2013} for CAS systems, \citealp{johns-krull1996, reiners2009, morin2010, afram2019, kochukhov2021} for M dwarfs, and \citealp{gregory2012} for YSOs). If the hypothesis of radio predominantly tracing magnetic field dynamics is true, then the predominant radio emission mechanism is likely to be ECMI. Since electron cyclotron maser emits mostly at the fundamental and the second harmonic of the cyclotron frequency: $\nu_B = 2.8B$\,MHz, where $B$ is the local magnetic field strength of the radio emission region \citep{melrose1982}, the main difference between the stellar population in LoTSS and VLASS would be that they have different magnetic field strength; $B \sim 50$\,G would peak at the LOFAR band, while $B \sim 1$\,kG at VLA band.

Currently, the LOFAR-detected sample size is severely limited by the incomplete sky coverage of LoTSS/V-LoTSS, as mentioned in Sect~\ref{sec:pre-main-sequence}. This makes interpreting the results of statistical test and drawing conclusions from the LOFAR sample more difficult due to small number statistics. By the completion of the LoTSS survey of the Northern Sky, this will effectively increase the current LoTSS/V-LoTSS sample size threefold. Having the new radio stellar detections will most certainly give us definite conclusion on e.g. whether there exists a discrepancy between the LOFAR-detected and VLA-detected stellar population. 
Moreover, as mentioned in Sect.~\ref{sec:v-lotss}, looking at the Stokes V sky for radio stellar sources is statistically motivated, since the chance-coincidence association is effectively zero in a circular-polarised radio sky survey. 
Looking at the Stokes V sky is also physically motivated: circular polarisation information may provide a measure of coronal density for stars, and magnetic field strength for brown dwarfs and exoplanets. Therefore, an equivalent of V-LoTSS for VLASS would be very useful for understanding the underlying radio emission mechanism of the VLA radio-bright population.

\begin{acknowledgements}
We thank S. Bloot, E. Ilin, A. A. Vidotto, T. J. Maccarone, and B. J. S. Pope for the useful discussions.
We also thank the participants in Lorentz Center Workshop 2022: Life Around a Radio star; and the attendees of the New Eyes on the Universe: SKA and ngVLA conference, for the fruitful discussions.
Additionally, we thank the referee of this paper J. L. Linsky for the useful comments.
TWHY and HKV acknowledge funding from EOSC Future (Grant Agreement no. 101017536) projects funded by the European Union’s Horizon 2020 research and innovation programme.
The National Radio Astronomy Observatory is a facility of the National Science Foundation operated under cooperative agreement by Associated Universities, Inc.
CIRADA is a partership between the University of Toronto, the University of Alberta, the University of Manitoba, the University of British Columbia, McGill University and Queen's University, in collaboration with National Research Council of Canada, the US National Radio Astronomy Observatory, and Australia's Commonwealth Scientific and Industrial Research Organization. CIRADA is funded by a grant from the Canadian Foundation for Innovation 2017 Innovation Fund (Project 35999) and by the Provinces of Ontario, British Columbia, Alberta, Manitoba and Quebec.
This work made use of the following: 
(i) the functionalities provided by \texttt{numpy}, \texttt{scipy}, \texttt{astropy}, and \texttt{matplotlib}, 
(ii) the SIMBAD database and the VizieR catalogue access tool, operated at CDS, Strasbourg, France, 
(iii) NASA’s Astrophysics Data System (ADS) Abstract Service; the ADS is operated by the Smithsonian Astrophysical Observatory under NASA Cooperative Agreement 80NSSC21M0056, and 
(iv) CARTA (Cube Analysis and Rendering Tool for Astronomy), an image visualisation and analysis tool designed for the VLA.

\end{acknowledgements}

\bibliographystyle{aa}
\bibliography{library}

\appendix

\section{Radio stellar samples}
\label{app:tables}

The radio $\times$ GCNS crossmatching samples are listed in Tables~\ref{tab:lotss-table} for LoTSS, \ref{tab:v-lotss-table} for V-LoTSS, and \ref{tab:vlass-table} for VLASS.

\section{Peculiar radio stellar systems}
\label{app:peculiar-matches}

The following are the interesting radio-bright stellar systems from Tables~\ref{tab:lotss-table}, \ref{tab:v-lotss-table}, and \ref{tab:vlass-table}.

\subsection{HD 220242}
\label{app:hd220242}

HD 220242 (\gaia DR3 2844641816070874240) is a main-sequence F5V star at a distance of 69.3 pc. As mentioned in Sect.~\ref{sec:vlotss-gcns}, it is the only main-sequence star other than an M dwarf in the V-LoTSS $\times$ GCNS population that is not in a known binary. The absence of double-lined spectrum nor an excess radial velocity signature strongly indicates the lack of a close stellar companion \citep{nordstrom2004}. Moreover, we already mentioned in Sect.~\ref{sec:gb} that unlike RS CVn and BY Dra systems, HD 220242 does not obey GBR, despite having a X-ray luminosity typical for its spectral type \citep{hd220242_x-ray}. And yet, HD 220242 is by far the most radio-luminous object in our sample, even more so than the chromospherically active stellar systems (recall Fig.~\ref{fig:gb-lotss}). This makes the F5 star an obvious anomaly since we did not expect any isolated main-sequence star earlier than an M dwarf to be radio emitter, let alone one that is so bright and highly circularly polarised ($78 \pm 16\%)$.
Furthermore, HD 220242 shows a significant tangential acceleration (i.e. motion beyond that owing to unaccelerated proper motion and parallactic shift) between Hipparcos and \gaia\ mean positions \citep{kervella2019, kervella2022}. Such a proper-motion anomaly implies that the source has a distant ($>3~\si{AU}$) stellar companion of a stellar mass (around 187 $M_J$ assuming an orbit of 10\,AU) consistent with an M dwarf\footnote{Note that the mass of the stellar companion is degenerate with its orbital radius $r$, i.e. its mass cannot be determined by proper motion anomaly alone. However, for the HD 220242 system, the companion always has a mass consistent with an M dwarf (but differs in spectral type) when assuming different but reasonable values of $r$ by \cite{kervella2022}.} \citep{kervella2022}. Therefore, it is more likely that the radio emission actually stems from the low-mass companion of HD 220242 rather than the host F5 star itself.

In an attempt to prove such a hypothesis, we observed HD 220242 with VLA (Project ID: 22A-501; PI: Vedantham) in C-band (4--8\,GHz) as LOFAR-VLBI (which stands for very-long-baseline interferometry) does not have sufficient resolution for this task. The intention was to perform a follow-up VLBI campaign in order to resolve the location of the radio emission compared to the target, given that the radio emission does extend to GHz frequencies. Unfortunately, we are unable to detect any radio emission in the vicinity of the target. The flux density 3$\sigma$ upper limit for HD 220242 in this observation is around 15\,$\mu$Jy. Note that, however, radio stellar sources are highly variable, as evidenced by the lack of radio detection for HD 220242 in LoTSS. Recall that unlike V-LoTSS, LoTSS creates mosaicked images which may wash out variable radio sources.

\subsection{$\alpha$~For B}

\begin{figure*}
    \centering
    \begin{subfigure}{.5\textwidth}
          \centering
          \includegraphics[width=\linewidth]{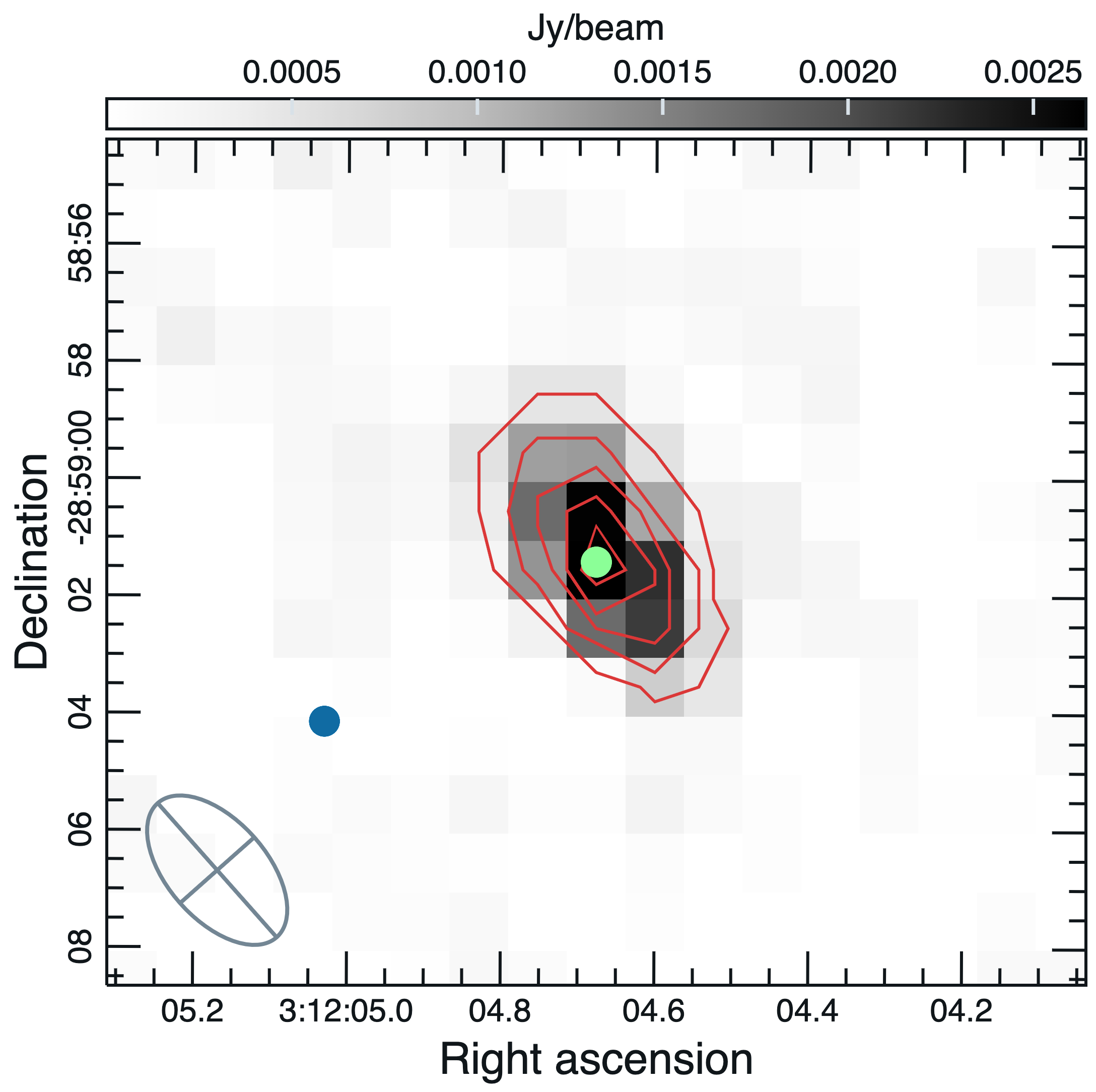}
    \end{subfigure}%
    \begin{subfigure}{.5\textwidth}
          \centering
          \includegraphics[width=\linewidth]{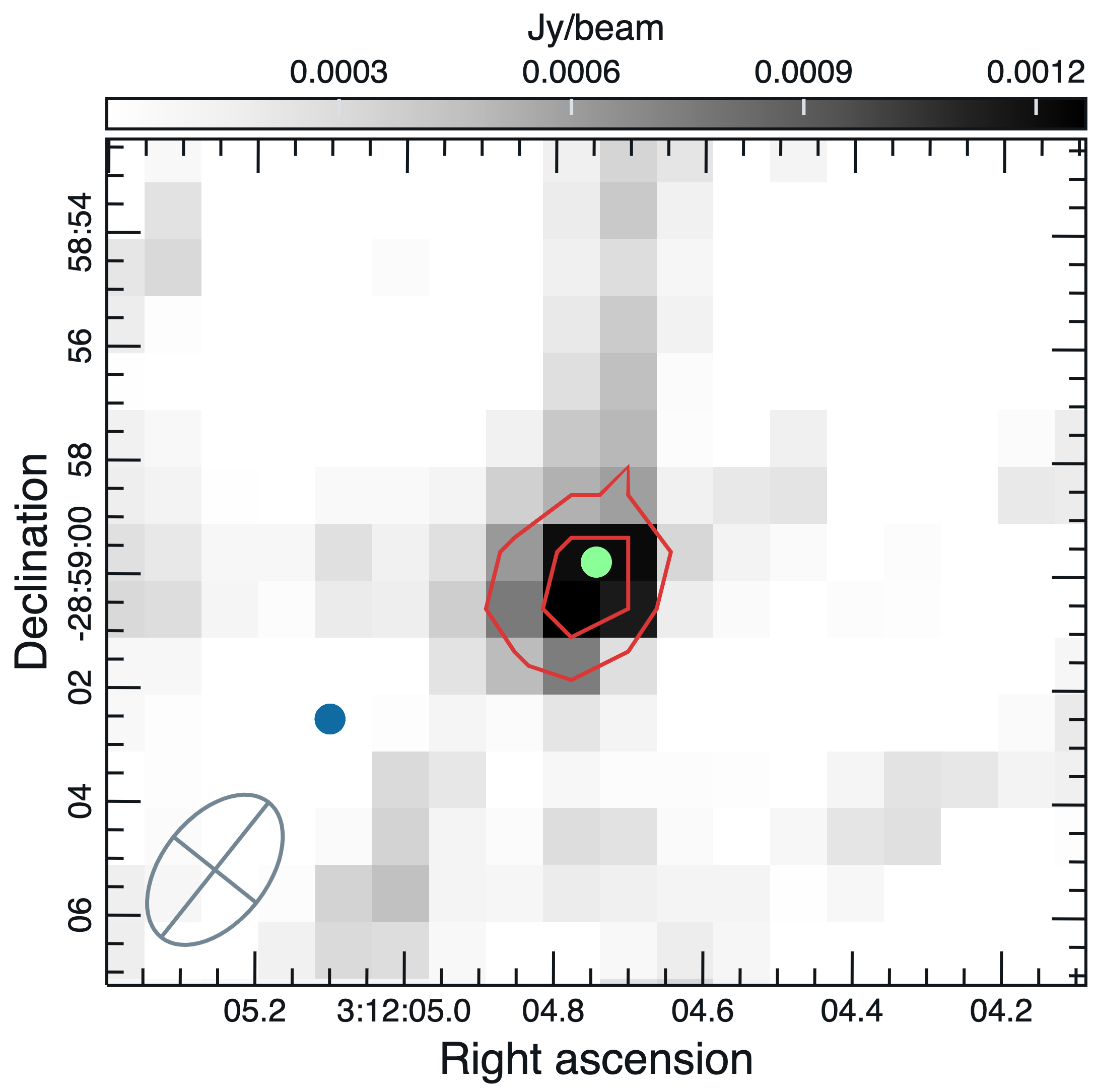}
    \end{subfigure}
    \caption[Caption for LOF]{
    VLASS Epoch 1 (left panel) and Epoch 2 (right panel) images using CIRADA cutout service\footnotemark, centred at the corresponding radio source. In both images, the lime dot represents the \gaia position of $\alpha$ For B with proper-motion correction, and the blue dot represents that of $\alpha$ For A. The FWHM beam size is shown in the bottom left corner of each image as a grey crossed circle. The rms noise $\sigma$ is around 140\,\si{\mu Jy} in both epochs. The contours are of levels with a increment of 4$\sigma$, i.e. 4$\sigma$, 8$\sigma$, 12$\sigma$, etc. The peak flux of the source is around 3.3\,mJy for Epoch 1 and 1.5\,mJy for Epoch 2.
    }
    \label{fig:alpha-for-vlass-image}
\end{figure*}

Alpha Fornacis B ($\alpha$ For B = \gaia DR3 5059349158319689344) is the secondary K2V component of the visual binary Alpha Fornacis ($\alpha$ For), where the primary component Alpha Fornacis A ($\alpha$ For A) is a F-type solar-age slow-rotating subgiant \cite{alpha-for_blue-straggler_fuhrmann2015} with a long orbital period of $280 \pm 11$ years \citep{alpha-for_izmailov2019}. The angular separation between $\alpha$ For A and B is $4.05 \pm 0.08\arcsec$ \citep{alpha-for_izmailov2019}. We clearly identify $\alpha$ For B as the radio-loud component in this system because the radio source is $<1\arcsec$ from B and $>4\arcsec$ from A in both epochs of VLASS (see Figure~\ref{fig:alpha-for-vlass-image}).

\cite{alpha-for_blue-straggler_fuhrmann2015} discovered that $\alpha$ For B shows significant chromospheric activity with strong H$\alpha$ and Ca II H\&K lines, whereas nothing alike can be seen on $\alpha$ For A. They argued that since $\alpha$ For A is about solar age and its slow-rotating inactive nature is confirmed by spectroscopy, this means $\alpha$ For B cannot be young and thus has reignited its activity, possibly due to some mass transfer or merger events \citep{alpha-for_blue-straggler_fuhrmann2015}.
Finally, \cite{alpha-for_white-dwarf_fuhrmann2016} followed up this system and found a close companion (dubbed $\alpha$ For Bb) around the K dwarf (dubbed $\alpha$ For Ba) using radial velocities. They determined $\alpha$ For Bb to be a white dwarf since there is significant barium enhancement in $\alpha$ For Ba compared to the primary A component. Barium is a red-giant nucleosynthesis product \citep{jorissen2019}, and thus the material must have been deposited on the K dwarf during a mass-transfer episode that occurred back when the now white dwarf was a giant. This mass transfer likely reignited the K dwarf's activity. At a distance of $d = 14.1$~pc and with a period of 3.75 days \citep{alpha-for_white-dwarf_fuhrmann2016}, this makes the subsystem $\alpha$ For Ba-Bb one of the nearest white dwarf in a close binary.

An important question, therefore, is the role of the white dwarf (WD) in generating the radio emission of $\alpha$ For B. Currently, there are only two known radio-bright WD systems, excluding a class of variable stars known as cataclysmic variables \citep{mason2007}. The first system discovered is AR Scorpii (AR Sco; \citealp{AR-Sco-nature_Marsh2016}), which is a WD/M-dwarf close binary with a period of 3.56 hours. The emission from AR Sco pulses over a broad range of wavelengths (from radio to X-ray) on a period of 1.97 min, corresponding to the beat frequency between the 1.95-min spin period of the white dwarf and the 3.56-hour orbital period, thus earning the moniker "white dwarf pulsar". 
Very recently, \cite{pelisoli2023} discovered another such system: J191213.72-441045.1 (J1912-4410) consists of a WD and an M dwarf in a 4.03-hour orbit, exhibiting similar pulsed emission with a period of 5.30 mins. Despite a large number of proposed models (e.g. \citealp{geng2016, katz2017, garnavich2019}) to understand its peculiar observed behaviour and follow-up observations of this system (e.g. \citealp{marcote2017,stanway2018}), however, the exact radio emission mechanism of AR Sco remains unknown. If $\alpha$ For B turns out to be a WD-pulsar system, this makes it an extremely unique system since localising the radio emission region within the $\alpha$ For Ba-Bb subsystem and detecting its orbital motion is possible with VLBI telescopes, unlike AR Sco and J1912-4410. This in turn would allow greater understanding of the underlying radio emission mechanism, validate the theoretical models, and show whether this is an unique phenomenon or rather pulsing radio emission is ubiquitous for close binaries that contain a WD.

Lastly, $\alpha$ For B being a radio source is not new -- \cite{gudel1995} observed this star with the VLA at 8.4 GHz in two different epochs (1993 and 1995; respective legacy ID: AG394 and AG458). There is a clear detection in each epoch despite strong variability: around 0.6 mJy in 1993 and 0.3 mJy in 1995, which is $\sim$$20 \sigma$ and $\sim$$10 \sigma$ respectively. More importantly, the radio emission shows strong circular polarisation ($\sim 30 \%$; \citealp{gudel1995}) and no significant linear polarisation ($< 1\%$), both of which are very similar to that seen in AR Sco \citep{stanway2018}. Previously, \cite{gudel1995} failed to ascertain if the radio emission stemmed from $\alpha$ For A or B. With \gaia astrometry and proper-motion values now available, we reimaged the data and can confidently conclude that the $\alpha$ For B is the radio emitter rather than A. Moreover, the K dwarf also exhibits strong X-ray luminosity \citep{1rxs_hunsch1999, 2rxs} compared to a typical isolated K dwarf (e.g. \citealp{fleming1994}), yet another similarity to AR Sco \citep{takata2018}.

\footnotetext{\url{http://cutouts.cirada.ca}}

\subsection{$\alpha^2$~CVn}
Alpha$^2$ Canum Venaticorum ($\alpha^2$~CVn = $\alpha$~CVn A = Cor Caroli = \gaia DR3 1517698716348324992) is the brighter component of the binary system $\alpha$~CVn at 30.6\,pc. This magnetic A0Vp star serves as a prototype for a class of variable stars appropriately named the $\alpha^2$~CVn variables \citep{maury1897}, which describes chemically peculiar main-sequence stars of spectral class B8 to A7 that exhibit photometric and spectroscopic variability \citep{hummerich2016, paunzen2021}.
Their strong magnetic field \citep{shultz2019} makes them capable of of generating both incoherent and coherent radio emission (e.g. \citealp{drake1987, linsky1992, hajduk2022, shultz2022}). In particular, strong circular-polarised electron cyclotron maser (ECM) emission has been detected from these systems (e.g. \citealp{trigilio2000, trigilio2011, leto2020, das2021}).

Here, $\alpha^2$~CVn is present in both LoTSS (with no significant circular polarisation) and VLASS in both epochs, but it was previously detected in other radio frequencies as well (e.g. \citealp{helfand1999, drake2006, leto2021}). \cite{hajduk2022} already investigated the radio emission of $\alpha^2$~CVn detected in LoTSS. They modelled its radio spectrum to constrain the distance at which co-rotation of plasma with the large-scale stellar magnetosphere breaks down, and confirmed that the spectral behaviour is consistent with gyrosynchrotron mechanism in $\alpha^2$~CVn. This is also consistent with previous studies by \cite{leto2021, owocki2022, shultz2022}, where centrifugal breakout is thought to drive magnetic reconnection, which is responsible the acceleration of radio-emitting gyrosynchrotron. However, the centrifugal breakout model has been challenged due to the lack of H$\alpha$ emission or shell absorption in $\alpha^2$~CVn \citep{pfeffer2022}.

One intriguing point regarding $\alpha^2$~CVn is that it deviates from GBR significantly in both 144-MHz and 3-GHz band; it has a radio spectral luminosity and X-ray luminosity very similar to a non-early M dwarf. This might be suggesting that its predominant radio emission mechanism may also be ECMI similar to that of UCDs. \cite{hajduk2022} determined the circular polarisation at 144\,MHz to be significantly below 60\% from the LoTSS detection, so $\alpha^2$~CVn may have mildly circularly polarised emission instead. UCDs have been observed to emit mildly circularly polarised quasi-quiescent radiation in the past \citep{berger2002, mclean2012, williams2014}, which is thought to be ECM emission that appears only mildly polarised because of depolarisation effect during propagation in the magnetosphere \citep{hallinan2008}. $\alpha^2$~CVn deviates from GBR in both LoTSS and VLASS, but the deviation is slightly more significant in the VLA band. This may imply that its cyclotron emission could be stronger in the GHz regime. This is not surprisingly given the strong organised magnetic field of $\alpha^2$~CVn \citep{kochukhov2010}. Stokes V information of this VLASS source would be very useful in discerning the underlying mechanism of $\alpha^2$~CVn's radio emission.

\begin{figure*}[h]
    \centering
    \begin{subfigure}{.5\textwidth}
          \centering
          \includegraphics[width=\linewidth]{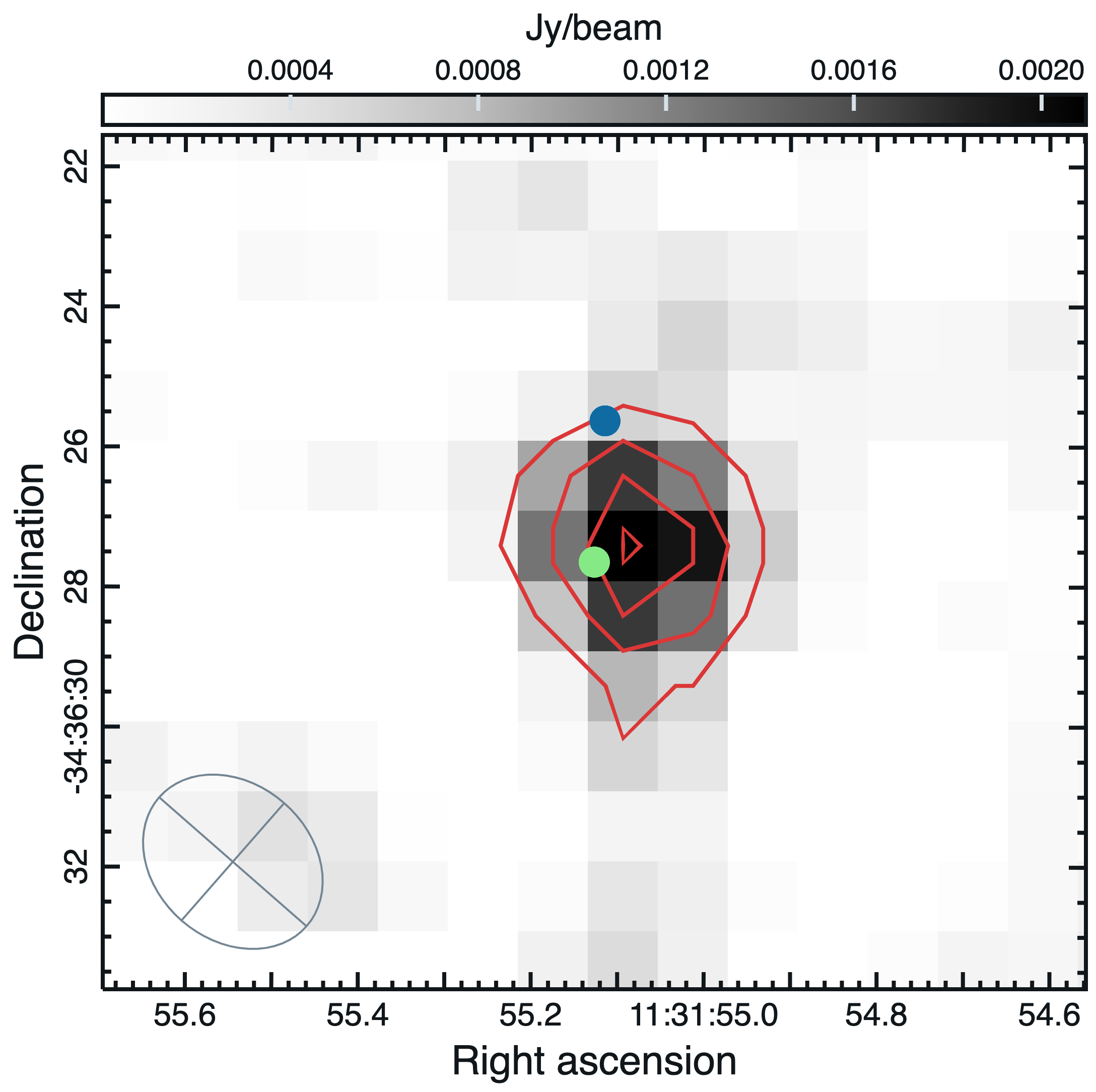}
    \end{subfigure}%
    \begin{subfigure}{.5\textwidth}
          \centering
          \includegraphics[width=\linewidth]{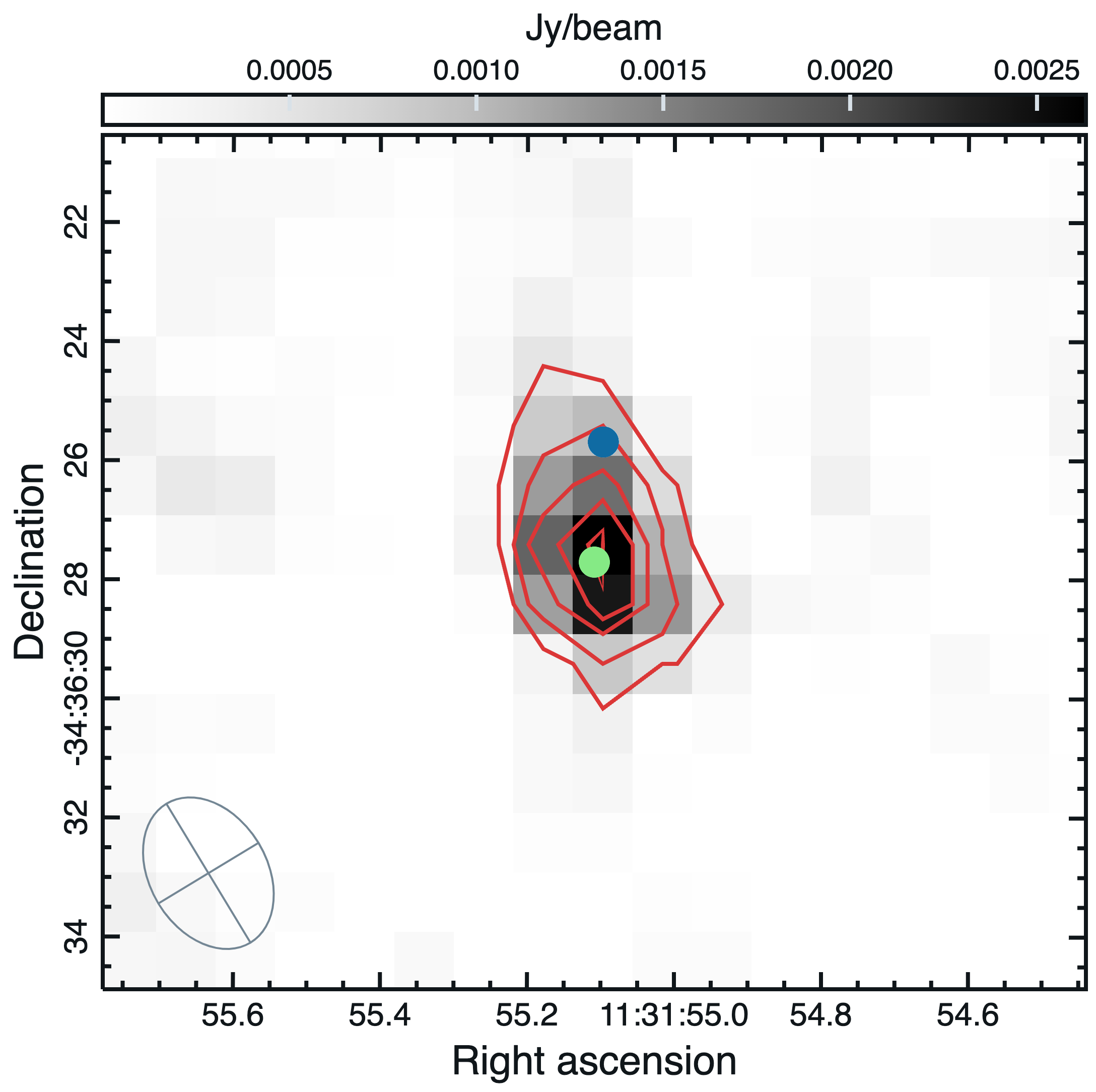}
    \end{subfigure}
    \caption[Caption for LOF]{
    VLASS Epoch 1 (left panel) and Epoch 2 (right panel) images using CIRADA cutout service\footnotemark, centred at the corresponding radio source. In both images, the lime dot represents the \gaia position of TWA 5 A with proper-motion correction, and the blue dot represents the position of TWA 5 B using astrometry by \cite{twa-5_neuhauser2000}. The FWHM beam size is shown in the bottom left corner of each image as a grey crossed circle. The rms noise $\sigma$ is around 150\,\si{\mu Jy} in Epoch 1 and 130\,\si{\mu Jy} in Epoch 2. The contours are of levels with a increment of 4$\sigma$, i.e. 4$\sigma$, 8$\sigma$, 12$\sigma$, etc. The peak flux of the source is around 2.5\,mJy for Epoch 1 and 2.7\,mJy for Epoch 2.
    }
    \label{fig:twa-5-vlass-image}
\end{figure*}

\subsection{TWA 5 A}
\label{app:twa-5}
The furthest radio source in our VLASS $\times$ GCNS sample (at 49.7\,pc) is discovered to be a stellar system called TWA 5 A (= CD-33$\deg$7795), which is one of the five original kinematic members of the TW Hydrae association (TWA; \citealp{kastner1997}): a group of very young ($\sim 5-15$ Myr) low-mass stars and substellar objects (e.g. \citealp{twa-neuh2010}). The system consists of the primary T Tauri star (TTS) and a brown dwarf companion (TWA 5 B) located approximately $2\arcsec$ away \citep{webb1999, lowrance1999}. TWA 5 B is of spectral type M8.5 and has an estimated mass between $\sim 15-40 M_J$ \citep{twa-5_neuhauser2000}. TWA 5 B is not present in the \gaia catalogues. Their very far ($>100$\,AU) separation precludes any magnetic interactions between them.
The VLASS images in both epochs show that the radio source is located much closer to A than to B (see Fig.~\ref{fig:twa-5-vlass-image}). Therefore, we conclude the TTS should indeed be the predominant radio emitter.

The primary TTS itself is a spectroscopic binary \citep{webb1999, muzerolle2000, torres2000}. \cite{macintosh2001} resolved the primary into a 60-mas binary system (TWA 5 Aab) using adaptive optics. \cite{konopacky2007} and later \cite{kohler2013} refined the semi-major axis to be $63.7 \pm 0.2$\,mas. Finally, \cite{kohler2016} improved the orbital parameters and estimated the mass ratio of Aa and Ab to be $M_{Ab}/M_{Aa} = 0.93 \pm 0.13$, making them very similar in mass. Therefore, these two components -- Aa and Ab -- are also equally likely to be radio emitters.

Intriguingly, there are previous attempts at detecting radio emission from this peculiar system. \cite{osten2006} searched for radio emission from young brown dwarfs in X band, with TWA 5 B being one of them, and did not detect any source in the map at the positions of TWA 5 B in 2005. They also looked at an archival observation made in 2002 and found no detection either. These two non-detections imply a radio luminosity upper limit of this system $L_{\nu, \rm{rad}} < 2-3 \times 10^{14} \si{erg~s^{-1} Hz^{-1}}$ at 8.4\,GHz \citep{osten2006}. And yet, TWA 5 A has very significant radio emission ($> 10$$\sigma$) in both epochs of VLASS, giving a radio luminosity $L_{\nu, \rm{rad}} \sim 10^{16} \si{erg~s^{-1} Hz^{-1}}$ at 3\,GHz. This radio information thus gives us two models for the radio emission of TWA 5 A: (i) gyrosynchrotron emission that peaks around 3\,GHz and quickly drops off beyond the turnover frequency; and (ii) ECM emission which cuts off before the X band.

\footnotetext{\url{http://cutouts.cirada.ca}}

\section{V-LoTSS survey completeness}
\label{app:incompleteness}

One peculiarity regarding the V-LoTSS $\times$ GCNS sample is the small amount of V-LoTSS stellar detections outside of the 50-parsec sphere; only 2 out of the 22 sources are beyond 50\,pc: EV Draconis which is a RS CVn system and the isolated F star HD 220242. One might simply conclude that a lack of detections is due to incompleteness of a flux-limited survey. However, since the GCNS is $>99\%$ complete within 100\,pc up to spectral type M6, the incompleteness must stem from V-LoTSS.

To test if that is the case, we assume the CAS systems (both RS CVn and BY Dra variables) from the V-LoTSS $\times$ GCNS sample to have prototypical luminosities. Then, we perform a Monte Carlo simulation by duplicating these stellar objects of particular luminosity and randomly distributing them in a 100-parsec sphere. Finally, we determine how many of them should be recovered given the survey's flux density detection threshold. The $\approx$100\% completeness of V-LoTSS was calculated to be 1.0\,mJy by injecting and recovering point sources \citep{v-lotss}, and thus we use this value as the detection threshold. The simulation gives us the expected ratio between detected sources within and beyond 50\,pc.

Finally, to determine how likely it is to only detect 1 CAS system beyond 50\,pc given that there are 6 detections within 50\,pc, we simply compute the $p$-value based on a Poisson distribution:
\begin{equation}
    \Pr(X_{\rm >50\,pc}=k) = \frac{\lambda^k e^{-\lambda}}{k!},
\end{equation}
where $k$ and $\lambda$ and the observed and expected number of radio detections beyond 50\,pc respectively. Here, $k=1$ and we compute $\lambda=13.3$. This gives us a probability of $0.0022\%$, which is the likelihood that incompleteness can explain the lack of sources beyond 50\,pc.
Following the same procedures for the case of M dwarfs, however, gives a result consistent with observation: we obtain $\lambda = 0$, i.e. we expect no radio detections from M dwarfs outside of the 50-pc sphere. 

In addition, we also perform the $V/V_{\max}$ statistical test to see if it agrees with the hypothesis that survey incompleteness alone cannot explain the lack of CAS detections. This test (also known as luminosity–volume test) was first introduced by \cite{v-vmax_schmidt1968} as a measure of the uniformity of the space distribution of radio sources. Originally, the test was to study the cosmic evolution prove that quasars are not at all uniformly distributed in space and the density of radio quasar evolves with distance. However, it is now also widely used also to test for volume completeness of a flux-limited population detected in a survey (e.g. \citealp{locatelli2019}).

The test is as follows:
let $V$ be the volume enclosed by a sphere with a radius equal to the distance $d$ to a radio source, and $V_{\max}$ be the volume enclosed within the maximum distance at which the source would be detectable. Then, for each source, one can calculate the ratio $V/V_{\max}$. If the sources are indeed uniformly distributed and their properties do not evolve with distance, the sample should have $V/V_{\max}$ uniformly distributed between 0 and 1 with an average value $\langle V/V_{\max} \rangle = 0.5$ in Euclidean space. In a flux-limited survey, this ratio can be expressed as
\begin{equation}
    \left\langle \frac{V}{V_{\max}} \right\rangle = \frac{1}{N} \sum^{N}_{i} \left( \frac{S_{\min}}{S_i} \right)^{1.5},
    \label{eq:v-vmax}
\end{equation}
where $N$ is the number of objects in the sample, $S_i$ is the flux density of the $i^{\rm th}$ source, and $S_{\min}$ is the detection threshold of the survey.
This powerful test makes no assumption on the luminosity function, and thus should work on any flux-limited population.

Now, we calculate the $\langle V/V_{\max} \rangle$ for the V-LoTSS sample.
For the non-CAS systems, we obtain $\langle V/V_{\max} \rangle = 0.355 \pm 0.075$, an inconsistency at the $\approx$2$\sigma$ level. For the CAS systems, we obtain $\langle V/V_{\max} \rangle = 0.450 \pm 0.116$ which is consistent with the expected value of $0.5$. The uncertainties are obtained via bootstrapping. In conclusion, there may be be some incompleteness in the V-LoTSS M-dwarfs population but the result is marginal. As such, we proceed with the assumption that we are not missing a significant population in our crossmatching.

\section{Table of Acronyms}

Table~\ref{tab:acronyms} contains the definitions of all the acronyms frequently used in this work.

\begin{table}
    \centering
    \begin{tabular}{c|l}
        \toprule
        Acronym & Definition \\
        \midrule
        CAS & Chromospherically active stellar \\
        CDF & Cumulative distribution function \\
        CvM & Cramér–von Mises \\
        ECM & Electron cyclotron maser \\
        ECMI & ECM instability \\
        FIRST & Faint Images of the Radio Sky at Twenty-Centimeters \\
        GB & G\"{u}del-Benz\\
        GBR & GB relationship\\
        GCNS & \gaia Catalogue of Nearby Stars \\
        LOFAR & Low-Frequency Array \\
        LoTSS & LOFAR Two-metre Sky Survey \\
        TESS & Transiting Exoplanet Survey Satellite \\
        UCD & Ultracool dwarf \\
        V-LoTSS & The circularly polarised component of LoTSS \\
        VLA & Karl G. Jansky Very Large Array \\
        VLASS & VLA Sky Survey \\
        YMG & Young moving group \\
        YSO & Young stellar object \\
        ZDI & Zeeman-Doppler imaging \\
        \bottomrule
    \end{tabular}
    \caption{Table of acronyms frequently used in this work.}
    \label{tab:acronyms}
\end{table}

\pagebreak

\setcounter{table}{0}
\renewcommand{\thetable}{A.\arabic{table}}

\begin{table*}[]
    \caption{Multi-wavelength properties of the 144-MHz detected LoTSS $\times$ GCNS population sorted by increasing distance $d$. The symbols $G_{BP}-G_{RP}$, $S_I$, and $\log L_X$ correspond to \gaia\ BP--RP colour, Stokes I flux density, and logarithmic soft (0.1--2.4 keV) X-ray luminosity respectively.}
    \centering
    \begin{tabular}{l l l r c c c}
        \hline
        \toprule
        Common & Object$^{(a)}$ & Spectral & \multicolumn{1}{c}{$d$\rlap{$\,^{(b)}$}} & $G_{BP}-G_{RP}$\rlap{$^{(b)}$} & \multicolumn{1}{c}{$S_I$\rlap{$^{(c)}$}} & \multicolumn{1}{c}{${\log L_X}$\rlap{$^{(d)}$}}\\
        name & type & type & \multicolumn{1}{c}{(pc)} & (mag) & \multicolumn{1}{c}{(mJy)} &  \multicolumn{1}{c}{($\si{erg~s^{-1}})$}\\
        \midrule
        WX UMa & Eruptive* & M6V$^{(1)}$ & 4.91 & 3.85 & $1.95 \pm 0.24$ & 27.56 \\
        EQ Peg A & Eruptive* & M3.5Ve$^{(2)}$ & 6.26 & 2.66 & $1.03 \pm 0.20$ & 28.94 \\
        GJ 1151 & * & M4.5V$^{(3)}$ & 8.04 & 3.14 & $0.48 \pm 0.11$ & 26.30\rlap{$^{(26)}$} \\
        V374 Peg & Eruptive* & M3.5Ve$^{(4)}$ & 9.10 & 2.87 & $1.09 \pm 0.33$ & 28.48 \\
        BL Lyn & BY Dra & M3.5V$^{(5)}$ & 11.99 & 2.77 & $0.80 \pm 0.21$ & 29.11 \\
        CW UMa & Eruptive* & M3.5V$^{(6)}$ & 13.35 & 2.69 & $0.75 \pm 0.11$ & 28.76 \\
        YY Gem & RS CVn & M1Ve+M1Ve$^{(7)}$ & 15.08 & 1.93 & $1.97 \pm 0.28$ & 29.88 \\
        GJ 856 A \& B$^\dag$ & * & M3.5V \& M3.5V$^{(8)}$ & 15.29 & 2.57 & $1.00 \pm 0.20$ & 29.42 \\
        LP 212-62 & * & M5V$^{(9)}$ & 18.19 & 3.40 & $5.42 \pm 0.19$ & 27.58\rlap{$^{(26)}$} \\
        GJ 3861 & Eruptive* & M3.0Ve$^{(10)}$ & 18.46 & 2.44 & $0.80 \pm 0.15$ & 28.53 \\
        CR Dra & * & M1.5Ve$^{(11)}$ & 20.15 & 2.17 & $0.83 \pm 0.23$ & 29.56 \\
        GJ 9603 & * & M2V$^{(12)}$ & 21.49 & 2.00 & $1.37 \pm 0.38$ & 27.98 \\
        $\sigma^2$ CrB (TZ CrB) & RS CVn & F6V+G0V$^{(13)}$ & 22.70 & 0.81 & $4.45 \pm 0.18$ & 30.71 \\
        BF Lyn & BY Dra & K3V+K3V$^{(14)}$ & 23.44 & 1.24 & $0.75 \pm 0.14$ & 30.22 \\
        GJ 3729 & * & M3.5V$^{(15)}$ & 23.51 & 2.76 & $0.60 \pm 0.11$ & 28.88 \\
        2MASS J10580751+5057028 & * & M3.5V$^{(16)}$ & 27.67 & 2.62 & $0.29 \pm 0.13$ & 28.68 \\
        $\alpha^2$~CVn & $\alpha^2$~CVn & A0VpSiCrEuHg$^{(17)}$ & 30.56 & 0.22 & $0.41 \pm 0.12$ & 28.89 \\
        V835 Her & BY Dra & K1V+K7V$^{(18)}$ & 31.29 & 1.09 & $0.45 \pm 0.12$ & 30.26 \\
        2MASS J09481615+5114518 & * & M4.5V$^{(19)}$ & 35.97 & 3.52 & $1.13 \pm 0.10$ & 27.45\rlap{$^{(26)}$} \\
        LP 259-39 & * & M4V$^{(20)}$ & 36.91 & 3.25 & $0.57 \pm 0.17$ & \llap{$<~$}29.27\rlap{$^{(26)}$} \\
        2MASS J13035648+4815197 & * & M3V$^{(21)}$ & 38.65 & 2.62 & $1.12 \pm 0.23$ & \llap{$<~$}28.33\rlap{$^{(26)}$} \\
        II Peg & RS CVn & K2IV+M0--3V$^{(22)}$ & 39.13 & 1.38 & $1.21 \pm 0.30$ & 31.17 \\
        LP 93-440 & * & M4.5V$^{(23)}$ & 42.22 & 3.38 & $0.44 \pm 0.12$ & \llap{$<~$}28.33\rlap{$^{(26)}$} \\
        BH CVn (HR 5110) & RS CVn & F2IV+K2IV$^{(24)}$ & 46.89 & 0.65 & $1.88 \pm 0.33$ & 30.75 \\
        2MASS J14333139+3417472 & * & M5V$^{(25)}$ & 48.20 & 3.75 & $0.95 \pm 0.13$ & 27.92\rlap{$^{(26)}$} \\
        \bottomrule
    \end{tabular}
    \label{tab:lotss-table}
    
\tablefoot{ 
\tablefoottext{a}{
Asterisk * denotes M dwarfs unless specified otherwise. ``Eruptive*'' denotes eruptive variable stars classified by SIMBAD.
}
\tablefoottext{b}{
Distance $d$ (derived from the parallaxes) and \gaia\ colour $G_{BP}-G_{RP}$ of each stellar systems are given by the \gaia\ Catalogue of Nearby stars \citep{gcns} which quotes measurements from \gaia\ Early Data Release 3 \citep{gaia-edr3}.
}
\tablefoottext{c}{
Stokes I flux density $S_I$ is given by LoTSS-DR2 \citep{lotss-dr2}.
}
\tablefoottext{d}{
X-ray luminosity $X_L$ of each stellar system is given by the Second ROSAT all-sky survey source catalogue (2RXS; \citealp{2rxs}), unless specified otherwise. The derivation of the $X_L$ upper-limit values for some stellar systems in our sample is described in detail by \cite{joe_nature}.
}
$^\dag$~Both M-dwarf components of the binary system GJ 856 have very similar angular separation ($\lesssim 1\arcsec$) with the LoTSS radio source. Since the GJ 856 A and GJ 856 B themselves are separated by around 1.25\arcsec, we cannot ascertain which one is (or if both are) emitting the radio signal. Their wide separation ($> 10$\,AU) precludes direct magnetic or chromospheric interactions between them (cf. CAS systems).
}
\tablebib{ 
(1) \cite{reid1995}; (2) \cite{davison2015}; (3) \cite{houdebine2019}; (4) \cite{davison2015}; (5) \cite{carmenes_alonso2015}; (6) \cite{carmenes_cifuentes2020}; (7) \cite{yygem_qian2002}; (8) \cite{fouque2018}; (9) \cite{carmenes_alonso2015}; (10) \cite{hejazi2020}; (11) \cite{petit2014}; (12) \cite{kharchenko2001}; (13) \cite{strassmeier2003}; (14) \cite{strassmeier1989}; (15) \cite{lepine2013}; (16) \cite{terrien2015}; (17) \cite{alpha2-cvn_bychkov2021}; (18) \cite{osten1998}; (19) \cite{cook2016}; (20) \cite{speculoo_sebastian2021}; (21) \cite{birky2020}; (22) \cite{ii-peg_berdyugina1998}; (23) \cite{cook2016}; (24) \cite{bh-cvn_eker1987}; (25) \cite{pickles2010}; (26) \cite{joe_nature}
}
\end{table*}

\begin{table*}[]
    \caption{Multi-wavelength properties of the 144-MHz detected V-LoTSS $\times$ GCNS population sorted by increasing distance $d$. The symbols $G_{BP}-G_{RP}$, $S_V$, $S_V/S_I$, and $\log L_X$ correspond to \gaia\ BP--RP colour, Stokes V flux density, fraction of circularly polarised radio emission, and logarithmic soft (0.1-2.4 keV) X-ray luminosity respectively.}
    \centering
    \begin{tabular}{l l l r c r c c}
        \hline
        \toprule
        Common & Object$^{(a)}$ & Spectral & \multicolumn{1}{c}{$d$\rlap{$\,^{(b)}$}} & $G_{BP}-G_{RP}$\rlap{$^{(b)}$} & \multicolumn{1}{c}{$S_V$\rlap{$^{(c)}$}} & \multicolumn{1}{c}{$S_V/S_I$\rlap{$^{(c)}$}} & \multicolumn{1}{c}{${\log L_X}$\rlap{$^{(d)}$}}\\
        name & type & type & \multicolumn{1}{c}{(pc)} & (mag) & \multicolumn{1}{c}{(mJy)} & \multicolumn{1}{c}{(\%)} &  \multicolumn{1}{c}{($\si{erg~s^{-1}})$}\\
        \midrule
        WX UMa & Eruptive* & M6V$^{(1)}$ & 4.91 & 3.85 & $-0.86 \pm 0.10$ & $40 \pm 9$ & 27.56 \\
        GJ 1151 & * & M4.5V$^{(2)}$ & 8.04 & 3.14 & $0.53 \pm 0.07$ & $40 \pm 8$ & 26.30\rlap{$^{(23)}$} \\
        GJ 450 & * & M1.5V$^{(3)}$ & 8.77 & 2.16 & $0.40 \pm 0.06$ & $54 \pm 23$ & 27.82 \\
        LP 169-22 & * & M5.5V$^{(4)}$ & 10.47 & 3.81 & $-0.64 \pm 0.13$ & $60 \pm 29$ & \llap{$<~$}26.48\rlap{$^{(23)}$} \\
        CW UMa & Eruptive* & M3.5V$^{(5)}$ & 13.35 & 2.69 & $-1.53 \pm 0.14$ & $77 \pm 10$ & 28.76 \\
        YY Gem & RS CVn & M1Ve+M1Ve$^{(6)}$ & 15.08 & 1.93 & $-0.87 \pm 0.13$ & $47 \pm 10$ & 29.88 \\
        HAT 182-00605 & * & M4V$^{(7)}$ & 17.90 & 2.92 & $-0.73 \pm 0.11$ & $82 \pm 25$ & 28.53 \\
        LP 212-62 & * & M5V$^{(8)}$ & 18.19 & 3.40 & $-6.63 \pm 0.13$ & $75 \pm 3$ & 27.58\rlap{$^{(23)}$} \\
        GJ 3861 & Eruptive* & M3.5V$^{(9)}$ & 18.46 & 2.44 & $0.73 \pm 0.08$ & $79 \pm 16$ & 28.53 \\
        CR Dra & * & M1.5Ve$^{(10)}$ & 20.15 & 2.17 & $0.34 \pm 0.07$ & $56 \pm 26$ & 29.56 \\
        $\sigma^2$ CrB (TZ CrB) & RS CVn & F6V+G0V$^{(11)}$ & 22.70 & 0.81 & $-6.07 \pm 0.19$ & $81 \pm 4$ & 30.71 \\
        BF Lyn & BY Dra & K3V+K3V$^{(12)}$ & 23.44 & 1.24 & $1.37 \pm 0.15$ & $72 \pm 13$ & 30.22 \\
        GJ 3729 & * & M3.5V$^{(13)}$ & 23.51 & 2.76 & $-1.61 \pm 0.17$ & $113 \pm 18$ & 28.88 \\
        2MASS J09481615+5114518 & * & M4.5V$^{(14)}$ & 35.97 & 3.53 & $1.86 \pm 0.09$ & $94 \pm 8$ & 27.45\rlap{$^{(23)}$} \\
        $\sigma$~Gem & RS CVn & K1IIIe+dG/K$^{(15)}$ & 36.87 & 1.35 & $1.27 \pm 0.26$ & $45 \pm 13$ & 31.07 \\
        LP 259-39 & * & M4V$^{(16)}$ & 36.91 & 3.25 & $-0.39 \pm 0.06$ & $71 \pm 24$ & \llap{$<~$}29.27\rlap{$^{(23)}$} \\
        II Peg & RS CVn & K2IV+M0--3V$^{(17)}$ & 39.12 & 1.38 & $2.66 \pm 0.14$ & $71 \pm 8$ & 31.17 \\
        2MASS J10534129+5253040 & * & M4V$^{(18)}$ & 45.08 & 2.78 & $-0.86 \pm 0.14$ & $83 \pm 25$ & 29.45 \\
        BH CVn (HR 5110) & RS CVn & F2IV+K2IV$^{(19)}$ & 46.89 & 0.65 & $1.12 \pm 0.09$ & $54 \pm 11$ & 30.75 \\
        2MASS J14333139+3417472 & * & M5V$^{(20)}$ & 48.21 & 3.75 & $0.96 \pm 0.11$ & $65 \pm 11$ & 27.92\rlap{$^{(23)}$} \\
        EV Dra & RS CVn & G5V+K0V$^{(21)}$ & 57.54 & 0.94 & $-2.59 \pm 0.15$ & $85 \pm 9$ & 30.44 \\
        HD 220242 & * (F-type) & F5V$^{(22)}$ & 69.33 & 0.54 & $2.19 \pm 0.29$ & $78 \pm 16$ & 29.00 \\
        \bottomrule
    \end{tabular}
    \label{tab:v-lotss-table}
    
\tablefoot{ 
\tablefoottext{a}{
Asterisk * denotes M dwarfs unless specified otherwise. ``Eruptive*'' denotes eruptive variable stars classified by SIMBAD.
}
\tablefoottext{b}{
Distance $d$ (derived from the parallaxes) and \gaia\ colour $G_{BP}-G_{RP}$ of each stellar systems are given by the \gaia\ Catalogue of Nearby stars \citep{gcns} which quotes measurements from \gaia\ Early Data Release 3 \citep{gaia-edr3}.
}
\tablefoottext{c}{
Stokes V flux density $S_V$ and fraction of circularly polarised radio emission $S_V/S_I$ are given by the V-LoTSS catalogue \citep{v-lotss}.
}
\tablefoottext{d}{
X-ray luminosity $X_L$ of each stellar system is given by the Second ROSAT all-sky survey source catalogue (2RXS; \citealp{2rxs}), unless specified otherwise. The derivation of the $X_L$ upper-limit values for some stellar systems in our sample is described in detail by \cite{joe_nature}.
}
}
\tablebib{ 
(1) \cite{reid1995}; (2) \cite{houdebine2019}; (3) \cite{brown2022}; (4) \cite{speculoo_sebastian2021}; (5) \cite{carmenes_cifuentes2020}; (6) \cite{yygem_qian2002}; (7) \cite{speculoo_sebastian2021}; (8) \cite{carmenes_alonso2015}; (9) \cite{hejazi2020}; (10) \cite{petit2014}; (11) \cite{strassmeier2003}; (12) \cite{strassmeier1989}; (13) \cite{lepine2013}; (14) \cite{cook2016}; (15) \cite{makarov2003}; (16) \cite{speculoo_sebastian2021}; (17) \cite{ii-peg_berdyugina1998}; (18) \cite{galgano2020}; (19) \cite{bh-cvn_eker1987}; (20) \cite{pickles2010}; (21) \cite{EV-Dra_fekel2005}; (22) \cite{kharchenko2001}; (23) \cite{joe_nature}
}
\end{table*}

\onecolumn
\begin{longtable}{l l l r c r c c}
    \caption{Multi-wavelength properties of the 3-GHz detected VLASS $\times$ GCNS population sorted by increasing distance $d$. The symbols $G_{BP}-G_{RP}$, $S_I$, and $\log L_X$ correspond to distance, \gaia\ BP--RP colour, Stokes I flux density, and logarithmic soft (0.1-2.4 keV) X-ray luminosity respectively. The last column refers to the VLASS epoch from which the radio data is used for this study, i.e. whether the value of $S_I$ is quoted from VLASS Epoch 1 or 2. \label{tab:vlass-table}}\\

    \hline
    \toprule
    Common & Object$^{(a)}$ & Spectral & \multicolumn{1}{c}{$d$\rlap{$\,^{(b)}$}} & $G_{BP}-G_{RP}$\rlap{$^{(b)}$} & \multicolumn{1}{c}{$S_I$\rlap{$^{(c)}$}} & \multicolumn{1}{c}{$\log L_X$\rlap{$^{(d)}$}} & Epoch\rlap{$^{(e)}$}\\
    name & type & type & \multicolumn{1}{c}{(pc)} & (mag) & \multicolumn{1}{c}{(mJy)} & \multicolumn{1}{c}{($\si{erg~s^{-1}})$} & \\[0.5ex] 
    \midrule
    \endfirsthead
    Continuation of Table~\ref{tab:vlass-table}\\
    \hline
    \toprule
    Common & Object$^{(a)}$ & Spectral & \multicolumn{1}{c}{$d$\rlap{$\,^{(b)}$}} & $G_{BP}-G_{RP}$\rlap{$^{(b)}$} & \multicolumn{1}{c}{$S_I$\rlap{$^{(c)}$}} & \multicolumn{1}{c}{$\log L_X$\rlap{$^{(d)}$}} & Epoch\rlap{$^{(e)}$}\\
    name & type & type & \multicolumn{1}{c}{(pc)} & (mag) & \multicolumn{1}{c}{(mJy)} & \multicolumn{1}{c}{($\si{erg~s^{-1}})$} & \\[0.5ex] 
    \midrule
    \endhead
    
    \hline
    Continued on the next page
    \endfoot

    \hline \\
    \multicolumn{8}{l}{
    \tablefoot{ 
    \tablefoottext{a}{
    Asterisk * denotes M dwarfs unless specified otherwise. ``Eruptive*'' denotes eruptive variable stars classified by SIMBAD.
    }
    \tablefoottext{b}{
    Distance $d$ (derived from the parallaxes) and \gaia\ colour $G_{BP}-G_{RP}$ of each stellar systems are given by the \gaia\ Catalogue of Nearby stars \citep{gcns} which quotes measurements from \gaia\ Early Data Release 3 \citep{gaia-edr3}.
    }
    \tablefoottext{c}{
    Total flux density $S_I$ is given by VLASS Quick Look (QL) catalogue \citep{vlass-ql-gordon2021}. $S_I$ from Epoch 1 is adjusted to $S_{\rm VLASS~Ep.1} = 0.87 S_I$, as the Epoch 1 flux densities in the VLASS Quick Look imaging are known to be systematically underestimated due to multiple calibration issues (e.g. an antenna pointing error that affected two-thirds of the antennas), which were fixed for subsequent campaigns of VLASS. See \cite{memo13_lacy2019} and \cite{vlass-ql-gordon2021} for more details. The Stokes I flux density from Epoch 2 is unaffected by these issues, i.e. $S_{\rm VLASS~Ep.2} = S_I$.
    }
    \tablefoottext{d}{
    X-ray luminosity $X_L$ of each stellar system is given by the Second ROSAT all-sky survey source catalogue (2RXS; \citealp{2rxs}), unless specified otherwise. The upper limits of $X_L$ for LP 732-30 and 2MASS J13582164-0046262 are derived assuming the 2RXS flux limit $\approx 10^{-13}~\si{erg~cm^{-2}~s^{-1}}$ \citep{2rxs}.
    }
    \tablefoottext{e}{
    In the case where a stellar radio source is detected in both Epochs 1 and 2, we choose the epoch which contains a greater value of $S_I$.
    }
    $^\dag$~For the TWA 5 system, there exists a very young brown dwarf companion TWA 5 B of the primary T Tauri star TWA 5 A, with an angular separation of around 2\arcsec. Their very far separation ($>100$\,AU) precludes any magnetic or chromospheric interactions between them. Therefore, the dominant radio emission must be isolated to either component. Due to their proximity and the fact that these two types of YSOs are both capable of generating strong radio emission, we note that it is possible that TWA 5 B may contribute to the radio emission as well.
    However, since the VLASS radio source is located much closer to A than to B (see Appendix~\ref{app:twa-5} for more details), we conclude the TTS is indeed the predominant radio emitter.
    } 
    } \\
    \multicolumn{8}{l}{
    \tablebib{ 
    (1) \cite{gcvs_samus2017}; (2) \cite{reid1995}; (3) \cite{newton2014}; (4) \cite{newton2014}; (5) \cite{deshpande2012}; (6) \cite{davison2015}; (7) \cite{davison2015}; (8) \cite{ozeren1999}; (9) \cite{davison2015}; (10) \cite{baron2019}; (11) \cite{baron2019}; (12) \cite{newton2014}; (13) \cite{gray2006}; (14) \cite{scholz2005}; (15) \cite{lepine2013}; (16) \cite{alpha-for_white-dwarf_fuhrmann2016}; (17) \cite{shkolnik2009}; (18) \cite{alfonso2012}; (19) \cite{carmenes_alonso2015}; (20) \cite{pokemon_clark2022}; (21) \cite{bowler2019}; (22) \cite{v833-tauro_naftilan1993}; (23) \cite{gigoyan2010}; (24) \cite{shkolnik2009}; (25) \cite{petit2014}; (26) \cite{shkolnik2009}; (27) \cite{v775-her_alekseev2000}; (28) \cite{strassmeier2003}; (29) \cite{gray2003}; (30) \cite{shkolnik2009}; (31) \cite{newton2014}; (32) \cite{kozhevnikova2015}; (33) \cite{v711-tau_garc2003}; (34) \cite{alpha2-cvn_bychkov2021}; (35) \cite{torres2006}; (36) \cite{v815-her_fekel2005}; (37) \cite{tokovinin2018}; (38) \cite{bowler2019}; (39) \cite{beta-pic_shkolnik2017}; (40) \cite{cutispoto1995}; (41) \cite{shkolnik2009}; (42) \cite{reid2007}; (43) \cite{cutispoto1999}; (44) \cite{carmenes_alonso2015}; (45) \cite{lepine2013}; (46) \cite{makarov2003}; (47) \cite{ii-peg_berdyugina1998}; (48) \cite{pourbaix2004}; (49) \cite{carmenes_alonso2015}; (50) \cite{riaz2006}; (51) \cite{vy-ari_favata1997}; (52) \cite{ar-lac_zboril2005}; (53) \cite{v1044-sco_strassmeier2012}; (54) \cite{reid2005}; (55) \cite{pribulla2009}; (56) \cite{ar-psc_fekel1996}; (57) \cite{gagne2015}; (58) \cite{riedel2017}; (59) \cite{avvakumova2013}; (60) \cite{bh-cvn_eker1987}; (61) \cite{patel2016}; (62) \cite{carmenes_alonso2015}; (63) \cite{wang2022}; (64) \cite{riaz2006}; (65) \cite{twa-5_neuhauser2000}; (66) \cite{freund2022}; (67) \cite{joe_nature}; (68) \cite{schmitt1995}; (69) \cite{schmitt1995}; (70) \cite{berger2008}; (71) \cite{4xmm-dr11_webb2020}; (72) \cite{joe_nature};
    }
    }
    \endlastfoot

        UV Cet & Eruptive* & M5.5Ve$^{(1)}$ & 2.67 & 3.83 & $2.32 \pm 0.31$ & 27.66\rlap{$^{(66)}$} & 2\\
        WX UMa & Eruptive* & M6V$^{(2)}$ & 4.91 & 3.85 & $7.64 \pm 0.22$ & 27.56\rlap{$^{(67)}$} & 2\\
        GJ 1116 B & * & M7V$^{(3)}$ & 5.10 & 4.04 & $4.99 \pm 1.07$ & 27.90\rlap{$^{(68)}$} & 1\\
        GJ 1116 A & * & M7V$^{(4)}$ & 5.15 & 3.78 & $1.32 \pm 0.21$ & 27.90\rlap{$^{(69)}$} & 2\\
        LSR J1835+3259 & * & M8.5V$^{(5)}$ & 5.69 & 5.15 & $1.30 \pm 0.35$ & \llap{$<~$}24.51\rlap{$^{(70)}$} & 1\\
        GJ 896 A & Eruptive* & M3.5Ve$^{(6)}$ & 6.26 & 2.66 & $3.43 \pm 0.24$ & $28.94$ & 2\\
        GJ 4274 & Eruptive* & M4.5Ve$^{(7)}$ & 7.23 & 3.30 & $1.04 \pm 0.26$ & $27.92$ & 2\\
        FK Aqr & BY Dra & M2Ve+M3Ve$^{(8)}$ & 8.90 & 2.27 & $4.20 \pm 0.28$ & $29.42$ & 2\\
        V374 Peg & Eruptive* & M3.5Ve$^{(9)}$ & 9.10 & 2.87 & $1.70 \pm 0.33$ & $28.48$ & 1\\
        AT Mic B & * & M4Ve$^{(10)}$ & 9.81 & 3.13 & $2.34 \pm 0.44$ & $29.54$ & 2\\
        AT Mic A & * & M4Ve$^{(11)}$ & 9.92 & 3.03 & $3.60 \pm 0.50$ & $29.55$ & 1\\
        LSR J0510+2713 & * & M7V$^{(12)}$ & 10.28 & 4.54 & $1.32 \pm 0.31$ & $27.88$ & 2\\
        $\delta$~Cap & Am star & kA5hF0mF2III$^{(13)}$ & 11.64 & 0.69 & $1.48 \pm 0.33$ & $29.33$ & 2\\
        2MASS J05172292-3521545 & * & M4V$^{(14)}$ & 11.70 & 2.81 & $2.17 \pm 0.32$ & $28.98$ & 2\\
        LP 326-38 & * & M4Ve$^{(15)}$ & 13.05 & 2.78 & $2.31 \pm 0.27$ & $28.32$ & 2\\
        $\alpha$~For B & * (K-type) & K2V$^{(16)}$ & 14.08 & 1.11 & $5.22 \pm 0.35$ & $29.75$ & 1\\
        GJ 4338 & * & M4V$^{(17)}$ & 15.05 & 2.94 & $4.08 \pm 0.25$ & $29.21$ & 2\\
        GJ 2123 B & BY Dra & M4Ve$^{(18)}$ & 15.74 & 2.96 & $2.23 \pm 0.57$ & $29.58$ & 1\\
        LP 86-173 & * & M4.5V$^{(19)}$ & 16.28 & 2.87 & $9.68 \pm 0.22$ & $28.65$ & 1\\
        V1274 Her & BY Dra & M3.5V+M4.5V$^{(20)}$ & 16.42 & 3.54 & $1.52 \pm 0.32$ & $28.39$ & 2\\
        HD 43162C & * & M5V$^{(21)}$ & 16.67 & 2.78 & $3.39 \pm 0.31$ & 24.80\rlap{$^{(71)}$} & 1\\
        V833 Tau & BY Dra & K5Ve$^{(22)}$ & 17.40 & 1.38 & $5.64 \pm 0.23$ & $29.90$ & 2\\
        2MASS J14172209+4525461 & * & M5V$^{(23)}$ & 18.99 & 3.06 & $1.96 \pm 0.23$ & $28.44$ & 2\\
        GJ 4185 & Eruptive* & M3.5V$^{(24)}$ & 19.70 & 2.79 & $3.44 \pm 0.31$ & $29.18$ & 1\\
        CR Dra & * & M1.5Ve$^{(25)}$ & 20.15 & 2.17 & $1.85 \pm 0.27$ & $29.56$ & 1\\
        WW PsA & BY Dra & M4.5V$^{(26)}$ & 20.87 & 2.78 & $4.45 \pm 0.26$ & $29.63$ & 1\\
        V775 Her & RS CVn & K0Ve+M3Ve$^{(27)}$ & 21.33 & 1.21 & $1.75 \pm 0.28$ & $29.95$ & 2\\
        $\sigma^2$~CrB (TZ CrB) & RS CVn & F9V+G0V$^{(28)}$ & 22.70 & 0.81 & $4.14 \pm 0.33$ & $30.71$ & 1\\
        $\lambda$~And & RS CVn & G8IV+?$^{(29)}$ & 25.92 & 1.27 & $3.88 \pm 0.20$ & $30.83$ & 2\\
        GJ 4282 A & * & M2.5V$^{(30)}$ & 26.14 & 2.62 & $1.56 \pm 0.26$ & $29.29$ & 2\\
        GJ 3237 & Eruptive* & M5V$^{(31)}$ & 26.78 & 3.03 & $1.91 \pm 0.26$ & $28.50$ & 1\\
        V478 Lyr & RS CVn & G8V+M3V$^{(32)}$ & 27.06 & 0.98 & $3.15 \pm 0.29$ & $30.18$ & 1\\
        V711 Tau & RS CVn & K1IV+G5V$^{(33)}$ & 29.43 & 1.24 & $31.27 \pm 0.28$ & $31.23$ & 2\\
        $\alpha^2$~CVn & $\alpha^2$~CVn & A0VpSiCrEuHg$^{(34)}$ & 30.56 & 0.22 & $1.77 \pm 0.26$ & $28.89$ & 1\\
        HD 314741 & * (K-type) & K5Ve$^{(35)}$ & 31.45 & 1.66 & $2.10 \pm 0.32$ & $29.74$ & 2\\
        V815 Her & RS CVn & G6V+M3V$^{(36)}$ & 32.09 & 0.92 & $2.13 \pm 0.31$ & $30.50$ & 2\\
        $\delta$~And B & * & M2V$^{(37)}$ & 32.34 & 2.29 & $1.43 \pm 0.29$ & $28.46$ & 2\\
        HD 159911 & * (K-type) & K7IV$^{(38)}$ & 33.21 & 1.54 & $3.06 \pm 0.29$ & $30.34$ & 2\\
        2MASS J21100461-1920302 & * & M4V$^{(39)}$ & 33.60 & 3.05 & $2.00 \pm 0.35$ & $29.74$ & 2\\
        V1198 Ori & RS CVn & G6IV+G1V$^{(40)}$ & 33.62 & 0.91 & $1.21 \pm 0.34$ & $30.24$ & 2\\
        2MASS J20003177+5921289 & * & M4V$^{(41)}$ & 34.43 & 2.87 & $0.93 \pm 0.22$ & $29.10$ & 2\\
        2MASS J09165078+2448559 & * & M4.5V$^{(42)}$ & 35.10 & 3.19 & $0.94 \pm 0.26$ & $28.04$ & 2\\
        GS Leo & BY Dra & G9V+K4V$^{(43)}$ & 35.43 & 1.06 & $1.93 \pm 0.29$ & $30.08$ & 1\\
        2MASS J21374019+0137137 & * & M4.5V$^{(44)}$ & 35.91 & 2.97 & $1.96 \pm 0.30$ & $29.40$ & 2\\
        IZ Boo & * & M3Ve$^{(45)}$ & 36.00 & 2.40 & $2.07 \pm 0.24$ & $29.35$ & 1\\
        $\sigma$~Gem & RS CVn & K1IIIe+dG/K$^{(46)}$ & 36.87 & 1.35 & $64.05 \pm 0.26$ & $31.07$ & 1\\
        II Peg & RS CVn & K2IV+M0--3V$^{(47)}$ & 39.13 & 1.38 & $18.51 \pm 0.29$ & $31.17$ & 1\\
        V376 Cep & RS CVn & G4p+?$^{(48)}$ & 39.93 & 0.91 & $4.06 \pm 0.29$ & $30.60$ & 2\\
        2MASS J03510078+1413398 & * & M4.5V$^{(49)}$ & 40.03 & 2.94 & $2.06 \pm 0.51$ & $29.30$ & 2\\
        2MASS J21103096-2710513 & * & M5V$^{(50)}$ & 40.26 & 3.75 & $2.72 \pm 0.35$ & $28.93$ & 1\\
        VY Ari & RS CVn & K3--4V/IV+dM$^{(51)}$ & 41.49 & 1.25 & $8.11 \pm 0.28$ & $31.05$ & 1\\
        AR Lac & RS CVn & G2IV+K0IV$^{(52)}$ & 42.51 & 0.96 & $8.14 \pm 0.21$ & $31.12$ & 2\\
        V1044 Sco & RS CVn & G9V+M0V$^{(53)}$ & 42.91 & 1.10 & $2.34 \pm 0.41$ & $30.28$ & 1\\
        LP 732-30 & * & M4.5V$^{(54)}$ & 44.16 & 3.12 & $11.71 \pm 0.34$ & $\llap{<~}28.37$ & 1\\
        MR Del & BY Dra & K2V+K6V$^{(55)}$ & 45.09 & 1.20 & $1.37 \pm 0.31$ & $30.29$ & 1\\
        AR Psc & RS CVn & KlIV+G7V$^{(56)}$ & 45.49 & 1.14 & $2.90 \pm 0.26$ & $30.98$ & 1\\
        2MASS J13582164-0046262 & * & M5.5$\gamma$$^{(57)}$ & 45.90 & 3.49 & $2.62 \pm 0.61$ & $\llap{<~}28.40$ & 1\\
        2MASS J20100002-2801410 & * & M3Ve$^{(58)}$ & 46.41 & 2.79 & $1.94 \pm 0.43$ & $29.58$ & 2\\
        V818 Tau & RS CVn & G6V+K6V$^{(59)}$ & 46.64 & 0.96 & $1.15 \pm 0.34$ & $29.76$ & 2\\
        BH CVn (HR 5110) & RS CVn & F2IV+K2IV$^{(60)}$ & 46.89 & 0.65 & $17.92 \pm 0.24$ & $30.75$ & 2\\
        V402 Hya & BY Dra & K0--1V$^{(61)}$ & 47.54 & 1.17 & $3.04 \pm 0.41$ & $30.34$ & 1\\
        2MASS J05184455+4629597 & * & M4.5V$^{(62)}$ & 47.85 & 3.02 & $1.07 \pm 0.26$ & $28.55$ & 2\\
        2MASS J14333139+3417472 & * & M7V$^{(63)}$ & 48.20 & 3.75 & $1.38 \pm 0.30$ & 27.92\rlap{$^{(72)}$} & 2\\
        2MASS J05082729-2101444 & * & M5V$^{(64)}$ & 48.30 & 3.37 & $4.42 \pm 0.29$ & $28.88$ & 1\\
        TWA 5 A$^\dag$ & T Tauri & M2Ve$^{(65)}$ & 49.67 & 2.40 & $3.54 \pm 0.27$ & $30.14$ & 2\\
    \bottomrule
\end{longtable}
\twocolumn

\end{document}